\documentclass [11pt, a4paper] {article}

\usepackage {amssymb,amsthm,mathrsfs}
\usepackage [toc,page] {appendix}
\usepackage {preview}
\usepackage {fullpage}
\usepackage [hyperfootnotes] {hyperref}
\newcounter {savecntr}

\newcounter {Hsavecntr}

\newtheorem {theorem}{Proposition}
\newtheorem {lemma}{Lemma} 
\newtheorem {corollary}{Corollary}
\newtheorem {remark}{Remark}
\newtheorem* {definition}{Definition}

\newcommand*\Laplace{\mathop{}\!\mathbin\bigtriangleup}

\def\bsk {\vspace {1.5 cm}}
\def\msk {\vspace {1.0 cm}}
\def\ssk {\vspace {0.2cm}}
\def\ni {\noindent}

\begin {document}

\makeatletter
\begingroup

 \renewcommand\thefootnote{\@fnsymbol\c@footnote}

\bsk
\centerline {\Large {Infrared Gupta-Bleuler Quantum Electrodynamics:}}
\ssk
\centerline {\Large {Solvable Models And Perturbative Expansion}}

\ssk

\msk

\setcounter{savecntr}{\value{footnote}} \setcounter{Hsavecntr}{\value{Hfootnote}}
\centerline {\Large {Simone Zerella}\footnote {email address:simone.zerella@gmail.com}}
\centerline {\emph {Dipartimento di Fisica dell'Universit\`a, I-56126 Pisa, Italy}} 
\endgroup
\setcounter{footnote}{0}

\ssk

\bsk\ni

\ni
\small {\textbf {Abstract}. 
We study two Hamiltonian models, based on infrared approximations which render them solvable, in 
order to obtain an operator formulation of the soft-photon corrections to the scattering of a single 
electron, as given in Quantum Electrodynamics by the method of Feynman's diagrams.
The first model is based on the same approximations of the Pauli-Fierz Hamiltonian, the second one
stems from an expansion in powers of the four-momentum transfer, along the lines of Bloch and
Nordsieck. 
For both models, the dynamics of the charge is accounted for by suitably chosen classical currents, 
interacting with the quantum e.m. potential.

M\"{o}ller operators, preserving respectively the Hilbert scalar product, for the Coulomb-gauge formulation
of the models, and an indefinite metric, for the formulation of the models in the Feynman-Gupta-Bleuler 
gauge, are obtained in the presence of an infrared cutoff, with the help of suitable renormalization 
counterterms.

We show that the soft-photon corrections to the electron scattering under consideration are reproduced by 
suitable matrix elements of the M\"{o}ller operators pertaining to the model ``of the Bloch-Nordsieck type'',
both in the $\, FGB$ gauge and in the Coulomb gauge. 

Further, we prove that if one assumes that the charged particle is non relativistic and employs a dipole 
approximation, the resulting low-energy radiative corrections admit an operator formulation as well, in terms 
of the M\"{o}ller operators of the model ``of Pauli-Fierz type'', but lack the invariance property with 
respect to the gauge employed in their calculation; spurious, yet infrared-relevant, contributions 
in fact arise, causing in particular a discrepancy between the corrections 
in the $\, FGB$ gauge and their Coulomb-gauge expression.
The reason why such a discrepancy occurs is finally traced back in full generality, also in connection 
with the Gupta-Bleuler formulation of non-relativistic models.}

\ssk\ni 

\thispagestyle {empty}

\newpage

\section* {Introduction}
\addcontentsline{toc}{section}{Introduction}

In Quantum Electrodynamics ($\, QED\, $), the description of states at asymptotic times and the 
derivation of the scattering matrix are still open issues. 

At the perturbative level, transition amplitudes between states containing a finite number of photons 
are ill-defined, since radiative corrections typically exhibit low-energy logarithmic divergences.
As a consequence, in contrast with ordinary quantum field theories, Dyson's $\, S$-matrix 
(\cite {Dys49a,Dys51}) is defined only in the presence of an infrared ($\, IR\, $) cutoff and the problem 
of a proper identification of asymptotic states arises.

As early as 1937, in their pioneering paper on the subject (\cite {BN37}), Bloch and Nordsieck pointed out 
that infrared singularities arise in perturbation theory because of fundamental physical facts; they in fact
argued that, on the basis of the correspondence principle, one has to expect a vanishing probability for 
the emission of a finite number of photons in any collision process involving electrically charged
 particles. 

Exponentiation of the infrared radiative corrections was conjectured by Schwinger (\cite {Schw49}) and 
proved by Yennie, Frautschi and Suura (\cite {YFS61}) within the framework of the local and covariant 
Feynman-Dyson formulation (\cite {Dys49b, Feyn49, Feyn50}) of $\, QED\, .\, $ 
The paper \cite {YFS61} laid the basis of the recipe currently adopted to cope with the soft-photon 
divergences: The $\, IR\, $ cutoff is removed after summing the transition rates over all final photon 
states with energy below the threshold of the detectors.
Finiteness of the resulting inclusive cross-sections is ensured by cancellations, at each perturbative 
order, between the infrared singularities associated respectively to low-energy radiative corrections 
and to soft-photon emission. 

As emphasized by Steinmann (\cite {Stei}), such a procedure involves an exchange of limits. 
Moreover, its relation with structural (non-perturbative) properties, such as the spontaneous breaking of the 
symmetry under Lorentz boosts in the charged superselection sectors (\cite {FMS79a, FMS79b, Buc86}) 
and the absence of a sharp eigenvalue for the mass operator of a particle carrying an electric charge 
(\cite {FMS79a, Buc86}), is still unclear.
In addition, local and covariant quantizations of abelian gauge theories, within which the perturbation-theoretic 
formulation of $\, QED$ is mainly developed, are incompatible with positivity (\cite {S67}) and require a suitable 
generalization of Wightman's axioms (\cite {SW74, MS80}).

Quite generally, the problem of filling the gap that separates the perturbative approach from a collision 
theory of ``infraparticles'' (\cite {Schr63}), in which the structural features of the $IR\, $ problem are 
taken into account, is a relevant issue, not only conceptually but also from a practical point of view; 
perturbation theory remains in fact the only source of detailed information on Quantum 
Electrodynamics and its local and covariant version is the best controlled one regarding 
renormalization procedures.

In the present paper we wish to take a small step forward, by carrying out an analysis of suitable 
Hamiltonian models, especially focused on a comparison with the perturbative treatment of the 
infrared divergences.

The aim of such an analysis is twofold.
First, we would like to give a mathematical formulation of the structures underlying the 
diagrammatic treatment of the soft-photon contributions.
Secondly, since the electric dipole approximation has been employed in non-relativistic models of 
infrared $\, QED\, $ in order to obtain \emph {non-perturbative} constructions, both in the Coulomb 
gauge (\cite {Blan69, Ara83, Greenb00}) and within the Gupta-Bleuler formulation (\cite {HiSu09}), 
we wish to check its reliability with respect to the \emph {perturbative} low-energy 
approximations and results.

We shall discuss two Hamiltonian models of a single charged particle, whose dynamics is accounted for 
by suitably chosen classical currents, interacting with the quantum e.m. potential. 
Such models are based on infrared approximations which render them solvable and might seem to be 
equally suited, from a physical point of view, for an investigation of soft-photon effects; the 
dipole approximation, following the treatment of  Pauli and Fierz, and the expansion 
in powers of the four-momentum transfer around a fixed (asymptotic) charged particle 
four-momentum, along the lines of Bloch and Nordsieck.

Our main result is that the soft-photon corrections to the scattering of a single charged particle, 
computed in perturbation theory under the standard infrared approximations 
(\cite {YFS61}), for a fixed value of a low-energy cutoff, admit an operator formulation 
in terms of the M\"{o}ller operators of the model ``of the Bloch-Nordsieck type'', both 
in the $FGB\, $ gauge  (\cite {Zer09}) and in the Coulomb gauge.

Likewise, we show that the infrared radiative corrections to the scattering of a non-relativistic electron, 
in the presence of a dipole approximation, are reproduced in terms of the M\"{o}ller operators
of the model ``of the Pauli-Fierz type''. 
However, we shall also see that the introduction of a dipole approximation turns out to be too strong 
an assumption, for its effect on current conservation, also recognized and studied in \cite {HiSu09}, 
leads to a discrepancy of the $FGB$-gauge soft-photon corrections, by a factor $\, 3/2\, ,$ with 
respect to the corresponding Coulomb-gauge expressions and thus to a violation of the 
invariance property of such corrections with respect to the gauge adopted in 
their calculation.

The content of the paper is as follows.
In Section \ref {sect:1} we first recall the basic features of the Pauli-Fierz model, introduced in \cite {PF38} and 
later reconsidered by Blanchard (\cite {Blan69}), who studied it in the interaction representation, focusing 
on the mathematical issues connected with the basic physical fact that an infinite number of photons is 
emitted in any collision process involving electrically charged particles.

Although our treatment parallels Blanchard's analysis to a large extent, we find it useful to give a  
complete presentation; this choice will in fact enable us to mainly devote Section \ref {sect:2} to the 
operator formulation of infrared $\, QED\, ,\, $ while omitting a thorough description of similar 
mathematical procedures and techniques employed therein.

The main differences with respect to the Pauli-Fierz-Blanchard setting stem from the fact that our aim is 
to reproduce the infrared-regularized Feynman amplitudes associated to a given scattering process, 
involving a single charged particle; a low-energy cutoff will thus be adopted throughout the 
discussion and a suitably chosen classical current, accounting for the asymptotic 
dynamics of the charge, will be introduced.
The question concerning the existence of the large-time limits of the evolution operators once the infrared 
regularization is removed will instead not be addressed, because it is already discussed in \cite {Blan69}.
In addition, an adiabatic switching of the interaction will be used as an auxiliary tool, in order to
keep contact with the perturbative procedures and calculations.

In order to set up a comparison with the Feynman-Dyson expansion, we then formulate the 
model in the $\, FGB$ gauge, define its dynamics and obtain M\"{o}ller operators, acting as 
isometries on an indefinite inner-product space, as weak asymptotic 
limits of the evolution operator (in the interaction representation), for a fixed $\, IR$ cutoff.
A preliminary discussion of the spurious effects induced by the dipole approximation is given
at the end of Section \ref {sect:1}.

In Section \ref {sect:2} we introduce a model based on an expansion already implicit in \cite {BN37},
hereafter referred to as Bloch-Nordsieck $(BN)$ model, and formulate it both in the Coulomb 
gauge and in the $FGB\, $ gauge.
The definition of the dynamics and the control of the asymptotic limits follow the same pattern of 
the first Section.
We prove that the soft-photon corrections to the scattering of an electron, which result from 
the diagrammatics of $\, QED$ in the $\, FGB\, $ gauge under the standard low-energy 
approximations, are given by suitable matrix elements of the M\"{o}ller operators of the 
$\, BN$ model in the same gauge.

In particular, we reproduce the exponentiation of the second-order soft-photon radiative corrections, including 
the interplay with renormalization procedures (Propositions \ref {proposizione 5}, \ref {proposizione 6}),
and the contributions due to the emission of soft photons 
(Lemma \ref {Lemma 3}, Corollary \ref {corollario 1}).
Moreover, we recover the independence of the expressions associated to soft-photon emission with 
respect to the low-energy radiative corrections (Proposition \ref {proposizione 7}). 
We then show that the $BN$ model also allows to reproduce the soft-photon corrections to the electron
scattering in the Coulomb gauge (Propositions \ref {proposizione 8}, \ref {proposizione 9}). 

The low-energy corrections of $\, QED\, ,\, $ obtained under the standard infrared approximations, enjoy 
an invariance property with respect to the gauge adopted in their calculation; in Section \ref {sect:3} we 
give an operator proof of such a property for the corrections discussed in Section \ref {sect:2} 
(Proposition \ref {proposizione 10} and Corollaries \ref {corollario 3}, \ref {corollario 4}). 
Afterwards, we show that the soft-photon radiative corrections to the scattering of a non-relativistic 
electron, evaluated with the aid of the diagrammatic rules in the presence of a dipole approximation, 
are not invariant with respect to the choice of a gauge.
Such corrections are likewise proved to admit an operator formulation (Corollaries \ref {corollario 6},
\ref {corollario 7}), in terms of the M\"{o}ller operators constructed in Section \ref {sect:1},
and the reason at the root of the discrepancy between their expressions in the 
$\, FGB\, $ gauge and in the Coulomb gauge is finally traced back 
(Lemma \ref {Lemma 8}).

An outlook for future research is finally given.
Appendix \ref {app:1} is devoted to the proof of the self-adjointness of the Pauli-Fierz Hamiltonian.
In Appendix \ref {app:2}, we describe the construction of a suitable indefinite-metric space, on 
which the evolution operators of the $FGB$-gauge models studied in this work are 
shown to be well defined and unique.
In Appendix \ref {app:3}, the main results about the Gupta-Bleuler quantization of the
free e.m. field and its relationship with the Coulomb-gauge quantization are 
given.
No pretension of completeness is made about the bibliography.

\subsection* {Notations}
\addcontentsline{toc}{subsection}{Notations}

\ni
The metric $\, g^{\: \mu\, \nu}=\, diag\; (\, 1\, ,\, -\, 1\, ,\, -\, 1\, ,\, -\, 1\, )\, $ is adopted and natural units are 
used $(\, \hbar=c=1\, )\, .$\\
We shall reserve the symbol $\, v\, $ for a four-vector or, alternatively, denote its components by greek 
indices, say $\, v^{\: \mu}=(\, v^{\; 0}\, ,\mathbf {\, v}\, )\, .\, $ 
A three-vector will be denoted by $\, \mathbf {v}\, $ or by labelling its components with latin indices, 
say $\mathbf {\, v}^{\; i}.\, $ 
The symbol $\mathbf {v}\, $ will also indicate the module of the three-vector; when confusion may arise, 
the notation $\, \vert\, \mathbf {v}\, \vert\, $ will be instead employed.
We use the symbol $\, c \cdot d\, $ for the indefinite inner product between four-vectors $\, c\, $ 
and $\, d\, ,\, $ and likewise for the scalar product between three-vectors.   

The Hilbert scalar product is denoted by $\, (\, .\, ,.\, )\, $ and indefinite inner products by 
$\, \langle\, .\, ,.\, \rangle\, ,\, $ all products being taken to be linear in the second factor.
The norm of $\, \phi\in L^{\, 2}\, $ is indicated by $\, \Vert\, \phi\, \Vert_{\, 2\, }^{}.\, $
The adjoint of an operator $\, A\, $ on a Hilbert space is denoted by $A^{\; *}$ and the symbol $\, B^{\; \dagger}$  
will stand for the hermitian conjugate, with respect to the inner product, of an operator $B$ defined on an 
indefinite-metric space.

The symmetric Fock space $\, \bigoplus_{\, n\, =\; 0}^{\, \infty}\, S_{\, n}^{}\, \mathscr {H}^{\, (\, n\, )}$ over 
a Hilbert space $\mathscr {H}$ will be denoted by $\mathscr {F}, $
the symbols $\, \mathscr {H}^{\, (\, n\, )},$ $S_{\: n}^{}$ standing respectively for the $\, n$-fold tensor 
product $\, \bigotimes_{\, k\, =\; 1}^{\; n}\mathscr {H}$ and for the symmetrization operator, defined in 
terms of the permutation group of degree $\, n\, .\, $ 
The norm in $\, \mathscr {F}$ will be denoted by $\, \Vert\, .\, \Vert\, ,$ the no-particle vector by 
$\Psi_{\, F}^{}$ and the number operator by $N.\, $ 
We let $\phi^{\, (n)}$ be the orthogonal projection of $\, \phi\in\mathscr {F}$ onto the $\, n$-particle 
subspace $\, S_{\: n}^{}\, \mathscr {H}^{\; (\, n\, )}\, .$\\
The symbols $\, \delta_{\; \mathbf {y}}^{}\, (\, \mathbf {x}\, )\, ,$ $\, \delta\, (\, \mathbf {x\, }-\mathbf {\, y}\, )\, $ 
will denote the Dirac delta measure at $\, \mathbf {y}\, .$
$\, G_{\, \mathbf {y}\, }^{}(\, \mathbf {x}\, )$ will stand for the Green's function for Poisson's 
equation in $\, \mathbb {R}^{\, 3}:\, $
\begin {equation}\label {funzione di Green Poisson}
-\Laplace\; G_{\: \mathbf {y}}^{}\, (\, \mathbf {x}\, )\, =\,\, \delta_{\; \mathbf {y}}^{}\, (\, \mathbf {x}\, )\; .
\quad\quad\quad\quad
\end {equation}
In the Coulomb gauge, $\, a_{\: s}^{}\, (\, a_{\; s}^{\; *}\, )$ will stand for the photon annihilation 
(creation) operator-valued distribution, fulfilling the canonical commutation relations $(\, CCR\, )\, $
\begin {equation}\label {CCR}
\quad [\,\, a_{\: s\; }^{}(\, \mathbf {k}\, )\, ,\: a_{\: s\, '\; }^{\; *}(\, \mathbf {k}\: '\, )\, ]
\: =\,\, \delta_{\: s\, s\, '}^{}\;\, \delta\, (\: \mathbf {k\, }-\, \mathbf {k}\: '\, )\; ,
\nonumber
\end {equation}
with $\, s\, $ and $\, s\, '\, $ polarization indices.\\
In the same gauge, we denote the free e.m. Hamiltonian by $\, H_{\: 0\, ,\; C}^{\; e.\, m.}\, $
and the free vector potential at time $\, t=0\, $ by
\begin {eqnarray}\label {potenziale vettore Coulomb}  
\mathbf {A}_{\, C\, }^{\: i}(\, \mathbf {x}\, )\, \equiv\; \sum_{s\; }\;\, \int\,\, \frac {\, d^{\,\, 3\, }k} {(\, 2\; \pi\, )^{\, 3\, /\, 2}
\; \sqrt {\; 2\,\, \mathbf {k}\, }\, }\,\,\, \mathbf {\epsilon}_{\; s\; }^{\,\, i}(\, \mathbf {k}\, )\;\, 
[\;\, a_{\, s\; }^{}(\, \mathbf {k}\, )\,\,\, e^{\; i\,\, \mathbf {k}\, \cdot\; \mathbf {x}\; }+\, 
a_{\: s\; }^{\; *}(\, \mathbf {k}\, )\,\,\, e^{\, -\, i\;\, \mathbf {k}\, \cdot\; \mathbf {x}}\;\, ]\;\;
\nonumber\\
\equiv\, \int\,\, \frac {\, d^{\,\, 3\, }k} {(\, 2\; \pi\, )^{\, 3\, /\, 2}\; \sqrt {\; 2\,\, \mathbf {k}\, }\, }\;\,\, 
[\,\, a_{\, C}^{}\, (\, \mathbf {k}\, )\,\,\, e^{\; i\,\, \mathbf {k}\, \cdot\; \mathbf {x}\; }+\, 
a_{\: C}^{\; *}\, (\, \mathbf {k}\, )\,\,\, e^{\, -\, i\;\, \mathbf {k}\, \cdot\; \mathbf {x}}\;\, ]\; ,
\end {eqnarray} 
with $\, \mathbf {\epsilon}_{\: s}^{}\, (\, \mathbf {k}\, )\, ,\, s=1,\, 2,$ orthonormal polarization vectors
satisfying $\, \mathbf {k}\, \cdot\, \mathbf {\epsilon}_{\; s}^{}\, (\, \mathbf {k}\, )=\, 0\, .\, $
The annihilation and creation operator-valued distributions in the $FGB\, $ gauge, denoted respectively 
by $\, a^{\, \mu\, }(\, \mathbf {k}\, )$ and $\, a^{\: \mu\; \dagger}\, (\, \mathbf {k}\, )\, ,\, $ fulfill the $\, CCR\, $ 
\begin {equation}\label {CCR indefinite}
[\,\, a^{\, \mu}\, (\, \mathbf {k}\, )\: ,\; a^{\: \nu\; \dagger\, }(\, \mathbf {k}\, '\, )\; ]\, =
\, -\; g^{\: \mu\, \nu}\;\, \delta\, (\: \mathbf {k}-\, \mathbf {k}\: '\, )\; .
\nonumber
\end {equation}
In the same gauge, the Hamiltonian of the free e.m. field is denoted by $H_{\; 0}^{\; e.\, m.}$ and
the free vector potential at time $\, t=0\, $ by
\begin {equation}\label {espansione quadripotenziale}
A^{\; \mu\; }(\, \mathbf {x}\, )\, \equiv\; \int\,\, \frac {\, d^{\,\, 3\, }k} 
{(\, 2\; \pi\, )^{\, 3\, /\, 2}\; \sqrt {\; 2\,\, \mathbf {k}\, }}\;\,\, 
[\,\, a^{\, \mu}\, (\, \mathbf {k}\, )\,\,\, 
e^{\; i\,\, \mathbf {k}\, \cdot\; \mathbf {x}}\, +\, 
a^{\, \mu\: \dagger}\, (\, \mathbf {k}\, )
\,\,\, e^{\, -\, i\,\, \mathbf {k\; }\cdot\; \mathbf {x}}\;\, ]\, 
\equiv\; A_{\, +}^{\; \mu\; }(\, \mathbf {x}\, )\, +\; A_{\, -}^{\; \mu\; }(\, \mathbf {x}\, )\: .
\nonumber
\end {equation}
The convolution with a form factor $\, \rho\, $ is indicated by
\begin {equation}\label {convoluzione covariante}
A^{\; \mu}\, (\, \rho\, ,\, \mathbf {x}\; )\, \equiv\; \int\; d^{\,\, 3}\, \xi\,\,\,\, \rho\: (\, \mathbf {\xi}\, )
\,\,\, A^{\; \mu}\; (\, \mathbf {x\, }-\, \mathbf {\xi}\; )\: ,
\nonumber
\end {equation}
and likewise for $\, \mathbf {A}_{\, C\, }^{}.\, $
We shall use the symbol
\begin {equation}\label {Coulomb smearing}
\mathbf {A}_{\, C}^{}\, (\, \mathbf {f}\, )\, \equiv\, \int\; d^{\,\, 4}\, x\,\,\, \mathbf {A}_{\, C\, }^{\, r}(\, x\, )
\;\,\, \mathbf {f}^{\; r\, }(\, x\, )\, =\; \mathbf {A}_{\, C}^{\, r}\, (\, \mathbf {f}^{\: r}\, )
\end {equation}
for the smearing of (\ref {potenziale vettore Coulomb}) with a vector test function $\, \mathbf {f}\, $
and the symbol $\, A\, (\, f\, )\equiv\, A^{\; \mu}\, (\, f_{\, \mu}^{}\, )$ for the smeared 
four-vector potential in the $\, FGB\, $ gauge.\\
We denote by $\tilde {f}\, (\, k\, )\, ,$ or simply by $\, f\, (\, k\, )\, $ when no confusion should arise, the 
Fourier transform of a function $\, f\, (\, x\, )\, $ on Minkowski space; we employ the conventions
\begin {equation}
f\, (\, x\, )\, =\, \frac {1} {(\, 2\,\, \pi\, )^{\; 4}\, }\; \int\,\, d^{\,\, 4}\, k\;\,\, e^{\, -\, i\,\, k\, \cdot\; x}
\;\,\, \tilde {f}\: (\, k\, )\: .
\end {equation}
For brevity we shall also write
\begin {equation}
a\, (\, f\, (\, t\, )\, )\, \equiv\; \int\; d^{\,\, 3\, }k\;\,\, a^{\, \mu}\, (\, \mathbf {k}\, )\,\,\, \tilde {f}_{\, \mu}^{}\: (\, t\, ,\: \mathbf {k}\, )\, ,
\quad\quad\nonumber
\end {equation}
with
\begin {equation}
f\, (\, t\, ,\: \mathbf {x}\, )\, =\, \frac {1} {(\, 2\,\, \pi\, )^{\; 3}\, }\; \int\,\, 
d^{\,\, 3}\, k\;\,\, e^{\; i\,\, \mathbf {k}\, \cdot\; \mathbf {x}}\;\,\, \tilde {f}\, (\, t\, ,
\: \mathbf {k}\, )\nonumber\, ,
\end {equation}
and denote the corresponding sum in the Coulomb gauge by $\, a_{\, C\, }^{}(\, \mathbf {f}\, (\, t\, ))\, .$\\ 
$\, \mathscr {S}\, (\, \mathbb {R}^{\, 3}\, )\, $ stands for the Schwartz space of $\, C^{\; \infty}\, $ 
functions of rapid decrease defined on $\, \mathbb {R}^{\, 3}.$\\
Throughout the paper we shall frequently encounter integrals of the type
\begin {equation}
I\, \equiv\,\, \int\, d^{\,\, 4\, }k\;\,\, \tilde {h}\: (\, -\, k\, )\,\,\, \tilde {G}\, (\, k\, )\,\,\, \tilde {h}\: (\, k\, )\: ,
\,\, J\: \equiv\,\, \int\, d^{\,\, 4\, }k\;\,\, \tilde {f}\, (\, k\, )\,\,\, \tilde {G}\, (\, k\, )\: ,\,\, 
\end {equation}
with 
$\, \tilde {G}\, (\, k\, )=\; \theta\, (\, k_{\, 0}^{}\, )\;\, \delta\, (\, k^{\; 2}\, )\; \tilde {F}\, (\,  \mathbf {k}\, )
\, ,\, $ $\tilde {F}\, (\,  \mathbf {k}\, )\in\mathscr {S}\, (\, \mathbb {R}^{\, 3}\, )\, ,\, $
with the symbol $\, \theta\, $ standing for the Heaviside step distribution.
As an infrared regularization for such expressions, we shall restrict integrations in momentum space 
outside a sphere of radius $\, \lambda\, .\, $ The regularized $\, I\, $ will be denoted by
\begin {equation}
I_{\; \lambda}^{}\, \equiv\;\, \int_{\; \mathbf {k}\, >\, \lambda}\,\, \frac { d^{\,\, 4\, }k} {(\, 2\; \pi\, )^{\: 4}\, }\;
\,\, \tilde {h}\: (\, -\, k\, )\,\,\, \tilde {G}\, (\, k\, )\,\,\, \tilde {h}\: (\, k\, )\, =\;\, \int_{\; \mathbf {k}\, >\, \lambda}
\,\, \frac {d^{\,\, 3\, }k} {(\, 2\; \pi\, )^{\: 3}\;\, 2\,\, \mathbf {k}\, }\;\,\,  \tilde {h}\; (\, -\, k\, )\,\,\, \tilde {F}\, 
(\, k\, )\,\,\, \tilde {h}\; (\, k\, )
\end {equation}
and its expression as a space-time integral by
\begin {equation}
I_{\; \lambda}^{}\, =\;\, \int_{\, \lambda}\,\, d^{\,\, 4\, }x\;\, d^{\,\, 4\, }y\;\,\, h\; (\, x\, )\;\, G\, (\, x\, -\, y\, )\;\, h\; (\, y\, )\, .
\end {equation}
The same notations will be used for the infrared regularization of $\, J\, .$

\section {Pauli-Fierz-Blanchard Models}
\label {sect:1}

In this Section we introduce a model, based on the approximations of the Pauli-Fierz-Blanchard $(PFB)$ 
Hamiltonian, and discuss its formulation both in the Coulomb gauge and in the $\, FGB\, $ gauge.
First, we introduce the infrared-regularized $\, PFB\, $ Hamiltonian, 
\begin {equation}\label {hamiltoniano PFB}  
H_{\, \lambda}^{\, (PFB)}\, =\,\, \frac {\,\,\, {\mathbf {p}}^{\, 2}} {2\; m}\, +\, H_{\: 0\, ,\,\, C}^{\; e.\, m.\, }+
\, H_{\, int\, ,\; C\, }^{}\equiv\,\, H_{\; 0}^{}+H_{\, int\, ,\; C}^{}\,\, ,
\end {equation}
\begin {equation}\label {interazione PFB}
H_{\, int\, ,\,\, C}^{}=\, -\, \frac {e} {m\, }\;\,\, \mathbf {p}\, \cdot\, \mathbf {A}_{\, C\, ,\;\, \lambda}^{}
\; (\, \rho\: ,\, \mathbf {x}\, =\; 0\; )\: .
\quad\quad\quad\quad\quad\quad\quad
\end {equation}
We will also call electron the particle of mass $\, m,$ charge $e$ and spherically symmetric distribution 
of charge $\, e\; \rho\in\mathscr {S}(\, \mathbb {R}^{\, 3}\, )\, .\, $
The subscript $\, \lambda\, $ on the left-hand side of (\ref {hamiltoniano PFB}) denotes the infrared 
regularization, the $\, \lambda$-dependence in $H_{\: 0\, ,\,\, C\, }^{\; e.\, m.}\, $ and in 
$H_{\, int\, ,\,\, C}^{}$ being understood.
The functional form of the interaction Hamiltonian (\ref {interazione PFB}) is dictated by the electric 
dipole approximation and implies that the electron momentum is conserved, while the total one is 
not.

The Hilbert space of states of the model is $\, \mathscr {H}=\, L^{\, 2}\, (\, \mathbb {R}^{\, 3}\, )\otimes\,
\mathscr {H}_{\; C}^{}\, ,$ with $\, L^{\, 2}$ the one-particle space and $\, \mathscr {H}_{\; C}^{}\equiv 
\mathscr {F}\, $ the Fock space associated to the Coulomb-gauge canonical photon variables.\\
The free Hamiltonian $\, H_{\; 0}^{}\, $ is self-adjoint, since it is the sum of two positive, commuting 
self-adjoint operators; further, it is essentially self-adjoint (e.s.a.) on $\, D_{\: 0}^{}
\equiv\mathscr {S\, }(\, \mathbb {R}^{\, 3}\, )\otimes\, D_{\, F_{\: 0}^{}},$ 
where $F_{\,\, 0}^{}\, $ is the dense set spanned by the 
finite-particle vectors of $\, \mathscr {F}$ and
$D_{\, F_{\: 0}^{}}^{}\equiv\, (\; \psi\in F_{\; 0\; }^{};\, \psi^{\, (n)}\in S_{\; n}^{}\, 
\bigotimes_{\, k\; =\; 1}^{\, n}\, \mathscr {S\, }(\, \mathbb {R}^{\, 3}\, )\, ,\, 
\forall\; n\, )\, .$ 

The essential self-adjointness of the (infrared-regularized) $PFB\, $ Hamiltonian (\ref {hamiltoniano PFB})
can be established by methods exploited for instance in \cite {Nels64,Ara81}; for the sake of completeness, 
the details are given in Appendix \ref {app:1}.
Since the $\, PFB\, $ Hamiltonian commutes with the electron momentum operator, it 
is e.s.a. on (almost) any of the subspaces, on which $\, \mathbf {p}\, $ takes a constant value\footnote 
{The e.s.a. of the $\, PFB\, $ Hamiltonian \emph {at fixed \textbf {p}} also follows by the Kato-Rellich 
theorem.}, obtained by decomposing $D_{\; 0}^{}\, $ on the joint spectrum of the components of 
$\, \mathbf {p}\, .$ 

With the aim of reproducing the soft-photon corrections to the scattering $\, \mathbf {p}_{\, in}^{}
\rightarrow\mathbf {p}_{\, out}^{}$ of a non-relativistic electron, with initial (final) state of definite 
three-momentum $\, \mathbf {p}_{\, in}^{}\, (\, \mathbf {p}_{\, out}^{}\, )\, ,$ in the presence of a 
dipole approximation, we introduce the Coulomb-gauge Hamiltonian
\begin {equation}\label {hamiltoniano PFB modificato}
H_{\: \lambda}^{\; (Coul)\, ,\; (dip)}\, \equiv\,\, H_{\: 0\, ,\,\, C}^{\; e.\, m.\, }+H_{\; int}^{\: (Coul)\, ,\,\, (dip)\, },
\quad\quad\quad\quad\quad\quad\quad
\end {equation}
\begin {equation}\label {interazione hamiltoniano Coulomb dipolare}
H_{\; int}^{\: (Coul)\, ,\,\, (dip)\, }\equiv\, -\; e\; \int_{\, \lambda}^{}\; d^{\,\, 3}\, x\;\;\, 
\mathbf {j}_{\; C\, ,\,\, nr\; }^{\: (dip)}(\, x\, )\, \cdot\, \mathbf {A}_{\, C\; }^{}(\, \mathbf {x}\, )\; ,
\quad\quad
\end {equation}
\begin {equation}\label {corrente dipolare Coulomb non rel}
\mathbf {j}_{\; C\, ,\,\, nr}^{\: (dip)\; l}\; (\: t\, ,\; \mathbf {x}\, )\, \equiv\,\, \int\; d^{\,\, 3}\, y\;\;\, \delta_{\; tr}^{\,\, l\, m}
\: (\, \mathbf {x}\, -\, \mathbf {y}\, )\,\,\, \mathbf {\, j}_{\; nr}^{\: (dip)\; m}\; (\: t\, ,\: \mathbf {y}\, )\; ,
\end {equation}
with
\begin {equation}\label {corrente dipolare non rel}
\mathbf {j}_{\; nr}^{\, (dip)}\; (\, x\, )\, \equiv\,\, \frac {\; \rho\, (\, \mathbf {x}\, )\, } {m\, }
\,\,\, [\:\, \theta\: (\, -\; t\, )\;\, \mathbf {p}_{\, in}^{}+
\, \theta\: (\, t\, )\;\, \mathbf {p}_{\, out}^{}\; ]
\quad\quad\quad\quad\;\;
\end {equation}
a non-relativistic classical current obeying the dipole approximation and 
\begin {equation}\label {delta trasversa}
\delta_{\; tr}^{\; l\, m}\; (\, \mathbf {x}\, -\, \mathbf {y}\, )\, \equiv\;\, \delta^{\,\, l\, m}\;\, \delta\; (\, \mathbf {x}\, -
\, \mathbf {y}\, )\, +\; \partial_{\; x}^{\;\, l}\; \partial_{\: x}^{\,\, m}\,\, G_{\: \mathbf {y}}^{}\, (\, \mathbf {x}\, )\; .
\quad\quad\quad\;
\end {equation}
The Fourier transform of the ``transverse Dirac delta'' (\ref {delta trasversa}) is given by the 
orthogonal projection onto the transverse components of a momentum-space vector:
\begin {equation}\label {trasformata delta trasversa}
\delta_{\; tr}^{\; l\, m\; }(\, \mathbf {x}\, -\, \mathbf {y}\, )\, =\; \frac {1\; } {(\, 2\; \pi\, )^{\: 3}\, }
\, \int\; d^{\,\, 3}\, k\;\;\, e^{\; i\,\, \mathbf {k}\; \cdot\; (\, \mathbf {x}\, -\, \mathbf {y}\, )}\;\,\, 
P^{\,\, l\: m}\, (\, \mathbf {k}\, )\; ,\quad\;\;\;
\end {equation}
\begin {equation}\label {proiettore trasverso}
P^{\; l\, m}\, (\, \mathbf {k}\, )\, \equiv\,\, \delta^{\,\, l\, m}-\,  \frac {\; \mathbf {k}^{\; l}\,\, \mathbf {k}^{\; m}} 
{\mathbf {k}^{\; 2}}\; \cdot\quad\quad\quad\quad\quad\quad\quad\quad\quad\quad\quad\; 
\end {equation}
The Hamiltonian (\ref {hamiltoniano PFB modificato}) is e.s.a. since the same property holds for 
$\, H_{\, \lambda}^{\; (PFB)}$ at fixed $\, \mathbf {p}\, ;\, $ existence and uniqueness of the dynamics 
of the model thus follow by Stone's theorem (\cite {RS72a}).

Next we discuss the formulation of the dynamics in the interaction picture.
In such a representation, the time-evolution operator corresponding to a Hamiltonian 
$\, H=\, H_{\; 0}^{}+H_{\, int\, }^{},$ with $\, H_{\; 0}^{}\, $ the free part and $\, H_{\, int}^{}\, $ 
a time-independent interaction, is
\begin {equation}\label {UI}
U_{\,\, I}^{}\; (\, t\, )\, \equiv\;\, \exp\; (\; i\; H_{\; 0}^{}\,\, t\; )\;\, \exp\; (\, -\, i\,\, H\,\, t\; )\;\;\, \quad\quad\quad\quad
\end {equation}
and satisfies
\begin {equation}\label {equazione di evoluzione interazione}
i\,\,\, \frac {\, d\,\, U_{\,\, I}^{}\; (\, t\, )} {d\,\, t}\, =\; H_{\; I}^{}\; (\, t\, )\;\,\, U_{\,\, I}^{}\; (\, t\, )\: ,
\quad\quad\quad\quad\quad\quad\quad\quad\;\;
\end {equation}
$$
H_{\; I\;\, }^{}(\, t\, )\, \equiv\;\, \exp\; (\; i\; H_{\; 0}^{}\,\, t\; )\,\,\, H_{\, int}^{}\,\, \exp\; (\, -\, i\; H_{\: 0}^{}\,\, t\; )\; .
\quad\quad\;
$$ 
Equation (\ref {equazione di evoluzione interazione}) identifies $\, U_{\; I\; }^{}(\, t\, )$ (a more 
general result, also applying to time-dependent interactions, is stated below).
If the commutator of $\, H_{\; I}^{}$ evaluated at different times is a multiple of the identity operator, one 
can solve (\ref {equazione di evoluzione interazione}) by making use of the formula (\cite {Wil67})
\begin {equation}\label {caso particolare formula BH}
e^{\: A\, +\, B}\, =\,\, e^{\, A}\;\,\, e^{\, B}\;\,\, e^{\, -\frac {1} {2}\,\, [\, A\, ,\: B\; ]}\; .\quad\quad
\end {equation}
The resulting evolution operator is 
\begin {equation}\label {operatore di evoluzione a tempi finiti}
U_{\,\, I\,\,\, }^{}(\, t\, )\, =\;\, \exp\; (\, -\; i\; \int_{\, 0}^{\; t}\, d\,\, t\; '
\,\,\, H_{\,\, I\,\, }^{}(\, t\; '\, )\, )\;\, \exp\; (\, -\, \frac {1} {2}\,\, 
\int_{\, 0}^{\; t}\, d\,\, t\; '
\int_{\, 0}^{\; t\: '} d\,\, t\; ''\,\, [\,\, H_{\; I\;\, }^{}(\, t\; '\, )\; ,
\, H_{\; I\;\, }^{}(\, t\; ''\, )\; ]\, )\; ,
\end {equation}
whence $\, U\: (\, t\, )\, =\;\, \exp\; (\, -\, i\,\, H_{\; 0}^{}\;\, t\; )\;\,\, U_{\; I\:\, }^{}(\, t\, )\, .\, $
Since the commutator of 
\begin {equation}
H_{\; I}^{\: (Coul)\, ,\,\, (dip)\, }\, (\, t\, )\, =\;\, \exp\; (\; i\; H_{\: 0\, ,\,\, C}^{\; e.\, m.\, }\,\, t\; )\,\,\, 
H_{\; int}^{\: (Coul)\, ,\,\, (dip)\, }\; \exp\; (\; -\, i\; H_{\: 0\, ,\,\, C}^{\; e.\, m.\, }\,\, t\; )
\end {equation}
at different times is a multiple of the identity operator at each definite 
momentum, the (interaction-picture) time-evolution operator of our model can be computed with the 
help of the formula (\ref {operatore di evoluzione a tempi finiti}).

By an explicit calculation, with the aid of the expansion (\ref {potenziale vettore Coulomb}) and of 
the Fourier transform of $\, \mathbf {j}_{\; C\, ,\,\, nr\; }^{\: (dip)}(\, x\, )\, $ with respect to 
$\, \mathbf {x}\, ,\, $ given by
\begin {equation}
\tilde\mathbf {j}_{\; C\, ,\,\, nr}^{\; (dip)\; l}\, (\, t\, ,\: \mathbf {k}\, )\, =
\,\, \frac {\; \tilde {\rho\; }(\, \mathbf {k}\, )\, } {m\, }\,\,\, P^{\,\, l\, r}\, (\, \mathbf {k}\, )\;\, [\,\, \theta\: (\, -\; t\, )\;\, \mathbf {p}_{\, in}^{\; r}+\, \theta\: (\, t\, )\;\, \mathbf {p}_{\, out}^{\; r}\; ]\; ,
\end {equation}
we obtain 
\begin {equation}\label {operatore di evoluzione non rin}
U_{\,\, I\, ,\,\, C\, ,\,\, \mathbf {p}}^{\: (\, \lambda\, )}\; (\, t\, )\, =\;\, c_{\; \mathbf {p}}^{}\; (\, t\, )
\,\,\, \exp\; (\,\, i\; e\,\, [\,\, a_{\; C\, ,\; \lambda\; }^{\; *}
(\; \mathbf {f}_{\; \mathbf {p}}^{}\; (\, t\, )\, )\, +\, a_{\: C\, ,\; \lambda}^{}\; (\: \overline
\mathbf {\, f\, }_{\mathbf {p}}^{}\; (\, t\, )\, )\, ]\, )\: ,
\quad\quad
\end {equation}
\begin {equation}\label {contributo numerico tempo fisso}
c_{\; \mathbf {p}}^{}\; (\, t\, )\, =\;\, \exp\; (\,\, \frac {\; i\; e^{\, 2}\; \mathbf {p}^{\, 2}} {3\; m^{\, 2}}\;\,\, d\; (\, t\, )\, )\: ,
\quad\quad\quad\quad\quad\quad\quad\quad\quad\quad\quad\quad\quad
\end {equation}
\begin {equation}
d\,\, (\, t\, )\, =\;\, \int_{\; \mathbf {k}\, >\, \lambda}\; \frac {\, d^{\,\, 3}\, k} {(\, 2\; \pi\, )^{\: 3}\;\, \mathbf {k}^{\; 2}\, }
\,\,\, {\tilde {\rho\; }^{2\; }(\, \mathbf {k}\, )}\,\,\, (\,\, t\, -\frac {\, \sin\; \mathbf {k}\; t\, } {\mathbf {k}}\,\, )\: ,
\quad\quad\quad\quad\;\,\,
\end {equation}
\begin {equation}\label {funzione di coerenza tempo fisso}
\mathbf {f}_{\; \mathbf {p}\, s\; }^{}(\, t\, ,\: \mathbf {k}\; )\, =\; \frac {\; \tilde {\rho\; }(\, \mathbf {k}\, )} 
{(\, 2\; \pi\, )^{\; 3\, /\, 2}\,\, \sqrt {\; 2\,\, \mathbf {k}\, }}\;\, \frac {\; \mathbf {p}
\, \cdot\, \mathbf {\epsilon}_{\; s}^{}\, (\, \mathbf {k}\, )\,} {m\; }\;\, 
\frac {\; e^{\; i\; \mathbf {k}\; t}-\, 1\, }{i\,\,\mathbf {k}^{}}\,\, ,
\quad\quad\quad\,\, 
\end {equation}
with $\, \mathbf {p}=\, \mathbf {p}_{\, out}^{}\, ,\: t>0\, ,$ and $\, \mathbf {p}=\, \mathbf {p}_{\, in}^{}\, ,\: t<0\, .\, $

Since $\, U_{\; I\, ,\,\, C\, ,\; \mathbf {p}\; }^{\: (\, \lambda\, )}(\, t\, )\, $ does not converge for large times, we
renormalize (\ref {hamiltoniano PFB modificato}) by introducing suitable counterterms, in order to 
get well-defined asymptotic limits for the corresponding time-evolution operator; further, we 
adopt an adiabatic switching of the interaction, in the form
$\, e^{\: (ad)}\, (\, t\, )\equiv\; e\,\, e^{\, -\, \epsilon\,\, \vert\, t\, \vert\; }.$\\
Precisely, we consider the Hamiltonian 
\begin {eqnarray}\label {hamiltoniano PFB rinormalizzato}
H_{\: \lambda\; ,\,\, ren}^{\; (Coul)\, ,\; (dip)\, ,\,\, (\, \epsilon\, )}\, \equiv\,\, H_{\: 0\, ,\,\, C}^{\: e.\, m.\, }-
\, e\; \int_{\, \lambda}^{}\; d^{\,\, 3}\, x
\;\,\, \mathbf {j}_{\; C\, ,\,\, nr\; }^{\, (dip)\, ,\,\, (\, \epsilon\, )}\, (\, x\, )\, \cdot\, \mathbf {A}_{\, C}^{}\, (\, \mathbf {x}\, )\, 
+\, e^{\: 2}\,\, [\;\, \theta\: (\, -\; t\, )\;\, \frac {\, \mathbf {p}_{\, in}^{\, 2}\, } {\; m^{\: 2}\; }
\;\,\, \nonumber\\
+\,\, \theta\: (\, t\, )\;\, \frac {\, \mathbf {p}_{\, out}^{\, 2}\, } {m^{\: 2}}\;\, ]\,\,\, z
\;\, e^{\, -\, 2\:\, \epsilon\; \vert\, t\, \vert}\,
\equiv\; H_{\: 0\: ,\:\, C}^{\; e.\, m.}\, +\, H_{\, int\, ,\; ren}^{\, (Coul)\, ,\; (dip)\, ,\,\, (\, \epsilon\, )\; },\; 
\end {eqnarray}
with 
\begin {equation}\label {corrente dipolare epsilon}
\mathbf {j}_{\; C\, ,\,\, nr\; }^{\, (dip)\, ,\,\, (\, \epsilon\, )}\, (\, x\, )\equiv\,\,
\mathbf {j}_{\; C\, ,\,\, nr\; }^{\, (dip)}\, (\, x\, )\,\,\, e^{\, -\, \epsilon\; \vert\, t\, \vert}\; ,\quad\quad\quad
\end {equation}
\begin {equation}\label {controtermine PFB}
z\; =\:\, \frac {1} {3}\,\, \int_{\; \mathbf {k}\, >\, \lambda}\;\, \frac {\, d^{\,\, 3\: }k} {(\, 2\; \pi\, )^{\: 3}\,\, \mathbf {k}^{\; 2}\, }
\;\,\, \tilde {\rho\; }^{2}\, (\, \mathbf {k}\, )\, .\quad\quad\quad\quad
\end {equation}
The $\, \lambda$-dependence in $\, H_{\, int\, ,\; ren}^{\, (Coul)\, ,\; (dip)\, ,\,\, (\, \epsilon\, )}$ is again
understood.
In the sequel, we state the results holding for positive times; the corresponding expressions for $\, t<0\, $ 
are obtained by replacing $\, \epsilon\, $ by $\, -\, \epsilon\, $ and $\, \mathbf {p}_{\: out}^{}$ by 
$\, \mathbf {p}_{\; in}^{}.$

Although (\ref {hamiltoniano PFB rinormalizzato}) is time dependent, existence and 
uniqueness of the time-evolution unitary operators follow from the independence of time of their 
selfadjointness domain and strong differentiability of $\, (\, H\, (\, t\, )-\, i\, )\, (\, H\, (\, 0\, )-
\, i\, )^{\, -\, 1}\, ,$ through the results of \cite {Kato56}.
Insertion of
\begin {equation}
H_{\; I\, ,\; ren}^{\: (Coul)\, ,\,\, (dip)\, ,\,\, (\, \epsilon\, )\, }\, (\, t\, )\, \equiv\;\, \exp\; (\; i\; H_{\: 0\, ,\,\, C}^{\: e.\, m.\, }\,\, t\; )
\,\,\, H_{\, int\, ,\; ren}^{\, (Coul)\, ,\; (dip)\, ,\,\, (\, \epsilon\, )\; }\, \exp\; (\; -\, i\; H_{\: 0\, ,\,\, C}^{\: e.\, m.\, }\,\, t\; )
\end {equation}
into the right-hand side (r.h.s.) of  (\ref {operatore di evoluzione a tempi finiti}) yields the time-evolution 
operator of the model in the interaction representation,
\begin {equation}\label {operatore di evoluzione PF rinormalizzato}
U_{\; I\, ,\;\, C\, ,\,\, \mathbf {p}_{\, out}^{}\, ,\; z}^{\, (\, \lambda\, )\, ,\; (\, \epsilon\, )}\; (\, t\, )\, =
\,\, c_{\; \mathbf {p}_{\, out}^{}\, ,\; z}^{\; (\, \epsilon\, )}\, (\, t\, )
\,\,\, \exp\; (\,\, i\; e\:\, [\,\, a_{\: C\, ,\; \lambda}^{\: *}\; (\; \mathbf {f}_{\; \mathbf {p}_{\, out}^{}}^{\: (\, \epsilon\, )}\, (\, t\, )\, )
\, +\, a_{\: C\, ,\; \lambda}^{}\; (\, \overline\mathbf {\, f\, }_{\mathbf {p}_{\, out}^{}}^{\, (\, \epsilon\, )}\, (\, t\, )\, )\, ]\, )\: ,
\end {equation}
where
\begin {equation}\label {contributo numerico tempo fisso regolarizzato}
c_{\; \mathbf {p}_{\, out}^{}\, ,\; z}^{\; (\, \epsilon\, )}\, (\, t\, )\, \equiv\;\, 
c_{\; \mathbf {p}_{\, out}^{}}^{\; (\, \epsilon\, )}\, (\, t\, )\,\,\, 
\exp\; (\,\, i\; e^{\, 2}\; z\;\, \frac {\, \mathbf {p}_{\, out}^{\: 2}\, } {m^{\: 2}}
\;\, \frac {\; e^{\, -\, 2\,\, \epsilon\:\, t}\, -\, 1\; } {2\,\, \epsilon\, }\; )\; ,
\quad\quad\quad\quad\quad\quad\quad\quad\;\;
\end {equation}
\begin {equation}
c_{\; \mathbf {p}_{\, out}^{}}^{\; (\, \epsilon\, )}\, (\, t\, )\, \equiv\;\, \exp\; (\,\, 
\frac {\; i\; e^{\, 2}\,\, \mathbf {p}_{\, out}^{\: 2}\, } {3\; m^{\, 2}}
\;\,\, d^{\,\, (\, \epsilon\, )}\, (\, t\, )\, )\; ,
\quad\quad\quad\quad\quad\quad\quad\quad\quad\quad\quad
\quad\quad\quad\quad\quad\quad\;\,\,\,
\end {equation}
\begin {equation}\label {contributo numerico esplicito}
d^{\,\, (\, \epsilon\, )}\; (\, t\, )\, \equiv\, -\int_{\; \mathbf {k}\, >\, \lambda}\, \frac {d^{\,\, 3\, }k\;\,\, 
{\tilde {\rho\; }}^{2\; }(\, \mathbf {k}\, )} {(\, 2\; \pi\, )^{\, 3}\;\, \mathbf {k}\,\, (\, \mathbf {k}^{\; 2}+
\, \epsilon^{\; 2}\, )\, }\,\,\, [\;\, e^{\, -\, \epsilon\,\, t}\,\, \sin\; \mathbf {k}
\,\, t\; +\frac {\, \mathbf {k}} {2\,\, \epsilon}\;\,
(\,\, e^{\, -\, 2\; \epsilon\,\, t\, }-\, 1\; )\, ]\; ,
\end {equation}
\begin {equation}\label {funzione di coerenza regolarizzata}
\mathbf {f}_{\; \mathbf {p}_{\, out}^{}\, s}^{\: (\, \epsilon\, )}\; (\, t\, ,\; \mathbf {k}\, )\, =\, \frac {\; \tilde {\rho\; }
(\, \mathbf {k}\, )} {(\, 2\; \pi\, )^{\; 3\, /\, 2}\; \sqrt {\; 2\,\, \mathbf {k}\, }}\;\, \frac {\; \mathbf {p}_{\, out}^{}\cdot\, 
\mathbf {\epsilon}_{\: s}^{}\, (\, \mathbf {k}\, )\, } {m\; }\;\, \frac {\; e^{\; (\: i\,\, \mathbf {k}\, -\; \epsilon\, )\;\, t}\, -\, 1\; } 
{i\,\, \mathbf {k\, }-\, \epsilon}\; \cdot\quad\quad\quad\quad\quad\quad\;\; 
\end {equation}
The asymptotic limits are controlled by the following
\begin {theorem}\label {proposizione 1}
Both the large-time limits and the adiabatic limit of the evolution operator (\ref {operatore di evoluzione 
PF rinormalizzato}), defining the M\"{o}ller operators, exist in the strong operator topology, for each 
value of the infrared cutoff $\, \lambda:$
\begin {eqnarray}\label {Moller PFB}
\Omega_{\; \pm\, ,\,\, C\, ,\,\, \mathbf {p}_{\, \mp}^{}}^{\: (\, \lambda\, )\, }=\;\, s\; -\, \lim_{\epsilon\; \rightarrow\; 0}\; 
\lim_{t\; \rightarrow\, \mp\, \infty}\,\,\, U_{\,\, I\, ,\;\, C\, ,\,\, \mathbf {p}_{\, \mp}^{}\, ,\; z}^{\, (\, \lambda\, )\, ,\,\, (\, \pm\, \epsilon\, )}
\; (\, -\; t\; )\, =\,\, s\; -\, \lim_{\epsilon\; \rightarrow\; 0}\; 
\lim_{t\; \rightarrow\, \mp\, \infty}\,\,\, U_{\,\, I\, ,\;\, C\, ,\,\, \mathbf {p}_{\, \mp}^{}\, ,\; z}^{\, (\, \lambda\, )\, ,\,\, (\, \mp\, \epsilon\, )}
\; (\, t\, )
\quad\; \nonumber\\
\equiv\,\, s\; -\, \lim_{\epsilon\; \rightarrow\; 0}\; 
\Omega_{\; \pm\, ,\,\, C\, ,\,\, \mathbf {p}_{\, \mp}^{}}^{\: (\, \lambda\, )\, ,\,\, (\, \mp\, \epsilon\, )}\,
=\;\, \exp\; (\, -\; i\; e\,\, [\,\, a_{\: C\: ,\; \lambda\; }^{\; *}(\; \mathbf {f}_{\; \mathbf {p}_{\, \mp\, }^{}}^{})\, +
\, a_{\: C\: ,\; \lambda\; }^{}(\, \overline\mathbf {\, f\, }_{\mathbf {p}_{\, \mp\, }^{}}^{})\; ]\; ,\quad
\end {eqnarray}
\begin {eqnarray}\label {limiti forti}
\mathbf {f}_{\; \mathbf {p}_{\, \pm}^{}\, s\,\, }^{}(\, \mathbf {k}\, )\, \equiv\,\, L^{\; 2\, }-\lim_{\epsilon\,\, \rightarrow\; 0\, }
\; \mathbf {f}_{\; \mathbf {p}_{\, \pm}^{}\, s\,\, }^{\, (\, \pm\, \epsilon\, )\; }(\, \mathbf {k}\, )\, \equiv\,\, L^{\; 2\, }-
\lim_{\epsilon\; \rightarrow\; 0}\, \lim_{t\, \rightarrow\, \pm\, \infty}\;
\mathbf {f}_{\; \mathbf {p}_{\, \pm}^{}\, s\,\, }^{\, (\,  \pm\, \epsilon\, )\; }(\, t\, ,\: \mathbf {k}\, )
\quad\quad\quad\; \nonumber\\
=\; \frac {\; \tilde {\rho\; }(\, \mathbf {k}\, )} {(\, 2\; \pi\, )^{\; 3\, /\, 2}\; \sqrt {\; 2\,\, \mathbf {k}\, }}
\;\, \frac {\; \mathbf {p}_{\, \pm}^{}\cdot\, \mathbf {\epsilon}_{\; s\; }^{}(\, \mathbf {k}\, )} {m\; }\;\, 
\frac {i} {\mathbf {\, k}\, }\,\, ,\;\, \mathbf {p}_{\, +}^{}\equiv\,\, \mathbf {p}_{\, out}^{}\; ,\,\, 
\mathbf {p}_{\, -}^{}\equiv\,\, \mathbf {p}_{\, in}^{}\, .\quad\;\;\,
\end {eqnarray}
\end {theorem}
\begin {proof} 
The field operator $\, \Phi_{\, \lambda}^{}\, (\, f_{\; \mathbf {p}}^{\, (\, \epsilon\, )}(\, t\, ,\, \mathbf {k}\, )\, )
\equiv\; a_{\, C\, ,\; \lambda}^{}\, (\, f_{\,\, \mathbf {p}}^{\, (\, \epsilon\, )}(\, t\, ,\, \mathbf {k}\, )\, )+
\, a_{\; C\, ,\; \lambda}^{\; *}\, (\, \overline {f\, }_{\mathbf {p}}^{\; (\, \epsilon\, )}
(\, t\, ,\, \mathbf {k}\, )\, )$ admits $\, F_{\,\, 0\, }^{}$ as a 
dense and invariant set of analytic vectors, hence 
it is e.s.a. therein due to Nelson's analytic vector 
theorem (Theorem X.39 in \cite {RS72b}).
The limits of $\, f_{\; \mathbf {p}\, s\;\, }^{\, (\, \epsilon\, )}(\, t\, ,\, \mathbf {k}\, )\, $ exist pointwise, hence, 
by the dominated convergence theorem, also in the strong $\, L^{\; 2}$ topology;
by linearity and standard Fock-space estimates, $\, \Phi_{\, \lambda}^{}\, (\, f_{\; \mathbf {p}\;\, }^{\, (\, \epsilon\, )\, }
(\, t\, ,\, \mathbf {k}\, )\, )\, $ thus converges strongly to $\, \Phi_{\, \lambda}^{}\, (\, f_{\; \mathbf {p}\; }^{\, (\, \epsilon\, )\, }
(\, \mathbf {k}\, )\, )\, $ on $\, F_{\,\, 0}^{}\, .\, $
Since $\, \Phi_{\, \lambda}^{}\, (\, f_{\; \mathbf {p}\; }^{\, (\, \epsilon\, )\, }(\, \mathbf {k}\, )\, )\, $ is e.s.a. on 
$\, F_{\,\, 0}^{}\, ,$ convergence is in the strong generalized sense and the existence of the 
time-limits in (\ref {Moller PFB}) follows by Theorem VIII.20 in \cite {RS72a}.
A similar proof allows to establish the strong convergence of 
$\, \Phi_{\, \lambda}^{}\, (\, f_{\; \mathbf {p}\;\, }^{\, (\, \epsilon\, )\, }
(\, \mathbf {k}\, )\, )\, $ for $\, \epsilon\rightarrow 0\, .$ 
\end {proof}
\ni

In order to recover the Coulomb-gauge soft-photon contribution to the process
$\, \mathbf {p}_{\, in}^{}\rightarrow\mathbf {p}_{\, out}^{}\, ,$ we define the 
scattering matrix
\begin {equation}\label {matrice S Coul dip}
S_{\: \lambda\, ,\,\, \mathbf {p}_{\, out}^{}\; \mathbf {p}_{\, in}^{}}^{\: (Coul)\, ,\,\, (dip)}\, =\;\,s\, -\lim_{\epsilon\,\, \rightarrow\; 0}
\,\, S_{\: \lambda\, ,\,\, \mathbf {p}_{\, out}^{}\; \mathbf {p}_{\, in}^{}}^{\: (Coul)\, ,\,\, (dip)\, ,\; (\, \epsilon\, )}\, \equiv\;\,
s\, -\lim_{\epsilon\,\, \rightarrow\; 0}\;\, 
\Omega_{\; -\, ,\;\, C\,, \,\, \mathbf {p\, }_{out}^{}}^{\: (\, \lambda\, )\, ,\,\, (\, \epsilon\, )\; *}
\;\, \Omega_{\; +\, ,\;\, C\,, \,\, \mathbf {p}_{\, in}^{}}^{\: (\, \lambda\, )\, ,\,\, (\, -\, \epsilon\, )}\; .
\end {equation}

The comparison of the infrared diagrammatics resulting from the Feynman-Dyson expansion of $\, QED\, $ 
with the expansion of M\"{o}ller's operators in powers of the electric charge requires to adopt a formulation
in a gauge employing four independent photon degrees of freedom, such as the $\, FGB\, $ gauge.

With this aim, we introduce the Hamiltonian
\begin {equation}\label {hamiltoniano FGB dipolare}
H_{\: \lambda}^{\: (FGB)\, ,\,\, (dip)}\, =\,\, H_{\: 0}^{\,\, e.\, m.\, }+\, H_{\; int}^{\; (FGB)\, ,\,\, (dip)}\; ,\quad\quad\;\;\,
\end {equation}
\begin {equation}\label {interazione hamiltoniano FGB dipolare}
 H_{\; int}^{\: (FGB)\, ,\,\, (dip)\, }\equiv\;\, e\, \int_{\, \lambda}^{}\,\, d^{\,\, 3}\, x\;\,\, j_{\, nr\, ,\,\, \mu}^{\, (dip)}\; (\, x\, )
\,\,\, A^{\; \mu}\, (\, \mathbf {x}\, )\; ,
\end {equation}
with
\begin {equation}\label {corrente dipolare quadrivettoriale}
j_{\; nr}^{\: (dip)\; \mu}\, (\, x\, )\, \equiv\,\, (\; \rho\: (\, \mathbf {x}\, )\, ,\; \mathbf {j}_{\; nr}^{\: (dip)}\: (\, x\, )\, )\; .
\quad\quad\quad\;\;\;\;
\end {equation}
Since (\ref {corrente dipolare quadrivettoriale}) is obtained by imposing a dipole approximation on the 
non-relativistic four-current $j_{\; nr}^{\; \mu}\, (\, x\, )\, ,$
\begin {eqnarray}\label {corrente quadrivettoriale non rel}
j_{\; nr}^{\; 0}\, (\, x\, )\, \equiv\;\, \theta\: (\, -\; t\, )\;\, \rho\; (\, \vert\; \mathbf {x\, }-\frac {\, \mathbf {p}_{\, in}^{}}
{m}\;\, t\,\, \vert\, )\, +\; \theta\: (\, t\, )\;\, \rho\; (\, \vert\; \mathbf {x\, }-\frac {\, \mathbf {p}_{\, out}^{}} {m}\;\, 
t\,\, \vert\, )\; ,
\quad\quad\quad\quad\;\;
\nonumber\\
\\
\mathbf {j}_{\; nr}^{}\, (\, x\, )\, \equiv\;\, \theta\: (\, -\; t\, )\;\, \rho\; (\, \vert\; \mathbf {x\, }-\frac {\, \mathbf {p}_{\, in}^{}}
{m}\;\, t\,\, \vert\, )\;\, \frac {\, \mathbf {p}_{\, in}^{}} {m}+\, \theta\: (\, t\, )\;\, \rho\; (\, \vert\; \mathbf {x\, }-
\frac {\, \mathbf {p}_{\, out}^{}} {m}\;\, t\,\, \vert\, )\;\, \frac {\, \mathbf {p}_{\, out}^{}} {m}\,\, ,
\nonumber
\end {eqnarray}
the Hamiltonian (\ref {hamiltoniano FGB dipolare}) shares the same approximations of 
(\ref {hamiltoniano PFB}) and (\ref {hamiltoniano PFB modificato}).

The photon space of states of the model, henceforth denoted by $\mathscr {\, G},\, $ is endowed with an 
indefinite inner product $\langle\, .\, ,.\, \rangle$ and contains a no-particle vector $\Psi_{\, 0\, }^{}.\, $ 
It will be regarded as a topological space with the weak topology $\, \tau_{\; w\, }^{},\, $ defined by the 
family of seminorms $\, p_{\, y}^{}\, (\, x\, )=\, \vert\langle\, y\, ,\, x\, \rangle\vert\, ,\, $ $y\in
\mathscr {G}.\, $

The space $\mathscr {\, G}\, $ results from the application to a non-positive linear functional 
and a ${}^{*}$-algebra, containing the Weyl exponentials of the photon canonical variables, 
of a generalization of the Gelfand-Naimark-Segal ($GNS$) reconstruction theorem 
(\cite {GN43, Se47}).
The representation of the Weyl operators \emph {defines} the corresponding exponentials of the creation 
and annihilation operators on $\, \mathscr {G}\, ;\, $ such exponentials are also given by their series, 
which are weakly converging on $\, \mathscr {G}\, .\, $ 
For the details, we refer to Appendix \ref {app:2}. 
By the same token, one can establish formula (\ref {caso particolare formula BH}) also in the indefinite-metric 
case, a fact which we will rely upon in the definition of the dynamics of the model.

The domain $\, \mathscr {G}_{\,\, 0}^{}\, $ of the Hamiltonian (\ref {hamiltoniano FGB dipolare}) is  
a weakly dense invariant subspace of $\, \mathscr {G}\, ;\, $ we again refer to Appendix \ref {app:2}
for the details of its construction.
The inner product of $\, \mathscr {G}_{\,\, 0}^{}$ is also denoted by $\, \langle\, .\, ,.\, \rangle\, ,\, $ 
since this should not be a source of confusion.  

Let $\, \mathscr {H}\, $ be the Hilbert space arising from the standard positive scalar product on 
$\, \mathscr {G}:$ 
\begin {equation}\label {prodotto scalare Hilbert}
(\, f\, ,\, g\; )\, \equiv\,\, (\, f^{\; 0},\, g^{\; 0}\, )\, +\, \sum_{i\, }\;\, (\, f^{\; i},\, g^{\; i}\, )\: .
\quad\quad
\end {equation}
The free and full dynamics of the model are determined by isometries of $\, \mathscr {G}$ which leave 
$\, \mathscr {G}_{\,\, 0}^{}$ invariant and are differentiable on $\, \mathscr {G}_{\,\, 0}\, $ in the 
strong topology of $\, \mathscr {H},\, $ with time-derivatives given respectively by 
$\, H_{\; 0}^{\; e.\, m.}\, $ and $\, H_{\: \lambda}^{\; (FGB)\, ,\,\, (dip)\, }.\, $
The formulation and the proof of uniqueness of such evolution operators are given in Appendix 
\ref {app:2} (Lemma \ref {Lemma 10}).

The dynamics of the model in the interaction representation is governed by a $\, \langle\, .\, ,.\, \rangle$-isometric 
operator $\, U_{\, I\, ,\,\, \mathbf {p}}^{\; (\, \lambda\, )}:\, \mathscr {G}\rightarrow\mathscr {G}\, ,$ $\mathbf {p}=
\mathbf {p}_{\, out\, }^{},\, t>0\, ,\, \mathbf {p}=\mathbf {p}_{\, in\, }^{},\, t<0\, ,$ which leaves 
$\, \mathscr {G}_{\,\, 0}$ invariant and is strongly differentiable on 
$\, \mathscr {G}_{\,\, 0}^{}\, .\, $ 
By employing the generalization of formula (\ref {caso particolare formula BH}) to the indefinite-metric 
case, the expansion (\ref {espansione quadripotenziale}) and the Fourier transform of $\, j_{\; nr}^{\: (dip)\; \mu}\, (\, x\, )\, $ with respect to $\, \mathbf {x}\, ,\, $ given by
\begin {equation}
\tilde {j\, }_{nr}^{\, (dip)\; \mu\, }(\, t\, ,\; \mathbf {k}\, )\, =\,\, \tilde {\rho\; }(\, \mathbf {k}\, )\;\, [\,\, \theta\: (\, -\; t\, )\;\, \tilde {v}_{\, in}^{\; \mu}+\, \theta\: (\, t\, )\;\, \tilde {v}_{\, out}^{\; \mu}\; ]\, ,\,\, 
{\tilde {v}}^{\; \mu}\equiv\, (\: 1\, ,\: \mathbf {p}\, /\, m\: )\: ,
\end {equation}
we obtain a formal solution
\begin {equation}\label {soluzione formale pauli fierz quadrivettoriale}
U_{\; I\, ,\,\, \mathbf {p}}^{\, (\, \lambda\, )}\; (\, t\, )\, =\;\, \tilde {c}_{\; \tilde {v}}^{}\; (\, t\, )
\;\, \exp\; (\, -\; i\; e\;\, [\,\, a_{\, \lambda}^{\; \dagger\; }(\, f_{\; \tilde {v}\; }^{}(\, t\, )\, )\, +
\, a_{\, \lambda}^{}\, (\, \overline {f\, }_{\tilde {v}\; }^{}(\, t\, )\, )\, ]\, )\: ,
\end {equation}
\begin {equation}
\tilde {c}_{\; \tilde {v}}^{}\; (\, t\, )\, =\;\, \exp\; (\, -\, \frac {\,\, i\; e^{\, 2}\; \tilde {v}^{\; 2}} 
{2\; }\,\,\, d\; (\, t\, )\, )\; ,\quad\quad\quad\quad\quad\quad\quad\quad\quad
\quad
\end {equation}
\begin {equation}\label {funzione di coerenza PFBR}
f_{\; \tilde {v}}^{\; \mu\; }(\, t\, ,\; \mathbf {k}\, )\, =\,\, \frac {\; \tilde {\rho\; }(\, \mathbf {k}\, )
\;\, {\tilde {v}}^{\,\, \mu}} {(\, 2\; \pi\, )^{\; 3\, /\, 2}\; \sqrt {\; 2\,\, \mathbf {k}\, }}\;\, 
\frac {\; e^{\; i\,\, \mathbf {k}\; t\; }-\, 1\, } {i\,\, \mathbf {k}}\,\, \cdot
\quad\quad\quad\quad\quad\quad\quad
\end {equation}
It can be immediately checked that the operator (\ref {soluzione formale pauli fierz quadrivettoriale}) fulfills 
equation (\ref {equazione di evoluzione interazione}) in the strong topology of $\, \mathscr {H},$ 
on the invariant subspace $\, \mathscr {G}_{\,\, 0}^{}\, .$

Proceeding as for the Coulomb-gauge formulation of the model, we consider the Hamiltonian resulting from
the introduction of suitable renormalization counterterms and of an adiabatic switching in (\ref {hamiltoniano 
FGB dipolare}):
\begin {eqnarray}\label {hamiltoniano FGB dipolare rinormalizzato}
H_{\: \lambda\, ,\; ren}^{\, (FGB)\, ,\,\, (dip)\, ,\,\, (\, \epsilon\, )}\, \equiv
\,\, H_{\: 0}^{\; e.\, m.\, }+
\, e\; \int_{\, \lambda}^{}\; d^{\,\, 3}\, x\;\,\, j_{\: nr\,\, \mu}^{\: (dip)\, ,\; (\, \epsilon\, )}\, (\, x\, )\,\,\, A^{\; \mu\; }
(\, \mathbf {x}\, )\, -\, e^{\: 2}\,\, [\:\, \theta\: (\, -\; t\, )\;\, 
\tilde {v}_{\, in}^{\; 2}\;\, 
\nonumber\\
+\; \theta\: (\, t\, )\;\, \tilde {v}_{\, out}^{\; 2}\; ]\,\,\, \tilde {z}\,\,\, e^{\, -\, 2\,\, \epsilon\; \vert\, t\, \vert}\, 
\equiv\,\, H_{\: 0}^{\; e.\, m.}\, +\, H_{\, int\, ,\; ren}^{\; (FGB)\, ,\; (dip)\, ,\,\, (\, \epsilon\, )\; },
\end {eqnarray}
with 
\begin {equation}\label {quadricorrente dipolare epsilon}
j_{\: nr\:\, \mu}^{\: (dip)\, ,\; (\, \epsilon\, )}\, (\, x\, )\, \equiv\; j_{\: nr\:\, \mu}^{\: (dip)}\; (\, x\, )\,\,\, e^{\, -\, \epsilon\; \vert\, t\, \vert}\; ,\quad\quad\quad
\end {equation}
\begin {equation}\label {controtermine FGB dipolare}
\tilde {z}\; =\;\, \frac {3} {2}\;\, z\; =\;\, \frac {1} {2}\,\, \int_{\; \mathbf {k}\, >\, \lambda}\; \frac {\, d^{\,\, 3\, }k} {(\, 2\; \pi\, )^{\: 3}
\:\, \mathbf {k}^{\: 2}\, }\,\,\, \tilde {\rho\; }^{2\; }(\, \mathbf {k}\, )\; .
\end {equation}
For positive times, the corresponding evolution operator in the interaction picture is 
\begin {equation}\label {operatore di evoluzione PFBR}
U_{\,\, I\, ,\,\, \tilde {v\, }_{out}^{}\, ,\,\, \tilde {z}}^{\, (\, \lambda\, )\, ,\; (\, \epsilon\, )}\; (\, t\, )\, =\:\, 
\tilde {c}_{\; \tilde {v}_{\, out}^{}\, ,\; \tilde {z}}^{\; (\, \epsilon\, )}\; (\, t\, )\;\, \exp\; (\, -\, i\; e\;\, 
[\:\, a_{\, \lambda}^{\; \dagger\; }(\, f_{\; \tilde {v}_{\, out}^{}}^{\: (\, \epsilon\, )}\, (\, t\, )\, )\, +\,
a_{\, \lambda}^{}\: (\, \overline {f\, }_{\tilde {v}_{\, out}^{}}^{\: (\, \epsilon\, )}\, (\, t\, )\, )\, ]\, )\: ,
\quad\quad\;\;\;\;\,\,\,
\end {equation}
with
\begin {equation}
\tilde {c}_{\; \tilde {v}_{\, out}^{}}^{\; (\, \epsilon\, )}\, (\, t\, )\, =\;\, \exp\; (\, -\, \frac {\,\, i\; e^{\, 2}\; \tilde {v}^{\; 2}} {2\; }\,\,\, d^{\; (\, \epsilon\, )}\; (\, t\, )\, )\, ,\quad\quad\quad\quad\quad\quad\;
\end {equation}
\begin {equation}
\tilde {c}_{\; \tilde {v}_{\, out}^{}\, ,\,\, \tilde {z}}^{\; (\, \epsilon\, )}\; (\, t\, )\, \equiv
\;\, \tilde {c}_{\; \tilde {v}_{\, out}^{}}^{\; (\, \epsilon\, )}\, (\, t\, )
\,\,\, \exp\; (\, -\; i\; e^{\, 2}\; \tilde {z}\,\,\, \tilde {v}_{\, out}^{\; 2}\,\,
\frac {\; e^{\, -\, 2\,\, \epsilon\:\, t}\, -\, 1\, } {2\,\, \epsilon}\,\, )\; ,
\end {equation}
\begin {equation}\label {funzioni di coerenza PFBR}
f_{\,\, \tilde {v}_{\, out}^{}}^{\; (\, \epsilon\, )\,\, \mu}\; (\: t\, ,\; \mathbf {k}\, )\, =\,\, \frac {\; \tilde {\rho}\; (\, \mathbf {k}\, )
\,\,\, \tilde {v}_{\, out}^{\; \mu}\, } {(\, 2\; \pi\, )^{\; 3\, /\, 2}\,\, \sqrt {\; 2\,\, \mathbf {k}\, }}\;\, 
\frac {\; e^{\; (\, i\; \mathbf {k}\, -\; \epsilon\, )\,\, t}\, -\, 1\, } 
{i\,\, \mathbf {k\, }-\, \epsilon}\; \cdot
\quad\quad\quad\quad\;\;\;
\end {equation}
For $\, t<0\, ,\, $ one has to replace $\, \epsilon\, $ by $\, -\, \epsilon\, $ and $\, \tilde {v}_{\, out}^{}\, $ by
$\, \tilde {v}_{\, in}^{}$ in the expressions above.\\
The asymptotic limits are given by the following
\begin {theorem}\label {proposizione 2}
The large-time limits and the adiabatic limit of the time-evolution operator (\ref {operatore di evoluzione PFBR}) 
\begin {equation}\label {Moller Feynman PFBR}
\Omega_{\; \pm\; ,\,\, \tilde {v\, }_{\mp}^{}}^{\, (\, \lambda\, )\, }=\;\, 
\tau_{\; w}^{}-\lim_{\epsilon\; \rightarrow\; 0}\;\, 
\Omega_{\; \pm\, ,\,\, \tilde {v\, }_{\mp}^{}}^{\: (\, \lambda\, )\, ,\,\, (\, \mp\, \epsilon\, )}\, 
=\;\, \exp\,\, (\; i\; e\;\, [\,\, a_{\, \lambda}^{\; \dagger}\: (\, f_{\,\, \tilde {v\, }_{\mp}^{}\, }^{})+
\, a_{\, \lambda}^{}\; (\, \overline {f\, }_{\tilde {v\, }_{\mp}^{}\, }^{})\, ]\, )\: ,
\end {equation}
\begin {eqnarray}\label {Moller Feynman PFBR epsilon}
\Omega_{\; \mp\, ,\,\, \tilde {v\, }_{\pm}^{}}^{\: (\, \lambda\, )\, ,\,\, (\, \pm\, \epsilon\, )}\, \equiv\,\, \tau_{\; w}^{}-
\lim_{t\,\, \rightarrow\, \pm\, \infty}\; U_{\,\, I\; ,\,\, \tilde {v\, }_{\pm}^{}\, ,\,\, \tilde {z}}^{\, (\, \lambda\, )\, ,\; (\, \pm\, \epsilon\, )}
\; (\, t\, )\, =\,\, \tilde {c}_{\; \tilde {v\, }_{\pm}^{}\, ,\,\, \tilde {z}}^{\; (\, \pm\, \epsilon\, )}
\quad\quad\quad\quad\quad\quad\quad\quad\, \nonumber\\
\times\; \exp\,\, (\; i\; e\;\, [\,\, a_{\, \lambda}^{\; \dagger}\: (\, f_{\,\, \tilde {v\, }_{\pm}^{}\, }^{\; (\, \pm\, \epsilon\, )})+
\, a_{\, \lambda}^{}\; (\, \overline {f\, }_{\tilde {v\, }_{\pm}^{}\, }^{\; (\, \pm\, \epsilon\, )})\, ]\, )\: ,
\quad\quad\quad\quad\quad\quad\quad\,
\end {eqnarray}
\begin {equation}
\tilde {c}_{\; \tilde {v\, }_{\pm}^{}\, ,\,\, \tilde {z}}^{\; (\, \pm\, \epsilon\, )}\, \equiv\,\, \lim_{t\,\, \rightarrow\, \pm\, \infty}
\; \tilde {c}_{\; \tilde {v\, }_{\pm}^{}\, ,\,\, \tilde {z}}^{\; (\, \pm\, \epsilon\, )}\; (\, t\, )\; ,
\quad\quad\quad\quad\quad\quad\quad\quad\quad\quad\quad\quad
\quad\quad\quad\quad\quad\quad\;
\end {equation}
\begin {equation}\label {f Moll PFBR}
f_{\; \tilde {v}_{\, \pm}^{}}^{\; \mu\; }(\, \mathbf {k}\, )\, =\, \frac {\; \tilde {\rho\;}(\, \mathbf {k}\, )
\,\,\, \tilde {v\, }_{\pm}^{\, \mu}\, } {(\, 2\; \pi\, )^{\; 3\, /\, 2}\; \sqrt {\; 2\,\, \mathbf {k}}}\;\, 
\frac {\, i} {\, \mathbf {k}^{}\, }\; ,\,\, \tilde {v}_{\: +}^{}\equiv\,\, \tilde {v}_{\, out\,\, }^{},
\,\, \tilde {v}_{\, -}^{}\equiv\,\, \tilde {v}_{\; in}^{}\, ,
\quad\quad\quad\quad\quad\quad\quad
\end {equation}
exist on $\, \mathscr {G}_{\,\, 0}^{}\, ,$ for a fixed value of the infrared cutoff $\, \lambda\, ,$ and define 
M\"{o}ller operators as isometries of $\, \mathscr {G}.$ 
\end {theorem}
\begin {proof}
By (\ref {commutatore W})-(\ref {aspettazione esponenziale}) in Appendix \ref {app:2}, the convergence 
of the coherence functions (\ref {funzioni di coerenza PFBR}) in $\, L^{\, 2}$ implies the weak 
convergence of the corresponding expectations on $\, \mathscr {G}_{\:\, 0}^{}$ of polynomials 
and exponentials of the smeared photon variables. 
The M\"{o}ller operators (\ref {Moller Feynman PFBR}) are invertible and preserve the inner 
product $\, \langle\, .\, ,.\, \rangle\, $ as a consequence of the $\, GNS\, $ representation of 
relations (\ref {proprieta' W}), (\ref {commutatore W}); they define therefore isometries of 
$\, \mathscr {G},\, $ uniquely determined\footnote {This result follows from 
Lemma \ref {Lemma 9} in Appendix \ref {app:3}.}  
by their restrictions to $\, \mathscr {G}_{\:\, 0\, }^{}.$
\end {proof}
As in the Coulomb-gauge formulation of the model, we define a scattering operator 
\begin {equation}\label {matrice di scattering PFBR}
S_{\; \lambda\, ,\,\, \tilde {v}_{\, out}^{}\,\, \tilde {v\, }_{in}^{}}^{\,\, (FGB)\, ,\,\, (dip)}\, =\;\, \tau_{\; w}^{}-\lim_{\epsilon\,\, \rightarrow\; 0}
\,\, S_{\; \lambda\, ,\,\, \tilde {v}_{\, out}^{}\,\, \tilde {v\, }_{in}^{}}^{\: (FGB)\, ,\,\, (dip)\, ,\; (\, \epsilon\, )}\, ,\;\;
\end {equation}
\begin {equation}\label {matrice di scattering PFBR adiabatica}
S_{\; \lambda\, ,\,\, \tilde {v}_{\, out}^{}\,\, \tilde {v\, }_{in}^{}}^{\: (FGB)\, ,\,\, (dip)\, ,\; (\, \epsilon\, )\, }
\equiv\;\, \Omega_{\; -\, ,\;\, \tilde {v}_{\, out}^{}}^{\: (\, \lambda\, )\, ,\,\, (\,\epsilon\, )\; \dagger}
\,\,\, \Omega_{\; +\, ,\:\, \tilde {v\, }_{in}^{}}^{\: (\, \lambda\, )\, ,\,\, (\, -\, \epsilon\, )}\; .\quad\quad
\end {equation}

We give here a preliminary discussion of the extent and limitations of the comparison of this model with 
the perturbative (infrared) structures and results.

In Section \ref {sect:3} we prove that the perturbative expressions for the overall soft-photon
radiative corrections to the process $\, \mathbf {p}_{\, in}^{}\rightarrow\mathbf {\, p}_{\, out}^{}\, ,\, $ in 
the presence of a dipole approximation, are indeed reproduced in terms of the M\"{o}ller operators 
of the $\, PFB\, $ model, \emph {both} in the Coulomb gauge and in the $FGB\, $ gauge.
In particular, the Coulomb-gauge expression for such corrections is given by the vacuum 
expectation 
\begin {equation}\label {elemento di matrice Coul}
(\, \Psi_{\, F}^{}\, ,\, S_{\, \lambda\, ,\; \mathbf {p}_{\, out}^{}\,\, \mathbf {p}_{\, in}^{}}^{\; (Coul)\, ,\,\, (dip)}\;\, \Psi_{\, F}^{}\, )\, =
\;\, \exp\; (\, -\, \frac {\, e^{\, 2}} {\, 3\; m^{\, 2}\, }\,\,\, \vert\; \mathbf {p}_{\, out}^{}-\, \mathbf {p}_{\, in}^{}\, \vert^{\,\, 2}
\int_{\; \mathbf {k}\, >\, \lambda}\; \frac {d^{\,\, 3\, }k} {(\, 2\; \pi\, )^{\, 3}\;\, 2\,\, \mathbf {k}^{\; 3}\, }\,\,\, 
{\tilde {\rho\; }}^{2\; }(\, \mathbf {k}\, )\, )\: ,
\end {equation}
for each 
value of the infrared cutoff $\, \lambda\, .\, $
Likewise, the soft-photon radiative corrections to the same process 
in the $FGB\, $ gauge are recovered in 
terms of the vacuum expectation
\begin {equation}\label {elemento di matrice rel}
\langle\; \Psi_{\, 0}^{}\, ,\, S_{\; \lambda\, ,\; \tilde {v\, }_{out}^{}\; \tilde {v\, }_{in}^{}}^{\; (FGB)\, ,\,\, (dip)}\;\, \Psi_{\, 0}^{}
\, \rangle\, =\;\, \exp\; (\, -\, \frac {\, e^{\, 2}} {\, 2\; m^{\, 2}\, }\,\,\, \vert\; \mathbf {p}_{\, out}^{}-\, \mathbf {p}_{\, in}^{}
\, \vert^{\:\, 2}\int_{\; \mathbf {k}\, >\, \lambda}\; \frac {d^{\; 3\, }k} {(\, 2\; \pi\, )^{\, 3}\;\, 2\,\, \mathbf {k}^{\; 3}\, }
\,\,\, {\tilde {\rho\; }}^{2\; }(\, \mathbf {k}\, )\, )\: .\quad
\end {equation}
However, the expression on the r.h.s. of (\ref {elemento di matrice rel}) is not equal to its Coulomb-gauge 
counterpart (\ref {elemento di matrice Coul}), while radiative corrections are not expected to depend upon 
the gauge employed in their calculations.

In order to clarify this apparent paradox, we recall that an invariance property with respect to 
gauge employed in their calculation\footnote {In the literature of theoretical physics, this 
property is generally referred to as 
the gauge-invariance of Feynman amplitudes. However, since gauge-invariance is actually a concept 
with much broader scope (\cite {SW74}), we prefer to avoid the use of such a terminology.} holds indeed 
for the expression of the soft-photon corrections, as given in the $FGB$ gauge (for a fixed infrared 
cutoff) by the \emph {standard} low-energy Feynman rules, introduced and employed for instance 
in \cite {Wein}, Chap. 13. 
If additional infrared approximations are adopted, such an invariance property is however no more 
guaranteed. 

In Section \ref {sect:3} we will see that for the $FGB$-gauge soft-photon corrections to the scattering of 
a single charged particle to be invariant in the above sense, it is crucial that
the classical four-current of the associated Hamiltonian model fulfills 
the continuity equation.
Furthermore, we shall also show that the discrepancy between the expression
(\ref {elemento di matrice rel}) and the Coulomb-gauge result 
(\ref {elemento di matrice Coul}) is quantitatively related to 
the non conservation of the four-current (\ref {corrente 
dipolare quadrivettoriale}).

\section {Bloch-Nordsieck Models And Feynman-Dyson Expansion}
\label {sect:2}

In the present Section we introduce a Hamiltonian model based on an approximation first devised by Bloch 
and Nordsieck (\cite {BN37}). 
As a useful starting point we recall the basic features of the original treatment, which in a nutshell amounts 
to a first-order expansion around a \emph {fixed} four-momentum of each charged particle with respect to 
the energy-momentum transfer. 
Consider the one-particle Dirac Hamiltonian with minimal coupling,
\begin {equation}\label {hamiltoniano BN}
\quad\quad\;  
H\, =\,\, \mathbf {\alpha}\; \cdot\: (\,\, \mathbf {p}\, -\, e\,\, \mathbf {A}\; )\, +\, \beta\,\, m\; +
\, e\,\, A^{\,\, 0\; }\equiv\; H_{\, D}^{}-\, e\;\,\, \mathbf {\alpha}\, \cdot\, \mathbf {A}\; +\, 
e\,\, A^{\; 0}\, ,\quad\quad\quad\quad
\end {equation}
and an eigenstate $\, \psi_{\: +\, ,\,\, p\,\, }^{}(\, x\, )=\; e^{\, -\, i\,\, p\, \cdot\, x}\;\,\, 
\mathcal {U}_{\; r}^{}\, (\, \mathbf {p}\, )\, $ of $\, H_{\, D}^{}\, $ with momentum 
$\mathbf {\, p}\, $ and (positive) energy $E_{\: \mathbf {p}\, }^{},$ 
$\mathcal {U}_{\; r\, }^{}(\, \mathbf {p}\, )\, $ being the corresponding 
momentum-space spinor with helicity $\, r\, .\, $ 
The Dirac equation and the algebraic relations for Dirac's matrices yield
\begin {eqnarray}\label {hamiltoniano BN dall'espansione}
H_{\, D\;\,\, }^{}\psi_{\; 0\, ,\,\, p\,\, }^{}(\, x\, )\, =\,\, [\;\, \mathbf {u}\, \cdot\, \mathbf {p}\; +
\, \sqrt {\, 1\, -\, \mathbf {u}^{\: 2\, }}\,\,\, m\,\, ]\;\, \psi_{\; 0\, ,\; p\,\, }^{}(\, x\, )
+\, O\, (\: \mathbf {p}-\mathbf {p}_{\; 0}^{}\, )\: ,
\end {eqnarray}
with $\, \psi_{\; 0\, ,\,\, p\,\, }^{}(\, x\, )\equiv\; e^{\, -\, i\; p\, \cdot\, x}\;\, \mathcal {U}_{\; r}^{}\, (\, \mathbf {p}_{\; 0}^{}\, )\, ,\, $
$\mathbf {u}\equiv\, \mathbf {p}_{\; 0\, }^{}/\, E_{\; \mathbf {p}_{\, 0}^{}},\, $ $p_{\; 0}^{}=\, (\, E_{\, \mathbf {p}_{\, 0}^{}}^{},
\, \mathbf {p}_{\: 0}^{}\, )\, $ being a fixed four-momentum.\footnote {We denote the four-velocity by the symbol u
in order to avoid possible ambiguities with the notations of Section \ref {sect:1}.}
More formally, the $\, \mathbf {u}$-dependent terms on the r.h.s. of (\ref {hamiltoniano BN dall'espansione}) 
may be obtained as a result of replacing $\, \mathbf {\alpha}\, $ and $\, \beta\, $ in (\ref {hamiltoniano BN})
respectively by the (diagonal in the spinor indices) matrices $\, \mathbf {u}\, $ and 
$\, \sqrt {\; 1-\mathbf {u}^{\: 2}\, }.\, $
Although the outcome of such an approximation 
may seem to depend upon the linearity of $\, H_{\, D}^{}$ 
with respect to the $\, \mathbf {\alpha}\, $ matrices, it is indeed more general; for instance, the same 
expression would be obtained starting from the eigenvalue equation for the Klein-Gordon 
Hamiltonian.

The above discussion leads to a model defined, respectively in the Coulomb-gauge and in 
the $\, FGB\, $ gauge, by the Hamiltonians
\begin {equation}\label {hamiltoniano coulomb bn}
H^{\; (Coul)}\, = \,\, {\mathbf {p}}\, \cdot\, \mathbf {u}\; +\, H_{\: 0\: ,\,\, C\; }^{\; e.\, m.\, } - 
\, e\;\; \mathbf {u}\, \cdot\mathbf {\, A}_{\, C}^{}\, (\, \rho\, ,\,  {\mathbf {x}}\, )\; ,\quad
\quad\quad\quad
\end {equation}
\begin {equation}\label {hamiltoniano Feynman}
H^{\: (FGB)}\, =\;\, {\mathbf {p}}\, \cdot\, \mathbf {u}\; +\, H_{\; 0\,\, }^{\; e.\, m.}+\, e\,\,\, u
\, \cdot\, A\,\, (\, \rho\: ,\, \mathbf {x}\, )\; ,
\quad\quad\quad\quad\quad\;\,
\end {equation}
with $\, u\equiv\, (\, 1\, ,\, \mathbf {u}\, )\, ,$ $\, \mathbf {u}\, $ being a triple of self-adjoint operators, 
to be identified as the observable associated to the asymptotic velocity of the particle, 
which commute with the Weyl algebra $\, \mathscr {A}_{\; ch}^{}$ generated by $\, \mathbf {x}\, $ and 
$\, \mathbf {p}\, $ and with the polynomial algebras generated by the photon canonical variables.
 
In the sequel we let $\, \alpha\rightarrow\beta\, $ denote the scattering of an electron by a 
potential, with the initial (final) particle-state being of definite momentum and four-velocity 
$\, u_{\, in\, }^{}(\, u_{\, out}^{}\, )\, .\, $
In order to establish an operator formulation of the soft-photon corrections to this process, the full 
structure of (\ref {hamiltoniano coulomb bn}), (\ref {hamiltoniano Feynman}) will not be needed; in fact 
it will suffice to suppose that the dynamics of the charge is governed by classical mechanics 
and that its time evolution is given by $\, \mathbf {x\, }(\, t\, )=\; \theta\, (\, -\, t\, )\; \mathbf {\, u}_{\, in}^{}
\,\, t\, +\, \theta\, (\, t\, )\; \mathbf {\, u}_{\, out}^{}\,\, t\, .$ 

Precisely, starting from the Hamiltonians
\begin {equation}
H_{\; \mathbf {u}}^{\: (Coul)}\, =\,\, H_{\: 0\: ,\,\, C\; }^{\; e.\, m.\, } - 
\, e\;\; \mathbf {u}\, \cdot\mathbf {\, A}_{\, C}^{}\, (\, \rho\, ,\, \mathbf {u}\; t\, )\; ,
\quad\quad\quad\quad\, 
\end {equation}
\begin {equation}
H_{\; \mathbf {u}}^{\: (FGB)}\, =\;\, H_{\; 0\,\, }^{\; e.\, m.}+\, e\,\,\, u
\, \cdot\, A\,\, (\, \rho\: ,\, \mathbf {u}\; t\, )\; ,
\quad\quad\quad\quad\quad\;\,
\end {equation}
indexed by the value of the (classical) velocity $\, \mathbf {u}\, ,$ we introduce the infrared-regularized 
Hamiltonians
\begin {eqnarray}\label {hamiltoniano coulomb bn cl}
H_{\: \lambda}^{\: (Coul)}\, =\; H_{\: 0\: ,\:\, C}^{\; e.\, m.\, }-\, e\; \int_{\, \lambda}^{}\; d^{\,\, 3}\, x
\;\;\, \mathbf {j}_{\; C\; }^{}(\, x\, )\, \cdot\, \mathbf {A}_{\, C\; }^{}(\, \mathbf {x}\, )\, 
\equiv\,\, H_{\: 0\: ,\:\, C}^{\; e.\, m.\, }+\, H_{\, int}^{\: (Coul)}\, ,\quad\;\,\,
\end {eqnarray}
\begin {equation}\label {hamiltoniano Feynman cl}
H_{\: \lambda}^{\: (FGB)}\, =\,\, H_{\; 0}^{\; e.\, m.}+\, e\; \int_{\, \lambda}^{}\; d^{\,\, 3}\, x\;\,\, j_{\; \mu\; }^{}(\, x\, )
\,\,\, A^{\; \mu\; }(\, \mathbf {x}\, )\, \equiv\,\, H_{\; 0\;\, }^{\; e.\, m.}+\, H_{\, int\,\, }^{\: (FGB)}\, .
\quad\quad\quad\quad\quad\;
\end {equation}
Again, the $\lambda$-dependence in $H_{\: 0\: ,\:\, C}^{\; e.\, m.\, },\, H_{\, int}^{\: (Coul)},\, H_{\; 0}^{\; e.\, m.},
\, H_{\, int\,\, }^{\, (FGB)}$ is understood.
The classical four-current 
\begin {eqnarray}\label {quadricorrente classica}
j^{\; \mu\,\, }(\, x\, )\, =\;\, \theta\: (\, -\; t\, )\;\, \rho\; (\, \vert\; \mathbf {x\, }-\mathbf {\, u}_{\, in}^{}\; t\; \vert\, )
\;\, u_{\, in}^{\: \mu}+\, \theta\: (\, t\, )\;\, \rho\; (\, \vert\; \mathbf {x\, }-\mathbf {\, u}_{\, out}^{}\; 
t\; \vert\, )\,\, u_{\, out}^{\: \mu\, }
\; \nonumber\\
=\: (\: j^{\: 0\; }(\, x\, )\, ,\mathbf {\; j}\,\, (\, x\, )\, )\, \equiv\,\, j_{\, in}^{\; \mu}\, (\, x\, )\, +\, j_{\, out}^{\; \mu}\, (\, x\, )\, 
\end {eqnarray}
and the classical current
\begin {equation}\label {corrente trasversa}
\mathbf {j}_{\; C}^{\:\, l}\, (\: t\, ,\: \mathbf {x}\, )\, =\, \int\; d^{\,\, 3}\, y\;\;\, \delta_{\; tr}^{\,\, l\, m}\: (\, \mathbf {x}\, -
\, \mathbf {y}\, )\,\,\, \mathbf {\, j}^{\; m\; }(\: t\, ,\: \mathbf {y}\, )\: \quad\quad\quad\,\, 
\end {equation}
fulfill the continuity equation 
$\, \partial_{\, \mu\; }^{}j^{\; \mu\; }(\, x\, )= 0\, $ and the Coulomb-gauge condition 
$\, \partial_{\: l}^{}\; \mathbf {j}_{\; C}^{\;\, l}\, (\, x\, )= 0\, ,\, $ respectively.
The quantum e.m. potentials occurring in the interaction Hamiltonians of (\ref {hamiltoniano coulomb bn cl}), 
(\ref {hamiltoniano Feynman cl}) will be interpreted as describing the 
soft-photon degrees of freedom. 

We first consider the model with (Coulomb-gauge) Hamiltonian (\ref {hamiltoniano coulomb bn cl}).
The Hilbert space of photon states of this model is $\, \mathscr {H}_{\; C}^{}\equiv\mathscr {F},$ with $\mathscr {F}\, $ 
the boson Fock space already considered in Section \ref {sect:1}.
Proceeding in a similar way as in the proof of self-adjointness of the Pauli-Fierz Hamiltonian, one finds 
that $H_{\, \lambda}^{\,\, (Coul)}\, $ is e.s.a. on $\, D_{\, F_{\: 0}^{}}^{}\subset {\mathscr {H}}_{\; C}^{}\, ;\, $ 
Stone's theorem then implies the existence and uniqueness of the dynamics.

With the aid of (\ref {caso particolare formula BH}) and employing the expansion (\ref {potenziale vettore 
Coulomb}) and the Fourier transform of $\, \mathbf {j}_{\; C\; }^{}(\, x\, )$ with respect to
$\, \mathbf {x}\, ,\, $ given by
\begin {equation}
\tilde\mathbf {j}_{\; C}^{\,\, r}\, (\, t\, ,\: \mathbf {k}\, )\, =\;\, \tilde {\rho\; }(\, \mathbf {k}\, )\;\, P^{\; r\, s}\, (\, \mathbf {k}\, )\;\, [\;\, \theta\: (\, -\; t\, )\;\, \mathbf {u}_{\, in}^{\: s}\,\, e^{\,-\, i\; \mathbf {u}_{\, in}^{}\cdot\; \mathbf {k}\,\, t}+\, \theta\: (\, t\, )\;\, \mathbf {u}_{\, out}^{\: s}\,\, e^{\,-\, i\; \mathbf {u}_{\, out}^{}\cdot\; \mathbf {k}\,\, t}\,\, ]\; ,
\end {equation}
we obtain a solution of the equations of motion in the interaction 
representation, in the form
\begin {equation}
\mathscr {U}_{\:\, I\, ,\;\, C\, ,\,\, u}^{\, (\, \lambda\, )}\; (\, t\, )\, =\;\, c_{\,\, u}^{}\; (\, t\, )\,\,\, \exp\; (\; i\; e
\:\, [\,\, a_{\; C\, ,\; \lambda}^{\; *}\; (\; \mathbf {f}_{\; u\; }^{}(\, t\, )\, )+\, a_{\: C\, ,\; \lambda}^{}\; 
(\, \overline \mathbf {\, f\, }_{u\; }^{}(\, t\, )\, )\, ]\, )\; ,
\quad\quad\quad\quad\quad\;
\end {equation}
\begin {eqnarray}
c_{\,\, u}^{}\; (\, t\, )\, =\;\, \exp\,\, (\,\, \frac {\; i\; e^{\: 2}
\; {\mathbf {u}}^{\, 2}} {3\; }\;\; d_{\; u}^{}\; (\, t\, )\, )\: , 
\quad\quad\quad\quad\quad\quad\quad\quad\quad\quad\quad\quad\quad\quad\, 
\nonumber\\
d_{\; u}^{}\: (\, t\, )\, =\; \int_{\; \mathbf {k}\, >\, \lambda}\,\, \frac {\, d^{\,\, 3\, }k} {(\, 2\; \pi\, )^{\: 3}\,\, \mathbf {k}
\,\,\, u\, \cdot\, k\; }\,\,\, {\tilde {\rho\; }}^{2\; }(\, \mathbf {k}\, )\;\, (\,\, t\, -\frac {\; \sin\; u\, \cdot\, k\;\, t\, } 
{\;\;\, u\, \cdot\, k\, }\;\, )\: ,\quad\quad
\end {eqnarray}
\begin {equation}
\mathbf {f}_{\; u\, s}^{}\; (\, t\, ,\: \mathbf {k}\, )\, =\,\, \frac {\; \tilde {\rho\; }(\, \mathbf {k}\, )}{(\, 2\; \pi\, )^{\; 3\, /\, 2}
\; \sqrt {\, 2\,\, \mathbf {k}\, }}\;\,\, \mathbf {u}\, \cdot\; \mathbf {\epsilon}_{\; s}^{}\; (\, \mathbf {k}\, )\,\,\, 
\frac {\; e^{\,\, i\; u\; \cdot\; k\;\, t}\, -\, 1\, }{i\;\, u\, \cdot\, k\, }\,\, ,
\quad\quad\quad\quad\quad\quad
\end {equation}
where $\, k_{\; 0}^{}=\, \mathbf {k}\, $ and $\, u= u_{\, out}^{}\, ,\; t>0\, ,$ $u= u_{\, in}^{}\, ,\; t<0\, .\, $ 

With the same motivations and following the same treatment as in the first Section, we introduce the 
renormalized adiabatic Hamiltonian 
\begin {eqnarray}\label {hamiltoniano fp con controtermine}
H_{\: \lambda\; ,\,\, ren}^{\; (Coul)\, ,\,\, (\, \epsilon\, )}\, =\,\, H_{\: 0\: ,\:\, C}^{\; e.\, m.\, }-\, e\; \int_{\, \lambda}^{}\; d^{\,\, 3\, }x
\;\;\, \mathbf {j}_{\; C}^{\: (\, \epsilon\, )}\, (\, x\, )\, \cdot\, \mathbf {A}_{\, C}^{}\; (\, \mathbf {x}\, )\, 
+\, e^{\: 2}\,\, [\;\, \theta\: (\, -\; t\, )\;\, z_{\; 1\; }^{}(\, u_{\, in}^{}\, )\,\,\, \mathbf {u}_{\, in}^{\: 2}
\;\; \nonumber\\ 
+\,\, \theta\: (\, t\, )\,\,\, z_{\; 1\; }^{}(\, u_{\, out}^{}\, )\,\,\, \mathbf {u}_{\, out}^{\, 2}\; ]\;\; e^{\, -\, 2\:\, \epsilon\; \vert\, t\, \vert}
\, \equiv\,\, H_{\: 0\: ,\:\, C}^{\; e.\, m.}\, +\, H_{\, int\, ,\; ren}^{\, (Coul)\, ,\,\, (\, \epsilon\, )\; },
\end {eqnarray}
with $\, \mathbf {j}_{\; C}^{\: (\, \epsilon\, )\, }(\, x\, )\, \equiv\;\, \mathbf {j}_{\; C}^{}\: (\, x\, )\,\,\, e^{\, -\, \epsilon\,\, \vert\, t\, \vert}\, $ and
\begin {equation}\label {controtermine BN Coulomb}
z_{\; 1\; }^{}(\, u\, )\, =\;\, \frac {1} {3}\;\, \int_{\; \mathbf {k}\, >\, \lambda}\; \frac {\, d^{\,\, 3}\, k} {(\, 2\; \pi\, )^{\: 3}
\,\, \mathbf {k}\; }\;\, \frac {\, \tilde {\rho\; }^{2\; }(\, \mathbf {k}\, )} {u\, \cdot\, k\, }\; \cdot\;\;\;\;
\end {equation}
Kato's result on time-dependent Hamiltonians applies as in Section \ref {sect:1} and, using 
(\ref {caso particolare formula BH}), one obtains the corresponding evolution operator in 
the interaction representation; for positive times
\begin {equation}\label {operatore di evoluzione Coulomb rinormalizzato}
\mathscr {U}_{\,\, I\, ,\,\, C\, ,\; u_{\, out}^{}\, ,\,\, z_{\, 1}^{}}^{\, (\, \lambda\, )\, ,\; (\, \epsilon\, )}\, (\, t\, )\, =\;\, 
c_{\; u_{\, out}^{}\, ,\; z_{\, 1}^{}}^{\: (\, \epsilon\, )}\, (\, t\, )\,\,\, \exp\; (\,\, i\; e\;\, [\,\, a_{\: C\, ,\,\, \lambda}^{\: *}
\; (\; \mathbf {f}_{\,\, u_{\, out}^{}}^{\, (\, \epsilon\, )}\, (\, t\, )\, )\, +\, 
a_{\; C\, ,\,\, \lambda}^{}\; (\, \overline\mathbf {\, f\, }_{u_{\, out}^{}}^{\, (\, \epsilon\, )}\, (\, t\, )\, )\, ]\, )\; ,
\quad
\end {equation}
\begin {equation}\label {esponenziale regolarizzato}
c_{\; u_{\, out}^{}\, ,\; z_{\, 1}^{}}^{\: (\, \epsilon\, )}\, (\, t\, )\, \equiv\;\, c_{\,\, u_{\, out}^{}}^{\: (\, \epsilon\, )}\, (\, t\, )
\;\, \exp\; (\; i\; e^{\: 2}\,\, z_{\; 1}^{}\; (\, u_{\, out}^{}\, )\,\,\, {\mathbf {u}_{\, out}^{\: 2}}\,\, 
\frac {\; e^{\, -\, 2\,\, \epsilon\:\, t}\, -\, 1\; } {2\,\, \epsilon\, }\,\, )\; ,
\quad\quad\quad\quad\quad\quad\;\;
\end {equation}
\begin {eqnarray}
d_{\; u_{\, out}^{}}^{\,\, (\, \epsilon\, )}\, (\, t\, )\, =\, -\, \int_{\; \mathbf {k}\, >\, \lambda}\,\, \frac {\, d^{\,\, 3\, }k
\;\,\, {\tilde {\rho\; }}^{2\; }(\, \mathbf {k}\, )} {(\, 2\; \pi\, )^{\; 3}\;\, \mathbf {k}\;\, 
[\, (\, u_{\, out}^{}\cdot\, k\, )^{\; 2}\, +\, \epsilon^{\; 2}\; ]\, }\;\,\, 
[\;\, e^{\, -\, \epsilon\,\, t\, }\; \sin\; u_{\, out}^{}\cdot\: k\;\, t\;
\quad\quad\; \nonumber\\
+\; \frac {\, u_{\, out}^{}\cdot\, k\, } {2\,\, \epsilon}\;\, (\; e^{\, -\, 2\,\, \epsilon\:\, t\, }-\, 1\, )\; ]\; ,
\end {eqnarray}
\begin {equation}
\mathbf {f}_{\,\, u_{\, out}^{}\: s}^{\, (\, \epsilon\, )}\; (\: t\, ,\; \mathbf {k}\, )\, =\,\, \frac {\; \tilde {\rho\; }(\, \mathbf {k}\, )} 
{(\, 2\; \pi\, )^{\; 3\, /\, 2}\; \sqrt {\; 2\,\, \mathbf {k}\, }}\;\,\, \mathbf {u}_{\, out}^{}\cdot\: \mathbf {\epsilon}_{\: s\; }^{}
(\, \mathbf {k}\, )\,\,\, \frac {\; e^{\,\, i\; u_{\, out}^{}\, \cdot\,\, k\,\, t\,\, -\,\, \epsilon\,\, t}\, -\, 1\, }
{i\,\, u_{\, out}^{}\cdot\, k\, -\, \epsilon}\,\, \cdot\quad\quad\quad\quad\quad\;\;\,
\end {equation}
The results holding for $\, t<0\, $ are obtained by replacing $\, u_{\, out\, }^{}$ and $\, \epsilon\, $ 
respectively by $u_{\, in\, }^{}$ and $\, -\, \epsilon\, $ in the expressions above.
One can then prove the following 
\begin {theorem}\label {proposizione 3}
The evolution operator (\ref {operatore di evoluzione Coulomb rinormalizzato}) in the interaction 
representation corresponding to the renormalized Hamiltonian (\ref {hamiltoniano fp con 
controtermine}) admits asymptotic limits, yielding the M\"{o}ller operators in the Coulomb 
gauge, for each value of the infrared cutoff $\, \lambda:$
\begin {eqnarray}\label {Moller BN Coulomb}
\Omega_{\; \pm\, ,\;\, C\, ,\,\, u_{\, \mp}^{}}^{\: (\, \lambda\, )}=\;\, s\; -\, \lim_{\epsilon\; \rightarrow\; 0}
\; \lim_{t\,\, \rightarrow\, \mp\, \infty}\; 
\mathscr {U}_{\,\, I\, ,\;\, C\, ,\; u_{\, \mp}^{}\, ,\,\, z_{\, 1}^{}}^{\, (\, \lambda\, )\, ,\; (\, \mp\, \epsilon\, )}\; (\, t\, )
\, =\;\, s\; -\, \lim_{\epsilon\; \rightarrow\; 0}\,\, \Omega_{\; \pm\, ,\;\, C\, ,\,\, u_{\, \mp}^{}}^{\; (\, \lambda\, )\, ,\; (\, \mp\, \epsilon\, )}
\; \nonumber\\
=\;\, \exp\; (\, -\; i\; e\;\, [\,\, a_{\: C\, ,\,\, \lambda}^{\: *}\: (\; \mathbf {f}_{\; u_{\, \mp}^{}}^{}\, )\, +
\, a_{\: C\, ,\,\, \lambda}^{}\: (\: \overline\mathbf {\, f\: }_{u_{\, \mp}^{}}^{}\, )\, ]\, )\; ,
\end {eqnarray}
\begin {equation}\label {coerenza Coulomb}
\mathbf {f}_{\; u_{\, \pm}^{}\, s}^{}\; (\, \mathbf {k}\, )\, =\,\, \frac {\; \tilde {\rho\; }(\, \mathbf {k}\, )} 
{(\, 2\; \pi\, )^{\; 3\, /\, 2}
\; \sqrt {\; 2\,\, \mathbf {k}\, }}\,\,\, \frac {\; i\:\, \mathbf {u}_{\, \pm}^{}\cdot\, \mathbf {\epsilon}_{\; s}^{}\, 
(\, \mathbf {k}\, )\, }{u_{\, \pm}^{}\cdot\, k\, }\,\, ,
\; u_{\, +}^{}\equiv\; u_{\, out\; }^{},\; u_{\, -}^{}\equiv\; u_{\, in}^{}\, .
\quad\quad\quad\quad\;
\end {equation}
The $\, t\, \rightarrow\pm\, \infty\, $ and $\, \epsilon\, \rightarrow\, 0\, $ limits in (\ref {Moller BN Coulomb}) 
exists in the strong topology. 
\end {theorem}
\begin {proof}
The control of the limits is very similar to the case discussed in Proposition \ref {proposizione 1}.
\end {proof}
\ni
We introduce the Coulomb-gauge photon scattering operator of the model,
\begin {equation}\label {matrice S Coulomb BN}
S_{\; \lambda\, ,\; u_{\, out}^{}\; u_{\, in}^{}}^{\; (Coul)}=\;\, s\, -\lim_{\epsilon\,\, \rightarrow\; 0}
\,\, S_{\; \lambda\, ,\; u_{\, out}^{}\; u_{\, in}^{}}^{\: (Coul)\, ,\; (\, \epsilon\, )}\; ,\quad\; 
\end {equation}
\begin {equation}\label {matrice S Coulomb BN adiabatica}
S_{\; \lambda\, ,\; u_{\, out}^{}\; u_{\, in}^{}}^{\: (Coul)\, ,\; (\, \epsilon\, )}\, =\;\, 
\Omega_{\; -\, ,\;\, C\, ,\,\, u_{\, out}^{}}^{\, (\, \lambda\, )\, ,\,\, (\, \epsilon\, )\; \dagger}
\,\,\, \Omega_{\; +\, ,\;\, C\, ,\,\, u_{\, in}^{}}^{\, (\, \lambda\, )\, ,\,\, (\, -\, \epsilon\, )}\, . 
\quad
\end {equation}

We shall now turn to the formulation of the model in the $FGB\, $ gauge.
Concerning the problems posed by the absence of a positive scalar product on the state space of
the model, we adopt the same choices as in Section \ref {sect:1}, following the same steps with 
identical results; in particular, as discussed in detail in Appendix \ref {app:2}, a 
(generalized) $\, GNS$ construction provides an indefinite space $\, \mathscr {G}^{}\, $ 
and a weakly dense domain $\, \mathscr {G}_{\,\, 0}^{}\, .\, $ 
Isometric evolution operators on $\, \mathscr {G}_{\,\, 0}^{}\, $ are constructed as in 
Section \ref {sect:1} and are unique in the same sense. 

With the aid of the expansion (\ref {espansione quadripotenziale}) and of the Fourier transform
of $\, j^{\: \mu}\, (\, x\, )\, $ with respect to $\, \mathbf {x}\, ,$ whose explicit expression is
\begin {equation}
 \tilde {j\, }^{\, \mu}\, (\, \, t\, ,\: \mathbf {k}\, )\, =\,\, \tilde {\rho\; }(\, \mathbf {k}\, )\;\, [\;\, \theta\: (\, -\; t\, )\;\, u_{\, in}^{\; \mu}\,\, e^{\,-\, i\; \mathbf {u}_{\, in}^{}\, \cdot\; \mathbf {k}\,\, t}+\, \theta\: (\, t\, )\;\, u_{\, out}^{\; \mu}\,\, e^{\,-\, i\; \mathbf {u}_{\, out}^{}\, \cdot\; \mathbf {k}\,\, t}\,\, ]\; ,
\end {equation}
we obtain the evolution operator in the interaction 
representation, in the form
\begin {equation}\label {operatore formale}
\mathscr {U}_{\,\, I\: ,\,\, u\,\, }^{\, (\, \lambda\, )}(\, t\, )\, =\;\, \mathscr {C}_{\,\, u}^{}\; (\, t\, )
\,\,\, \exp\,\, (\, -\; i\; e\;\, [\,\, a_{\, \lambda}^{\; \dagger}\, (\, f_{\; u}^{}\, (\, t\, )\, )\, +\,
a_{\, \lambda\; }^{}(\; \overline {f\, }_{u}^{}\, (\, t\, )\, )\, ]\, )\: ,\;
\end {equation}
\begin {eqnarray}
\mathscr {C}_{\,\, u}^{}\; (\, t\, )\, =\;\, \exp\,\, (\, -\, \frac {\;\, i\; e^{\, 2}\, u^{\: 2}} {2\, }
\;\,\, d_{\; u\,\, }^{}(\, t\, )\, )\: ,
\quad\quad\quad\quad\quad\quad\quad\quad\;\;\; \nonumber\\
f_{\; u}^{\; \mu}\; (\, t\, ,\: \mathbf {k}\, )\, =\,\, \frac {\; \tilde {\rho\; }(\, \mathbf {k}\, )\;\, u^{\, \mu}\, } 
{(\, 2\; \pi\, )^{\, 3\, /\, 2}\,\, \sqrt {\; 2\,\, \mathbf {k}\, }}\;\, \frac {\; e^{\; i\,\, u\; \cdot\,\, k\,\,\, t}\, -\, 1\; } 
{i\;\, u\, \cdot\, k}\; , 
\quad\quad\quad\;\;\;\;\;\;\; \nonumber
\end {eqnarray}
where $k_{\; 0}^{}=\mathbf {k}\, $ and $\, u= u_{\, out}^{}\, ,\, t>0\, ,$ $u= u_{\, in}^{}\, ,\, t<0\, .\, $
 
In order to construct M\"{o}ller operators, we introduce the adiabatic Hamiltonian
\begin{equation}\label {hamiltoniano Feynman adiabatico}
H_{\; \lambda}^{\; (FGB)\, ,\,\, (\, \epsilon\, )\; }=\;\, H_{\; 0}^{\; e.\, m.}\, +\, e\; \int_{\, \lambda}^{}
\; d^{\,\, 3}\, x\;\,\, j_{\; \mu}^{\: (\, \epsilon\, )}\; (\, x\, )\,\,\, A^{\; \mu\; }(\, \mathbf {x}\, )\; ,
\quad\quad\quad\quad
\end {equation}
\begin {equation}\label {decomposizione quadricorrente adiabatica}
j_{\; \mu}^{\: (\, \epsilon\, )}\: (\, x\, )\, \equiv\,\, j_{\; \mu}^{}\: (\, x\, )\,\,\, e^{\, -\, \epsilon\,\, \vert\, t\, \vert}\, =
\,\, j_{\: in\, ,\,\, \mu}^{\: (\, -\, \epsilon\, )}\: (\, x\, )\, +\, j_{\: out\, ,\,\, \mu}^{\: (\, \epsilon\, )}\: (\, x\, )\; ,\;\;
\end {equation}
and renormalize it via suitable counterterms; the resulting expression is 
\begin {eqnarray}\label {hamiltoniano Feynman rinormalizzato}
H_{\; \lambda\; ,\,\, ren}^{\; (FGB)\, ,\,\, (\, \epsilon\, )\; }=\;\, 
H_{\; \lambda}^{\; (FGB)\, ,\,\, (\, \epsilon\, )}\, -\, e^{\, 2}\,\, [\;\, \theta\: (\, -\; t\, )\,\,\, z_{\; 2}^{}\; (\, u_{\, in}^{}\, )\;\, 
u_{\, in}^{\: 2}
+\, \theta\: (\, t\, )\,\,\, z_{\; 2}^{}\; (\, u_{\, out}^{}\, )\;\, u_{\, out}^{\: 2}\; ]
\,\,\, e^{\, -\, 2\,\, \epsilon\; \vert\, t\, \vert}\; ,
\end {eqnarray}
with
\begin {equation}\label {coefficiente controtermine covariante BN}
z_{\; 2\,\, }^{}(\, u\, )\, =\;\, \frac {3} {2}\;\,\, z_{\; 1}^{}\; (\, u\, )\, =\;\, \frac {1} {2}
\;\, \int_{\; \mathbf {k}\, >\, \lambda}\; \frac {\, d^{\,\, 3\, }k}
{(\, 2\; \pi\, )^{\; 3}\,\, \mathbf {k}\, }\;\, \frac {\; \tilde {\rho\; }^{2\; }(\, \mathbf {k}\, )} 
{u\, \cdot\, k\, }\,\, \cdot\quad\,\, 
\end {equation}
For $\, t>0\, $, the corresponding evolution operator in the interaction representation is
\begin {equation}\label {operatore di evoluzione BN rinormalizzato}
\mathscr {U}_{\,\, I\, ,\; u_{\, out}^{}\, ,\; z_{\, 2}^{}}^{\, (\, \lambda\, )\, ,\; (\, \epsilon\, )}\; (\, t\, )\, =
\;\, \mathscr {C}_{\; u_{\, out}^{}\, ,\; z_{\, 2}^{}}^{\; (\, \epsilon\, )}\: (\, t\, )\,\,\, 
\exp\; (\; i\; e\:\, [\,\, a_{\, \lambda}^{\; \dagger}\; (\, f_{\; u_{\, out}^{}}^{\, (\, \epsilon\, )}\, (\, t\, )\, )\, +
\, a_{\, \lambda}^{}\: (\, \overline {f\, }_{u_{\, out}^{}}^{\; (\, \epsilon\, )}\, (\, t\, )\, )\, ]\, )\; ,
\quad\;\;
\end {equation}
\begin {equation}\label {coefficiente rinormalizzato}
\mathscr {C}_{\; u_{\, out}^{}\, ,\; z_{\, 2}^{}}^{\; (\, \epsilon\, )}\: (\, t\, )\, \equiv\;\, 
\mathscr {C}_{\,\, u_{\, out}^{}}^{\; (\, \epsilon\, )}\, (\, t\, )\;\, \exp\; (\, -\, i\; e^{\: 2}
\,\, z_{\; 2}^{}\; (\, u_{\, out}^{}\, )\;\, u_{\, out}^{\: 2}\,\, 
\frac {\,\, e^{\, -\, 2\,\, \epsilon\,\, t}\, -\, 1\; } {2\,\, \epsilon}\,\, )\; ,
\quad\quad\quad\quad\,\,\,\,
\end {equation}
\begin {equation}
f_{\; u_{\, out}^{}}^{\: (\, \epsilon\, )\, ,\:\, \mu}\; (\: t\, ,\: \mathbf {k}\, )\, =\; \frac {\; \tilde {\rho\; }(\, \mathbf {k}\, )
\,\,\, u_{\, out}^{\, \mu}\, } {(\, 2\; \pi\, )^{\; 3\, /\, 2}\; \sqrt {\; 2\,\, \mathbf {k}\, }}\;\, 
\frac {\,\, e^{\,\, i\; u_{out\, }^{}\cdot\,\, k\,\, t\;\, -\; \epsilon\,\, t}\; -\, 1\; } {i\,\, u_{\, out}^{}\cdot\, k\, -
\, \epsilon}\,\, \cdot
\quad\quad\quad\quad\quad\quad\quad\quad\quad
\quad\quad
\end {equation}
The results holding for $\, t<0\, $ are obtained by replacing $\, \epsilon\, $ by $\, -\, \epsilon\, $ 
and $\, u_{\, out}^{}$ by $\, u_{\, in}^{}$ in the expressions above.
The asymptotic limits are controlled through the following
\begin {theorem}\label {proposizione 4}
The large-time limits and the adiabatic limit of the evolution operator 
(\ref {operatore di evoluzione BN rinormalizzato}), 
corresponding to the renormalized adiabatic Hamiltonian (\ref {hamiltoniano Feynman rinormalizzato}),
\begin {equation}\label {Moller Feynman BN}
\Omega_{\; \pm\, ,\,\, u_{\, \mp}^{}}^{\: (\, \lambda\, )}\equiv\;\, \tau_{\; w}^{}\, -\, \lim_{\epsilon\; \rightarrow\; 0}
\;\, \Omega_{\; \pm\, ,\,\, u_{\, \mp}^{}}^{\, (\, \lambda\, )\, ,\,\, (\, \mp\, \epsilon\, )\, }=\;\, \exp\; (\,\, i\; e\;\, [\,\, 
a_{\, \lambda}^{\; \dagger}\; (\, f_{\; u_{\, \mp\, }^{}}^{})\, +\, a_{\, \lambda}^{}\, (\, \overline {f\, }_{u_{\, \mp}^{}}^{}\, )\; ]\, )\: ,
\end {equation}
\begin {eqnarray}\label {Moller Feynman BN epsilon}
\Omega_{\; \mp\, ,\,\, u_{\, \pm}^{}}^{\, (\, \lambda\, )\, ,\,\, (\, \pm\, \epsilon\, )\, }\equiv
\;\, \tau_{\; w}^{}\, -\, \lim_{t\,\, \rightarrow\,\, \pm\, \infty}\; 
\mathscr {U}_{\,\, I\: ,\,\, u_{\, \pm}^{}\, ,\; z_{\, 2}^{}}^{\, (\, \lambda\, )\, ,\,\, (\, \pm\, \epsilon\, )}
\; (\, t\, )\, =\;\, \mathscr {C}_{\; u_{\, \pm}^{}\, ,\,\, z_{\, 2}^{}}^{\; (\, \pm\, \epsilon\, )}
\quad\quad\quad\quad\quad\quad\;\;\;
\nonumber\\
\times\; \exp\; (\,\, i\; e\:\, [\,\, a_{\, \lambda}^{\; \dagger}\: (\, f_{\; u_{\, \pm}^{}}^{\: (\, \pm\, \epsilon\, )}\, )\, +
\, a_{\, \lambda}^{}\, (\, \overline {f\, }_{u_{\, \pm}^{}}^{\; (\, \pm\, \epsilon\, )}\, )\; ]\, )\: ,\;\;\,
\quad\quad\quad\quad\quad\quad
\end {eqnarray}
\begin {equation}
\mathscr {C}_{\; u_{\, \pm}^{}\, ,\,\, z_{\, 2}^{}}^{\; (\, \pm\, \epsilon\, )}\, \equiv\,\, \lim_{t\,\, \rightarrow\,\, \pm\, \infty}
\;\, \mathscr {C}_{\;  u_{\, \pm}^{}\, ,\,\, z_{\, 2}^{}}^{\; (\, \pm\, \epsilon\, )}\; (\, t\: )\; ,
\quad\quad\quad\quad\quad\quad\quad\quad\quad\quad\quad\quad\quad\quad\quad\quad\quad
\quad\quad\;\,
\end {equation}
\begin {equation}\label {f Moll BN}
f_{\; u_{\, \pm}^{}}^{\; \mu}\, (\, \mathbf {k}\, )\, =\; \frac {\; \tilde {\rho\: }(\, \mathbf {k}\, )\,\,\, u_{\, \pm}^{\, \mu}\, } 
{(\, 2\; \pi\, )^{\; 3\, /\, 2}\; \sqrt {\; 2\,\, \mathbf {k}\, }}\;\, \frac {\;\;\;\, i\, } {u_{\, \pm}^{}\cdot\, k\, }\,\, ,
\,\, {u}_{\, +}^{}\equiv\; {u}_{\, out}^{}\; ,\; {u}_{\, -}^{}\equiv\; {u}_{\, in}^{}\, ,
\quad\quad\quad\quad\quad\quad\;\;\,\,\,
\end {equation}
exist on $\, \mathscr {G}_{\,\, 0}^{}$ and define the M\"{o}ller operators of the four-vector $\, BN$ model
as isometries in $\, \mathscr {G}, $ for each value of the infrared cutoff $\, \lambda\, .$
\end {theorem}
\begin {proof}
The proof is the same as that of Proposition \ref {proposizione 2}.
\end {proof}
We turn now to the comparison of our model with the soft-photon contributions of $\, QED\, .\, $
As regards infrared approximations and results in perturbation theory, we refer to the streamlined 
approach of \cite {JR}, Chap. 16, \cite {Wein}, Chap. 13, while a more detailed analysis can be 
found in the classic paper of Yennie, Frautschi and Suura (\cite {YFS61}; for the convenience of 
the reader, the relevant results of such treatments are recalled below.

The effect to all orders of radiative corrections due to ``virtual soft-photons'' is to multiply the 
$S$-matrix element for a process $\, \gamma\rightarrow\delta\, ,$ involving a fixed number 
of charged particle and photons, by an exponential factor.
Precisely, in presence of an infrared cutoff $\, \lambda\, $ one has
\begin {equation}\label {fattorizzazione QED IR}
S_{\; \lambda\, ,\; \delta\, \gamma}^{}=\,\,\, \exp\; (\; e^{\: 2}\, \mathscr {M}_{\, \lambda\, ,\; \delta\, \gamma}^{}\, )
\,\,\, S_{\; \lambda\, ,\,\, \delta\, \gamma}^{\; (hard)}\; ,
\end {equation}
with $\, S_{\, \lambda\, ,\; \delta\, \gamma}^{\; (hard)}\, $ yielding the overall infrared-finite contributions 
to the transition amplitude $\, \gamma\rightarrow\delta\, .\, $ From a diagrammatic point of
view, it is given by the sum of a family of connected Feynman diagrams, sharing the 
same number of external lines and possibly different internal structure; it is customary 
to represent such a sum by a single diagram, in which the external lines common to 
the diagrams of the family are attached to a bubble.

The exponential on the r.h.s. of (\ref {fattorizzazione QED IR}) gives those radiative corrections 
which are associated to the insertion of soft-photon propagators on the external lines of the 
bubble; such corrections are singular as the low-energy cutoff is removed.
The $\, FGB$-gauge expression of the corresponding exponent is 
\begin {equation}\label {correzioni radiative perturbative}
\mathscr {M}_{\, \lambda\, ,\; \delta\, \gamma\, }^{\; (FGB)}\, =\,\, \frac {1} {2\; (\, 2\; \pi\, )^{\: 3}\, }\, \sum_{m\; ,\,\, n}\; 
\eta_{\; m}^{}\,\, \eta_{\,\, n}^{}\;\;\, p_{\: m}^{}\cdot\, p_{\; n}^{}\; \int_{\; \lambda}^{\,\, \Lambda}\, \frac {\, d^{\,\, 3}\, k} 
{\, 2\,\, \mathbf {k}\,\,\, p_{\: m}^{}\cdot\, k\;\,\, p_{\; n}^{}\cdot\, k\,\, }\;\, ,
\end {equation}
with the sum running over the external four-momenta of the charged particles involved in the process 
$\, \gamma\rightarrow\delta\, ;\, $ in (\ref {correzioni radiative perturbative}), $\, \eta=1\, $ for the outgoing 
electronic lines, $\, \eta=-\, 1\, $ for the incoming ones and $\, \Lambda\, $ is an energy scale, chosen 
in such a way that the approximations made in the analysis of the soft-photon contributions 
to the perturbative series of $\, QED\, $ are justified for photons with energy below 
$\, \Lambda\, .\, $

Concerning the corrections due to soft-photon emission, one has to take into account the contributions 
of additional vertices and electron propagators, which arise by virtue of the insertion of photon legs, 
representing the emitted photons, on the external lines of the bubble.
Since such contributions give rise to infrared divergences in the calculations of inclusive cross-sections, 
a low-energy cutoff on the photon momenta must be introduced. 
As explained in \cite {Wein}, $\S$ 13.3, one has to adopt the same infrared cutoff both for the 
momentum-space integrals which yield the low-energy radiative corrections and for the 
momenta of the emitted photons, in order to preserve unitarity at each stage of the 
calculations.

The Feynman rules which result from the standard low-energy approximations are as follows.\\
A vertex on an external charged line carrying four-velocity $\, u=p\, /\, E_{\: \mathbf {p}}^{}$ 
contributes a 
factor $-\, i\; e\; u^{\, \mu}.$\\ 
An electron propagator carrying four-momentum $\, p+k\, $
is associated to a term $\, i\, /\, (\, u\, \cdot\, k\, +\, i\; \epsilon\, )\, ,$ 
where $\, k\, $ is the 
overall photon contribution to the four-momentum carried by the propagator and $\, i\: \epsilon\, $ is the Feynman prescription for Green's 
functions (for its definition we refer for instance to \cite {MaSh}, $\S$ 3.4).
The rule for the photon propagator is the standard one: Letting 
\begin {equation}\label {propagatore associato}
\Delta_{\, F}^{}\, (\, x\, )\, =\;\, \theta\: (\, x_{\: 0}^{}\, )\,\,\, \Delta^{\, +}\; (\, x\, )\, 
+\, \theta\: (\: -\, x_{\; 0}^{}\, )\,\,\, \Delta^{\, +}\; (\: -\, x\, )\; ,
\quad\quad\;\;\;
\end {equation}
\begin {equation}\label {commutatore Pauli-Jordan}
g^{\: \mu\, \nu}\,\, \Delta^{\, +\; }(\, x\, -\, y\: )\, \equiv\;\, i\,\,\, [\; A_{\, +}^{\; \mu}\; (\, x\, )\, ,
\, A_{\, -}^{\; \nu}\; (\, y\, )\; ]\; ,\quad\quad\quad\quad\quad\quad\;\;\,\,
\end {equation}
\begin {equation}
\Delta^{\, +}\, (\, x\, )\: =\,\, \frac {1} {(\, 2\; \pi\, )^{\: 4}\, }\, \int\, d^{\,\, 4\: }k
\;\;\, e^{\, -\, i\,\, k\; \cdot\,\, x}\;\,\, \Delta^{\, +}\, (\, k\, )\, ,
\quad\quad\quad\quad\quad\quad
\end {equation}
\begin {equation}\label {supporto commutatore Pauli-Jordan}
\Delta^{\, +}\; (\, k\, )\, =\, -\, 2\; \pi\; i\;\,\, \theta\: (\, k^{\: 0}\, )\,\,\, \delta\: (\: k^{\: 2}\, )\, ,
\quad\quad\quad\quad\quad\quad\quad\quad\quad\quad\quad\;\;\;\,
\end {equation}
the space-time expression of the photon propagator is
\begin {equation}\label {propagatore FGB}
i\; \Delta_{\, F}^{\: \mu\, \nu}\, (\, x\, )\: \equiv\,\, \frac {i} {(\, 2\; \pi\, )^{\: 4}\, }\, \int\, d^{\,\, 4\: }k
\;\;\, e^{\, -\, i\,\, k\; \cdot\,\, x}\;\,\, \Delta_{\, F}^{\: \mu\, \nu}\, (\, k\, )\, 
=\, -\, i\; g^{\: \mu\, \nu}\,\, \Delta_{\, F\; }^{}(\, x\, )
\end {equation}
and the term $\, i\; \Delta_{\, F}^{\; \mu\, \nu}\, (\, k\, )\, $ is associated to a (photon) line carrying four-momentum 
$\, k\, $ and joining two vertices with indices $\, \mu\, $ and $\, \nu\, .\, $ 
In (\ref {commutatore Pauli-Jordan}), the positive-frequency part of the Pauli-Jordan function $\, \Delta\: (\, x\, )\, $
of a free massless scalar field has been introduced.
We shall also make use of the well-known relation
\begin {equation}\label {commutatore potenziale}
[\; A^{\; \mu}\; (\, x\, )\, ,\, A^{\; \nu}\; (\, y\, )\; ]\,=\, -\: i\; g^{\: \mu\, \nu}\,\, \Delta\; (\, x\, -\, y\: )\; .
\quad\quad\quad\quad\quad\;
\end {equation}
\par
\bigskip
\ni
Let
\begin {equation}\label {matrice S BN}
S_{\; \lambda\, ,\,\, u_{\, out}^{}\; u_{\, in}^{}}^{\: (FGB)}=\;\, \tau_{\; w}^{}-\lim_{\epsilon\,\, \rightarrow\; 0}
\;\, S_{\;  \lambda\, ,\; u_{\, out}^{}\; u_{\, in}^{}}^{\: (FGB)\, ,\; (\, \epsilon\, )}\; ,
\quad\quad\quad\quad\quad\quad\;\;\;\;
\end {equation}
\begin {equation}\label {matrice S BN adiabatica}
S_{\;  \lambda\, ,\; u_{\, out}^{}\; u_{\, in}^{}}^{\: (FGB)\, ,\; (\, \epsilon\, )}\, =\;\, 
\Omega_{\; -\, ,\,\, u_{\, out}^{}}^{\, (\, \lambda\, )\, ,\,\, (\, \epsilon\, )\; \dagger}
\,\,\, \Omega_{\; +\, ,\,\, u_{\, in}^{}}^{\, (\, \lambda\, )\, ,\,\, (\, -\, \epsilon\, )}\; .
\quad\quad\quad\quad\quad\quad
\quad\quad
\end {equation}
be the $\, S$-matrix for the $\, BN\, $ model in the $\, FGB\, $ gauge.
We wish to show that suitable matrix elements of (\ref {matrice S BN}) reproduce the soft-photon contributions 
to the process $\, \alpha\rightarrow\beta\, ,$ as given by the diagrammatic rules discussed above.
\begin {remark}\label {remark 1}
We shall recover the perturbative expressions corresponding to mass-shell (in the sequel referred to as 
on-shell) renormalization prescriptions, in accordance with the analysis carried out in \cite {JR,Wein}, 
and to the choice of a form factor $\, \tilde {\rho\: }(\, \mathbf {k}\, )$ as an ultraviolet cutoff in 
momentum-space integrals. 
This leads to the following minor change for the vertex rule: A
contribution $-\, i\; e\; \tilde {\rho\: }(\, \mathbf {k}\, )\; u^{\, \mu}$ will be assigned 
to a vertex with index $\, \mu\, $ on a fermion line carrying four-velocity $\, u\, .$ 
\end {remark}
In the following, we let $\, \mathcal {D}\, $ and $\, \mathscr {M}_{\; \beta\, \alpha}^{\; (FGB)}$ denote respectively 
the bubble diagram associated to the basic process $\, \alpha\rightarrow\beta\, $ and the overall 
soft-photon radiative corrections at second order in the $\, FGB\, $ gauge.
Further, we let $\, \mathcal {D}_{\, 1}^{}$ be the diagram obtained by attaching a photon line to the outgoing 
fermion leg of $\, \mathcal {D}\, .\, $
As a result of the extra photon line, $\, \mathcal {D}_{\, 1}^{}$ contains one more vertex and an additional 
electron propagator with respect to $\, \mathcal {D}\, .\, $

We begin our analysis by writing down the amplitudes which contribute to 
$\mathscr {M}_{\; \beta\, \alpha}^{\; (FGB)\, }.\, $ 
By virtue of the (infrared) diagrammatic rules discussed above, the correction associated to a soft 
photon, emitted from the incoming fermion leg of $\, \mathcal {D}\, $ and absorbed from the 
outgoing one, is given by
\begin {eqnarray}\label {inserzione di fotone tra linee}
e^{\, 2}\;\, \Gamma_{\, u_{\, out}^{}\; u_{\,in}^{}}^{}=\, \lim_{\epsilon\,\, \rightarrow\; 0}\,\, (\: i\; e\, )^{\: 2}
\int_{\; \mathbf {k}\, >\, \lambda}\; \frac {\,\, d^{\,\, 4}\, k\;\, \tilde {\rho\; }^{2\; }(\, \mathbf {k}\, )\, } 
{(\, 2\; \pi\, )^{\: 4}\, }\;\, \frac {\; i\; u_{\, out\, ,\,\, \mu}^{}} {\, -\, u_{\, out}^{}\cdot\, k\, +\, i\; \epsilon\, }
\:\,\, i\; \Delta_{\, F}^{\: \mu\, \nu}\, (\, k\, )\:\, \frac {\; i\; u_{\, in\, ,\,\, \nu}^{}} {\, -\, u_{\, in}^{}\cdot\, k
\, +\, i\; \epsilon\, }\; 
\nonumber\\
=\; e^{\: 2}\,\, u_{\, out}^{}\cdot\, u_{\, in}^{}\; \int_{\; \mathbf {k}\, >\, \lambda}\; \frac {\,\, d^{\,\, 4}\, k
\;\, \tilde {\rho\; }^{2\; }(\, \mathbf {k}\, )\, } {(\, 2\; \pi\, )^{\: 4}\, }\;\, \frac {\quad\;\;\;\;\; i} 
{\, -\, u_{\, out}^{}\cdot\, k\, }\:\,\, i\; \Delta^{\; +}\; (\, k\, )\;\, \frac {\quad\;\;\;\, i} {\, -\, 
u_{\,in}^{}\cdot\, k\, }\: ,
\end {eqnarray}
where the method of residues has also been employed.

A relevant feature of the perturbation-theoretic treatment of infrared $\, QED\, $ is that if the renormalization
procedure is performed by means of on-shell prescriptions, it yields an additional
low-energy divergent contribution for each external (charged) line (\cite {JR}, $\S$ 16.1).
The order-$e^{\, 2}$ expression is obtained as follows. 
Letting $\, \Sigma\, (\, u\, )$ be the self-energy function associated to the insertion of a second-order 
self-energy subgraph on an external fermion line carrying four-velocity $\, u\, ,$ the coefficient $B$ 
of the linear term in the expansion of $\, \Sigma\, (\, u\, )$ around the electron mass-shell 
(\cite {MaSh}, $\S$ 9.3) depends logarithmically upon the low-energy cutoff; the resulting 
correction takes the form $-\, e^{\, 2}\, B_{\, IR}^{}\, /\, 2\, ,$ with $B_{\, IR}^{}$ the infrared 
part of $\, B\, .$ 

Considering for definiteness the expression associated to the outgoing line of $\, \mathcal {D}\, ,\, $ 
the perturbation-theoretic result that we wish to reproduce (as can be inferred by taking $\, m=\, n\, $ 
in (\ref {correzioni radiative perturbative}) and recalling Remark \ref {remark 1}) is 
\begin {equation}\label {correzione linea esterna}
-\, \frac {\,\, e^{\: 2}} {2}\;\, B_{\, IR\, }^{\, out\, }=\, \frac {\,\, e^{\: 2}\; u_{\, out}^{\: 2}} {2\; }\int_{\; \mathbf {k}\, >\, \lambda}
\; \frac {d^{\,\, 3\: }k} {(\, 2\; \pi\, )^{\: 3}\:\, 2\,\, \mathbf {k}\, }\,\,\, \tilde {\rho\; }^{2\; }(\, \mathbf {k}\, )\;\, \frac {\,\, 1} 
{(\, u_{\, out}^{}\cdot\, k\: )^{\; 2\, }}\,\, \cdot\,\, 
\end {equation}

Next Lemma shows that the infrared radiative corrections at second order can be expressed as a functional 
of the four-current (\ref {quadricorrente classica}).
\begin {lemma}\label {Lemma 1}
The order-$e^{\, 2}$ infrared radiative corrections to the process $\, \alpha\rightarrow\beta\, ,$ obtained in 
the $\, FGB\, $ gauge with the aid of the low-energy Feynman rules, can be written as
\begin {equation}\label {correzioni radiative secondo ordine}
e^{\, 2}\; \mathscr {M}_{\; \lambda\, ,\,\, \beta\, \alpha}^{\; (FGB)}\, =\,\, \frac {\; i\; e^{\: 2}} {2\;\, }
\, \int_{\, \lambda}^{}\; d^{\; 4}\, x\;\; d^{\; 4}\, y\;\,\, j_{\: \mu}^{}\; (\, x\, )\;\; 
\Delta^{\, +\; }(\, x\, -\, y\: )\,\,\, j^{\; \mu}\; (\, y\, )\: .
\end {equation}
\end {lemma}
\begin {proof}
We have to evaluate
\begin {equation}\label {somma correzioni secondo ordine}
\mathscr {M}_{\; \lambda\, ,\,\, \beta\, \alpha}^{\; (FGB)\, }\, =\;\, \frac {1} {2}\;\, (\; \Gamma_{\, u_{\, out}^{}\, u_{\, in}^{}}^{}+
\: \Gamma_{\, u_{\, in}^{}\, u_{\, out}^{}\, }^{})-\frac {1} {2}\;\, (\, B_{\, IR\, }^{\; in}+\, B_{\, IR\, }^{\: out}\: )\; .
\quad\quad\quad\;
\end {equation}
It suffices to prove that
\begin {equation}\label {correzioni secondo ordine corrente}
e^{\, 2}\, \mathscr {M}_{\; \lambda\, ,\,\, \beta\, \alpha}^{\; (FGB)}\, =\,\, \frac {\; i\; e^{\: 2}} {2\,\, }
\, \int_{\; \mathbf {k}\, >\, \lambda}\,\, \frac {\; d^{\,\, 4\; }k\;\, } {(\, 2\; \pi\, )^{\: 4}\, }\;\,\, 
\tilde {j\, }_{\mu}^{}\; (\, -\, k\, )\,\,\, \Delta^{\, +\; }(\, k\, )\,\,\, 
\tilde {j\, }^{\: \mu\; }(\, k\, )\; ,\quad\quad
\end {equation}
with 
\begin {equation}\label {trasformata di Fourier quadricorrente}
\tilde {j\, }^{\, \mu}\, (\, k\, )\, =\;\, i\,\, \tilde {\rho\,\, }(\, \mathbf {k}\, )\,\, 
(\; -\, \frac {u_{\, in}^{\; \mu}} {\, u_{\, in}^{}\cdot\, k\, }+
\frac {u_{\, out}^{\; \mu}} {\, u_{\, out}^{}\cdot\, k\, }\,\, )\, 
\equiv\;\, \tilde {j\, }_{in}^{\, \mu}\; (\, k\, )+\, 
\tilde {j\, }_{out}^{\, \mu}\; (\, k\, )\quad\quad\;\;
\end {equation}
the Fourier transform of the four-current (\ref {quadricorrente classica}).
We can express (\ref {correzione linea esterna}) as
\begin {equation}\label {correzione linea esterna corrente}
-\, \frac {\; e^{\, 2}} {2\, }\,\,\, B_{\, IR\, }^{\: out\, }=\; \frac {\; i\; e^{\, 2}} {2\;\, }\, \int_{\; \mathbf {k}\, >\, \lambda}\,
\, \frac {\; d^{\,\, 4\; }k\;\, } {(\, 2\; \pi\, )^{\: 4}\, }\;\,\, \tilde {j\: }_{out\, ,\,\, \mu}^{}\; (\, -\, k\, )\,\,\, \Delta^{\, +}\; 
(\, k\, )\,\,\, \tilde {j\, }_{out}^{\: \mu}\; (\, k\, )\: ,\quad
\end {equation}
recalling (\ref {supporto commutatore Pauli-Jordan}), 
and write (\ref {inserzione di fotone tra linee}) as
\begin {equation}\label {inserzione di fotone tra linee corrente}
e^{\, 2}\;\, \Gamma_{\, u_{\, out}^{}\; u_{\,in}^{}}^{}=\,\, i\; e^{\, 2}\int_{\; \mathbf {k}\, >\, \lambda}\,\, \frac {\; d^{\,\, 4\; }k\;\, } 
{(\, 2\; \pi\, )^{\: 4}\, }\;\,\, \tilde {j\, }_{out\, ,\,\, \mu}^{}\; (\, -\, k\, )\,\,\, \Delta^{\, +\; }(\, k\, )\,\,\, \tilde {j\, }_{in}^{\: \mu}\; 
(\, k\, )\: ;\quad\;\;\;
\end {equation}
hence (\ref {correzioni secondo ordine corrente}) immediately follows.
\end {proof}
\begin {remark}\label {remark 2}
The contributions of the infrared diagrammatic rules to the r.h.s. of (\ref {correzioni secondo 
ordine corrente}) which are specific to the $\, FGB\, $ gauge are those given by the vertex, 
by the electron propagator and by the coefficient $\, -\, g^{\: \mu\, \nu},$ arising from 
the tensor structure of the photon propagator.
Therefore, in order to compute the amplitude associated to the same Feynman diagrams in a different 
gauge it suffices to replace the above contributions by those pertaining to that gauge.
\end {remark}
Consider the insertion of an order-$e^{\, 2}$ unrenormalized self-energy loop on the outgoing line 
of $\, \mathcal {D}\, $ and let $\, e^{\, 2}\, {\Sigma}_{\; unr\; }^{\: (\, \epsilon\, )}(\, u_{\, out}^{}\, )\, $ be the 
associated amplitude, obtained by adopting the expression $\, i\, /\, (\, u\, \cdot\, k\, +
\, i\; \epsilon\, )\, $ for the fermion propagator of the loop; the infrared contribution 
(\ref {correzione linea esterna corrente}) is related to the renormalization 
counterterms introduced in (\ref {hamiltoniano Feynman rinormalizzato}) 
through the following
\begin {lemma}\label {Lemma 2}
The $\, FGB$-gauge second-order infrared radiative corrections to the process 
$\, \alpha\rightarrow\beta\, ,$ which arise from on-shell renormalization
prescriptions and are associated to the outgoing line of $\, \mathcal {D\, },$ 
can be obtained as the limit
\begin {equation}\label {autoenergia epsilon}
\lim_{\epsilon\,\, \rightarrow\; 0}\;\, \frac {\; i\; e^{\, 2}} {2\,\, \epsilon\,\, }\;\,\, \tilde {\Sigma}^{\: (\, \epsilon\, )}
\; (\, u_{\, out}^{}\, )\, =\, -\, \frac {\,\, e^{\, 2}} {2\, }\;\, \frac {\; \partial\,\, \Sigma_{\; unr}^{\, (\, \epsilon\, )}\; 
(\, u_{\, out}^{}\, )\, } {\partial\; (\, i\; \epsilon\, )\, }\,\, |_{\,\, \epsilon\; =\; 0}^{}\, =\, -\, \frac {\,\, e^{\, 2}} 
{2\, }\;\, B_{\, IR\, }^{\: out}\; ,
\end {equation}
\begin {equation}\label {inserzione di autoenergia elettronica rinormalizzata}
\tilde {\Sigma}^{\: (\, \epsilon\, )}\; (\, u_{\, out}^{}\, )\, \equiv\,\, \Sigma_{\; unr}^{\: (\, \epsilon\, )}\; (\, u_{\, out}^{}\, )\, +
\, u_{\, out}^{\: 2}\;\, z_{\; 2}^{}\; (\, u_{\, out}^{}\, )\; .\quad\quad\quad\quad\quad\quad\quad\quad\quad\quad
\quad
\end {equation}
\end {lemma}
\begin {proof}
The $\, FGB$-gauge infrared diagrammatic rules and the method of residues yield\footnote {Regarding the 
definition of the $\, \Sigma$-function, we adopt the convention of \cite {MaSh}.}
\begin {equation}\label {inserzione di autoenergia elettronica}
i\; e^{\, 2}\;\, \Sigma_{\; unr}^{\: (\, \epsilon\, )}\; (\, u_{\, out}^{}\, )\, \equiv\,\, e^{\: 2}\; u_{\, out}^{\: 2}
\, \int_{\; \mathbf {k}\, >\, \lambda}\; \frac {d^{\,\, 4\: }k} {(\, 2\; \pi\, )^{\: 4}\, }\,\,\, \tilde {\rho\; }^{2\; }
(\, \mathbf {k}\, )\;\, \frac {\quad\quad\quad\;\, i} {-\: u_{\, out}^{}\cdot\, k\, +\, i\; \epsilon\, }
\;\,\, i\; \Delta^{\, +\; }(\, k\, )\; .
\end {equation} 
It then suffices to compute the sum (\ref {inserzione di autoenergia elettronica rinormalizzata}) 
and to take the limit (\ref {autoenergia epsilon}).
\end {proof}
The role of the renormalization counterterms in reproducing the infrared radiative corrections 
of $\, QED\, $ to all orders is conveniently displayed by explicitly writing down their 
contributions to the $\, S$-matrix (\ref  {matrice S BN adiabatica}).
Notice first that, recalling (\ref {operatore di evoluzione BN rinormalizzato}) and 
(\ref {coefficiente rinormalizzato}), we can write
\begin {equation}\label {operatore di evoluzione non rin adiab}
\mathscr {U}_{\:\, I\, ,\,\, u_{\, out}^{}}^{\, (\, \lambda\, )\, ,\; (\, \epsilon\, )}\; (\, t\, )\, =
\;\, \exp\; (\, -\, i\; e^{\: 2}\,\, z_{\; 2}^{}\; (\, u_{\, out}^{}\, )\;\, u_{\, out}^{\: 2}\,\, 
\frac {\; e^{\, -\, 2\,\, \epsilon\,\, t}\, -\, 1\; } {2\,\, \epsilon}\,\, )\;\,\,  
\mathscr {U}_{\:\, I\, ,\,\, u_{\, out}^{}\, ,\,\, unr}^{\, (\, \lambda\, )\, ,\; (\, \epsilon\, )}\, (\, t\, )\; ,
\end {equation}
where $\, \mathscr {U}_{\;\, I\, ,\; u_{\, out}^{}\, ,\,\, unr\; }^{\, (\, \lambda\, )\, ,\; (\, \epsilon\, )}(\, t\, )\, $ 
is the evolution operator for positive times, associated to the adiabatic (unrenormalized)
Hamiltonian (\ref {hamiltoniano Feynman adiabatico}). In order to obtain the corresponding
result for $\, t < 0\, ,$ it suffices to replace $u_{\, out}^{}$ by 
$u_{\, in}^{}$ and $\, \epsilon\, $ by $\, -\, \epsilon\, $ in the above expressions.

With the same notations employed in Proposition \ref {proposizione 4}, we introduce the 
unrenormalized M\"{o}ller operators, for a fixed value of $\, \epsilon\, ,$ 
\begin {equation}\label {Moller non rin}
\Omega_{\; \mp\, ,\,\, u_{\, \pm}^{}\, ,\,\, unr}^{\, (\, \lambda\, )\, ,\,\, (\, \pm\, \epsilon\, )\, }\, \equiv\;\, \tau_{\; w}^{}\, -\, 
\lim_{t\,\, \rightarrow\,\, \pm\, \infty}\; \mathscr {U}_{\;\, I\, ,\,\, u_{\, \pm}^{}\, ,\,\, unr}^{\, (\, \lambda\, )\, ,\; (\, \pm\, \epsilon\, )}
\; (\, t\: )\: .\quad\quad\quad\;
\end {equation}
The renormalized M\"{o}ller operators (\ref {Moller Feynman BN epsilon}) can then be written as
\begin {equation}\label {Moller rinormalizzati}
\Omega_{\; \mp\, ,\,\, u_{\, \pm}^{}}^{\, (\, \lambda\, )\, ,\,\, (\, \pm\, \epsilon\, )\, }=\;\, \exp\; (\, \pm\, i\; e^{\, 2}
\; u_{\, \pm}^{\: 2}\,\, \frac {\: z_{\; 2}^{}\; (\, u_{\, \pm}^{}\, )\, } {2\,\, \epsilon\;\; }\,\, )\;\,\, 
\Omega_{\; \mp\, ,\,\, u_{\, \pm\, }^{},\,\, unr}^{\, (\, \lambda\, )\, ,\,\, (\, \pm\, \epsilon\, )\, }\; , 
\end {equation}
whence
\begin {equation}\label {fattorizzazione e rin}
S_{\;  \lambda\, ,\; u_{\, out}^{}\; u_{\, in}^{}}^{\: (FGB)\, ,\,\, (\, \epsilon\, )}\, =\;\, \exp\; (\, -\, i\; e^{\, 2}\; u_{\, out}^{\: 2}
\,\, \frac {\: z_{\; 2}^{}\; (\, u_{\, out}^{}\, )\, } {2\,\, \epsilon\; }\,\, )\,\,\, 
S_{\;  \lambda\, ,\; u_{\, out}^{}\; u_{\, in}^{}\, ,\,\, unr}^{\: (FGB)\, ,\,\, (\, \epsilon\, )}\;
\exp\; (\, -\, i\; e^{\, 2}\; u_{\, in}^{\: 2}\; \frac {\: z_{\; 2}^{}\; (\, u_{\, in}^{}\, )\, } {2\,\, \epsilon\; }\,\, )\; ,
\end {equation}
with 
\begin {equation}\label {matrice S non rin}
S_{\;  \lambda\, ,\; u_{\, out}^{}\: u_{\, in}^{}\, ,\; unr}^{\: (FGB)\, ,\,\, (\, \epsilon\, )}\, \equiv\;\, 
\Omega_{\; -\, ,\,\, u_{\, out}^{}\, ,\,\, unr}^{\, (\, \lambda\, )\, ,\,\, (\, \epsilon\, )\,\, \dagger}
\,\,\, \Omega_{\; +\, ,\,\, u_{\, in}^{}\, ,\,\, unr}^{\, (\, \lambda\, )\, ,\,\, (\, -\, \epsilon\, )}
\quad\quad\quad\quad
\end {equation}
the corresponding unrenormalized scattering matrix.

Lemma \ref {Lemma 2} suggests that it should be possible to recover the exponentiation of the 
correction (\ref {correzione linea esterna}) within a Hamiltonian framework, by taking 
into account the effect to all orders of the 
counterterm with coefficient $\, z_{\; 2\; }^{}(\, u_{\, out}^{}\, )\, .\, $
The following Proposition shows that this is indeed the case.
\begin {theorem}\label {proposizione 5}
One has
\begin {equation}\label {correzioni non rinormalizzate}
\langle\; \Psi_{\, 0\; }^{},\; \Omega_{\; -\, ,\,\, u_{\, out}^{}\, ,\; unr}^{\; (\, \lambda\, )\, ,\,\, (\, \epsilon\, )\; \dagger}
\; \Psi_{\, 0}^{}\: \rangle\, =\;\, \exp\; (\; \frac {\; i\; e^{\: 2}} {2\;\, }\, \int_{\, \lambda}^{}\; d^{\; 4}\, x\;\; 
d^{\; 4}\, y\;\,\, j_{\; out\, ,\,\, \mu}^{\: (\, \epsilon\, )}\; (\, x\, )\,\,\, \Delta_{\, F\; }^{}(\; x\, -\, y\; )\,\,\, 
j_{\; out}^{\: (\, \epsilon\, )\; \mu}\; (\, y\, )\, )\; .
\end {equation}
Upon including the contribution of the renormalization counterterm, as given in (\ref {Moller 
rinormalizzati}), evaluating the vacuum expectation value of the resulting operator
and taking the adiabatic limit, one obtains 
the overall contribution to the infrared radiative corrections to 
the process 
$\, \alpha\rightarrow\beta\, ,$ which arises from on-shell renormalization prescriptions and is 
associated to the outgoing line of $\, \mathcal {D}:$
\begin {equation}\label {correzioni radiative linea esterna}
\, \lim_{\epsilon\,\, \rightarrow\,\, 0\, }\,\, \langle\, \Psi_{\, 0}^{}\, ,\; 
\Omega_{\; -\: ,\,\, u_{\, out}^{}}^{\, (\, \lambda\, )\, ,\; (\, \epsilon\, )\,\, \dagger}
\,\, \Psi_{\, 0}^{}\, \rangle\, =
\;\, \exp\; (\; -\, \frac {\; e^{\: 2}} {2\, }\;\, B_{\, IR\, }^{\, out\, }\, )\; .
\end {equation}
\end {theorem}
\begin {proof}
By evaluating the evolution operator $\, \mathscr {U}_{\,\, I\, ,\; u_{\, out}^{}\, ,\,\, unr\; }^{\, (\, \lambda\, )\, ,\; (\, \epsilon\, )}
(\, t\, )\, $ for positive times, in terms of the interaction Hamiltonian of the model as given in 
(\ref {hamiltoniano Feynman adiabatico}), and taking the asymptotic limit, we obtain
\begin {equation}\label {Moller non rin spaz}
\Omega_{\; -\, ,\,\, u_{\, out}^{}\, ,\; unr}^{\, (\, \lambda\, )\, ,\,\, (\, \epsilon\, )}\, =\;\,
\mathscr {C}_{\,\, u_{\, out}^{}}^{\; (\, \epsilon\, )}\;\, \exp\; (\; i\; e\int_{\, \lambda}^{}
\; d^{\,\, 4}\, x\;\,\, j_{\; out\, ,\,\, \mu}^{\; (\, \epsilon\, )}\: (\, x\, )\;\, 
A^{\; \mu}\; (\, x\, )\, )\; ,\quad\quad\quad\quad\quad\quad
\quad\quad\;\,
\end {equation}
\begin {eqnarray}
\mathscr {C}_{\,\, u_{\, out}^{}}^{\: (\, \epsilon\, )}\, \equiv\,\,  \lim_{t\,\, \rightarrow\; +\, \infty}\, 
\mathscr {C}_{\,\, u_{\, out}^{}}^{\: (\, \epsilon\, )}\; (\, t\: )\, =\,\, \exp\; (\; -\, \frac {\; i\; e^{\: 2}} {2\;\, }\, 
\int_{\, \lambda}^{}\,\, d^{\,\, 4}\, x\;\; d^{\,\, 4}\, y\;\,\, j_{\; out\, ,\,\, \mu\; }^{\: (\, \epsilon\, )}(\, x\, )\,\,\, 
\theta\: (\: x_{\: 0}^{}-\, y_{\; 0}^{}\, )
\;\,\, \nonumber\\
\times\; 
\Delta^{}\: (\, x\, -\, y\, )\,\,\, j_{\; out}^{\: (\, \epsilon\, )\: \mu\; }(\, y\, )\, )\; ,
\end {eqnarray}
whence
\begin {eqnarray}\label {aspettazione Moller non rin spaz}
\langle\; \Psi_{\, 0\; }^{},\: \Omega_{\; -\, ,\,\, u_{\, out}^{}\, ,\; unr}^{\, (\, \lambda\, )\, ,\,\, (\, \epsilon\, )\; \dagger}
\; \Psi_{\, 0}^{}\: \rangle\, =\;\, \exp\; (\; \frac {\; i\; e^{\: 2}} {2\;\, }\int_{\, \lambda}^{}\,\, d^{\,\, 4}\, x
\;\; d^{\,\, 4}\, y\;\,\, j_{\; out\, ,\,\, \mu\; }^{\: (\, \epsilon\, )}(\, x\, )\;\, \Delta^{\, +\; }(\, x\, -\, y\, )\;\, 
j_{\; out}^{\: (\, \epsilon\, )\; \mu\; }(\, y\, )\, )
\;\;\; \nonumber\\
\times\; \exp\; (\; \frac {\; i\; e^{\: 2}} {2\;\, }\, \int_{\, \lambda}^{}\,\, d^{\,\, 4}\, x\;\; d^{\,\, 4}\, y
\;\,\, j_{\; out\, ,\,\, \mu\; }^{\: (\, \epsilon\, )}(\, x\, )\,\,\, \theta\: (\: x_{\: 0}^{}-\, y_{\; 0}^{}\, )\;\, 
\Delta^{}\: (\, x\, -\, y\, )\,\,\, j_{\; out}^{\: (\, \epsilon\, )\; \mu\; }(\, y\, )\, )\: .\;  
\end {eqnarray}
Since
\begin {equation}\label {combinazione commutatore}
\theta\: (\, x_{\: 0}^{}-\, y_{\; 0}^{}\, )\;\, \Delta\: (\; x\, -\, y\; )\, =\,\, \theta\: (\, x_{\: 0}^{}-\, y_{\; 0}^{}\, )
\;\, \Delta^{\, +}\, (\; x\, -\, y\; )\, +\, [\:\, \theta\: (\, y_{\; 0}^{}-\, x_{\: 0}^{}\, )\, -\, 1\; ]\,\,\, \Delta^{\, +}\, 
(\, y\, -\, x\, )\; ,
\end {equation}
one gets (\ref {correzioni non rinormalizzate}).

By virtue of (\ref {Moller Feynman BN}), in order to prove (\ref {correzioni radiative linea esterna})
it is enough to show that
\begin {equation}\label {calcolo limite}
\langle\; \Psi_{\, 0}^{}\, ,\; \Omega_{\; -\: ,\,\, u_{\, out}^{}}^{\, (\, \lambda\, )\; \dagger}\,\, \Psi_{\, 0}^{}
\, \rangle\, =\;\, \exp\; (\; -\, \frac {\; e^{\: 2}} {2\, }\;\, B_{\, IR\, }^{\, out\, }\, )\: .\quad\quad\quad
\end {equation}
Recalling (\ref {Moller rinormalizzati}), we can write
\begin {equation}
\Omega_{\; -\, ,\,\, u_{\, out}^{}}^{\, (\, \lambda\, )\, ,\; (\, \epsilon\, )}\, =\;\, 
\mathscr {C}_{\,\, u_{\, out}^{}}^{\; (\, \epsilon\, )}\;\, \exp\; (\; i\; e^{\, 2}\; u_{\, out}^{\: 2}
\,\, \frac {\: z_{\; 2}^{}\; (\, u_{\, out}^{}\, )\, } {2\,\, \epsilon\; }\,\, )\;\, 
\exp\; (\; i\; e\int_{\, \lambda}^{}
\; d^{\,\, 4}\, x\;\,\, j_{\; out\, ,\,\, \mu}^{\; (\, \epsilon\, )}\: (\, x\, )\;\, A^{\; \mu}\; (\, x\, )\, )\; ,
\end {equation}
whence
\begin {equation}\label {calcolo limite parziale}
\langle\; \Psi_{\, 0}^{}\, ,\; \Omega_{\; -\: ,\,\, u_{\, out}^{}}^{\, (\, \lambda\, )\; \dagger}\,\, 
\Psi_{\, 0}^{}\, \rangle\, =\;\, \exp\; (\; \frac {\; i\; e^{\: 2}} {2\;\, }\int_{\, \lambda}^{}
\,\, d^{\,\, 4}\, x\;\; d^{\,\, 4}\, y\;\,\, j_{\; out\, ,\,\, \mu}^{}\; (\, x\, )\,\,\, 
\Delta^{\, +\; }(\; x\; -\: y\; )\,\,\, j_{\: out\; }^{\; \mu}(\, y\, )\, )\: .
\end {equation}
Taking the Fourier transform of the r.h.s. of (\ref {calcolo limite parziale}), we obtain
(\ref {correzioni radiative linea esterna}).
\end {proof}
\begin {remark}\label {remark 3}
Proposition \ref {proposizione 5} and Lemma \ref {Lemma 2} make it clear that the contribution of the 
renormalization counterterm is instrumental in order to obtain an operator formulation of the infrared 
corrections related to the outgoing line. 
The same reasoning applies to the incoming-line expressions.
\end {remark}
Employing (\ref {Moller non rin spaz}) and the corresponding
expression for $\,  \Omega_{\; +\, ,\,\, u_{\, in}^{}\, ,\; unr}^{\, (\, \lambda\, )\, ,\,\, (\, -\, \epsilon\, )}\, ,\, $ 
the scattering matrix (\ref {matrice S BN adiabatica}) can be written as
\begin {eqnarray}\label {matrice S BN spaziale adiabatica}
S_{\; \lambda\, ,\; u_{\, out}^{}\; u_{\, in}^{}}^{\; (FGB)\, ,\; (\, \epsilon\, )}\, =\,
\,\, \mathscr {C}_{\; u_{\, out}^{}\, ,\,\, z_{\, 2}^{}}^{\; (\, \epsilon\, )\, -\, 1}\;\,\, 
\mathscr {C}_{\;  u_{\, in}^{}\, ,\,\, z_{\, 2}^{}}^{\; (\, -\, \epsilon\, )}\;\, 
\exp\; (\, -\ i\; e^{\, 2}\; u_{\, out}^{\: 2}
\,\, \frac {\: z_{\; 2}^{}\; (\, u_{\, out}^{}\, )\, } {2\,\, \epsilon\; }\,\, )
\; \nonumber\\
\times\;
\exp\; (\, -\ i\; e^{\, 2}\; u_{\, in}^{\: 2}
\,\, \frac {\: z_{\; 2}^{}\; (\, u_{\, in}^{}\, )\, } {2\,\, \epsilon\; }\,\, )
\;\,\, S_{\; \lambda}^{}\; [\,\, j^{\; (\, \epsilon\, )}\; ]\; ,
\end {eqnarray}
\begin {equation}\label {matrice S BN spaziale adiabatica exp}
S_{\: \lambda}^{}\; [\,\, j^{\, (\, \epsilon\, )}\; ]\, \equiv\;\, \exp\; (\, -\, i\; e\int_{\, \lambda}^{}\,\, 
d^{\,\, 4}\, x\;\,\, j_{\; \mu}^{\; (\, \epsilon\, )}\: (\, x\, )\,\,\, A^{\; \mu}\; (\, x\, )\, )\: ,\quad\quad\quad\quad\;
\end {equation}
whence 
\begin {equation}\label {matrice S BN spaziale}
S_{\; \lambda\, ,\; u_{\, out}^{}\; u_{\, in}^{}}^{\: (FGB)}=\,\, S_{\; \lambda}^{}\; [\; j\,\, ]\; .
\quad\quad\quad\quad\quad\quad\quad\quad\quad\quad
\quad\quad\quad\quad\quad\quad\quad\quad\quad
\end {equation}

We are in a position to recover the \emph {overall} soft-photon radiative corrections within an operator 
framework.
\begin {theorem}\label {proposizione 6}
Taking the vacuum expectation value of the unrenormalized scattering operator (\ref {matrice S non rin}),
for fixed $\, \epsilon\, $ and $\, \lambda\, ,$ one obtains 
\begin {equation}\label {elemento di matrice non rin}
\langle\; \Psi_{\, 0\; }^{},\; 
S_{\;  \lambda\, ,\; u_{\, out}^{}\: u_{\, in}^{}\, ,\; unr}^{\: (FGB)\, ,\; (\, \epsilon\, )}
\,\, \Psi_{\, 0}^{}\; \rangle\, =
\;\, \exp\; (\; \frac {\,\, i\; e^{\: 2}} {2\,\, }\int_{\, \lambda}^{}\,\, d^{\,\, 4}\, x\;\; d^{\; 4}\, y\;\,\, 
j_{\; \mu}^{\: (\, \epsilon\, )\; }(\, x\, )\,\,\, \Delta_{\, F\; }^{}(\; x\; -\; y\; )\,\,\, 
j^{\: (\, \epsilon\, )\; \mu}\; (\, y\, )\, )\, .
\end {equation}
The overall soft-photon radiative corrections to the process $\, \alpha\rightarrow\beta$ 
in the $FGB$ gauge are given by 
\begin {equation}\label {correzioni radiative hamiltoniane}
\lim_{\epsilon\; \rightarrow\; 0\, }\,\, \langle\; \Psi_{\, 0\; }^{},\: 
S_{\;  \lambda\, ,\; u_{\, out}^{}\; u_{\, in}^{}}^{\: (FGB)\, ,\,\, (\, \epsilon\, )} 
\;\, \Psi_{\, 0}^{}\: \rangle\, 
=\;\, \exp\; (\; e^{\, 2}\, \mathscr {M}_{\; \lambda\, ,\,\, \beta\, \alpha}^{\; (FGB)}\, )\: ,\,
\end {equation}
for each value of the low-energy cutoff $\, \lambda\, .$
\end {theorem}
\begin {proof}
The proof of (\ref {elemento di matrice non rin}) requires to repeatedly apply the indefinite-metric generalization 
of (\ref {caso particolare formula BH}) and to recall (\ref {correzioni non rinormalizzate}),
(\ref {decomposizione quadricorrente adiabatica}) and (\ref {propagatore associato}).
By virtue of (\ref {matrice S BN}) and (\ref {matrice S BN spaziale}), in order to establish (\ref {correzioni 
radiative hamiltoniane}) it suffices to prove
\begin {equation}\label {calcolo correzioni radiative hamiltoniane}
\langle\; \Psi_{\, 0\; }^{},\: S_{\; \lambda}^{}\; [\,\, j\; ]\;\, \Psi_{\, 0}^{}\: \rangle\, =\;\,
\exp\; (\; e^{\, 2}\, \mathscr {M}_{\; \lambda\, ,\,\, \beta\, \alpha}^{\; (FGB)}\, )\: .\,
\end {equation}
Proceeding as for the evaluation of  (\ref {calcolo limite parziale}), we obtain
\begin {eqnarray}
\langle\; \Psi_{\, 0\; }^{},\: S_{\; \lambda}^{}\; [\,\, j\; ]\;\, \Psi_{\, 0}^{}\: \rangle\, =\;\, \exp\; (\,\, \frac 
{\; i\; e^{\: 2}} {2\,\, }\int_{\, \lambda}^{}\; d^{\; 4}\, x\;\; d^{\; 4}\, y\;\,\, j_{\; \mu\; }^{}(\, x\, )\,\,\, 
\Delta^{\, +\; }(\; x\; -\; y\; )\,\,\, j^{\; \mu}\; (\, y\, )\, )\; .
\end {eqnarray}
Recalling Lemma \ref {Lemma 1}, the result is proved.
\end {proof}
\begin {remark}\label {remark 4}
As it might already be clear from Proposition \ref {proposizione 5}, the renormalization counterterms 
do not play a role in reproducing the exponentiation of the $\, \Gamma$-amplitudes which contribute 
to (\ref {somma correzioni secondo ordine}).
We may also directly check this fact by showing that the term on the r.h.s. of (\ref {correzioni radiative 
hamiltoniane}) which is associated to the amplitude (\ref {inserzione di fotone tra linee})
equals the limiting expression 
of the contribution
$$ \exp\; (\; \frac 
{\; i\; e^{\: 2}} {2\,\, }\int_{\, \lambda}^{}\,\, d^{\; 4}\, x\;\; d^{\; 4}\, y\;\,\,
j_{\; out\, ,\,\, \mu\; }^{\: (\, \epsilon\, )}(\, x\, )\,\,\, \Delta_{\, F\; }^{}(\; x\, -\, y\; )
\,\,\, j_{\; in}^{\: (-\, \epsilon\, )\: \mu\; }(\, y\, )\, )\quad\quad\quad $$
and is thus not affected by renormalization.
In this respect, it suffices to employ the equality 
$$j_{\; out\, ,\; \mu}^{}\; (\, x\, )\,\,\, \Delta^{\; +}\, (\; x\, -\, y\; )\,\,\, j_{\; in}^{\,\, \mu}\; (\, y\, )\, =\,\, j_{\; out\, ,\; \mu}^{}\; (\, x\, )\,\,\, \Delta_{\, F\; }^{}(\; x\, -\, y\; )\,\,\, j_{\; in}^{\,\, \mu}\; (\, y\, )\; .
$$ 
\end {remark}

We turn now to the analysis of the infrared corrections to the process $\, \alpha\rightarrow\beta\, $ which 
are associated to the emission of soft photons.
Let $\mathscr {H}_{\; phys}$ be the Hilbert space of (physical) free-photon 
states\footnote {The definition and construction of $\mathscr {H}_{\; phys\, }^{}$ and the characterization 
of its state content are recalled in Appendix \ref {app:3}.} and let
$\, {\Psi\, }_{f}^{}\equiv\, a^{\, \dagger}\, (\, f\, )\,\, \Psi_{\, 0}^{}\in\mathscr {H}_{\; phys}^{}$
describe a ``soft-photon state''; namely, $\, f^{\, \mu\, }(\, \mathbf {k}\, )\, $ is selected
in such a way that the approximations leading to the infrared Feynman rules are justified; 
for instance, we may choose $\, f^{\, \mu}$ 
with support contained in a sphere of radius $\, r\leq\Lambda\, ,$ $\Lambda\, $ 
being the energy scale introduced in (\ref {correzioni radiative perturbative}).

We first prove a Lemma concerning the operator formulation of the correction due to the emission 
of a single soft photon.
\begin {lemma}\label {Lemma 3}
The $\, FGB$-gauge correction to the process $\, \alpha\rightarrow\beta\, $ associated to the emission 
of one photon in the state $\, \Psi_{\, f}$ is reproduced by the matrix element
\begin {equation}\label {ampiezza emissione un fotone}
\langle\; {\Psi\, }_{f\: }^{},\,\, \exp\; (\, -\, i\; e\int_{\, \lambda}^{}\,\, d^{\,\, 4}\, x
\;\,\, j_{\; \mu\; }^{}(\, x\, )\,\,\, A_{\, -}^{\: \mu}\; (\, x\, )\, )\,\, 
\Psi_{\, 0}^{}\: \rangle\: .
\end {equation}
\end {lemma}
\begin {proof}
With the aid of the commutation relations (\ref {CCR indefinite}), we obtain
\begin {equation}\label {calcolo ampiezza emissione un fotone} 
\langle\; {\Psi\, }_{f}^{}\, ,\,\, \exp\; (\, -\, i\; e\int_{\, \lambda}^{}\,\, d^{\,\, 4}\, x
\,\,\, j_{\; \mu}^{}\: (\, x\, )\,\,\, A_{\, -}^{\; \mu\; }(\, x\, )\, )\,\, 
\Psi_{\, 0}^{}\: \rangle\,
=\; -\int_{\; \mathbf {k}\, >\, \lambda}\; d^{\:\, 3\; }k\;\;\, \Gamma_{\, \mu\; }^{}(\, \mathbf {k}\, )
\,\,\, \frac {\bar {f}^{\: \mu}\, (\, \mathbf {k}\, )} {(\, 2\: \pi\, )^{\; 3\, /\, 2}\; \sqrt {\; 2\,\, \mathbf {k}\, }}\,\, \cdot
\end {equation}
The expression 
$$\, \Gamma_{\, \mu}^{}\, (\, \mathbf {k}\, )\, \equiv\,  -\, i\; e\;\, \tilde {j\: }_{\mu\; }^{}(\, k\, )
\, |_{\; k^{\, 0}\, =\,\, \mathbf {k}\, }$$ 
gives the infrared contribution due to the new vertex and to the extra electron propagator of 
$\, \mathcal {D}_{\; 1}^{}$ (as evaluated for instance in \cite {JR}, $\S$ 16.1).
Taking into account the function $\, \bar {f}^{\: \mu}\, (\, \mathbf {k}\, )\, /\, [\, (\, 2\: \pi\, )^{\; 3\, /\, 2}
\: \sqrt {\; 2\; \mathbf {k}\, }\, ]\, $ associated to the emitted photon and carrying out the 
momentum-space integration, one obtains the perturbative result.
\end {proof} 
\ni
In the same sense as in the one-photon case, let $\, \Psi_{\, f_{\, 1}^{}\, ...\, f_{\, n}^{}}\equiv\;
a^{\, \dagger}\, (\, f_{\, 1}^{}\, )\, ...\; a^{\, \dagger}\, (\, f_{\, n}^{}\, )\,\, \Psi_{\, 0}^{}\in\mathscr {H}_{\; phys}^{}$ 
be a state describing $\, n\, $ soft photons.
The contribution associated to emission in such a state is recovered, in the presence of 
an infrared cutoff $\, \lambda\, $ on the momentum of the photon, through the following
\begin {corollary}\label {corollario 1}
The $\, FGB$-gauge correction to the process $\, \alpha\rightarrow\beta\, $ due to the emission of 
$\, n$ photons in the state $\, {\Psi\, }_{f_{\, 1}^{}\, ...\; f_{\, n}^{}}$ is reproduced by the 
matrix element
\begin {equation}
\langle\; {\Psi\, }_{f_{\, 1}^{}\, ...\; f_{\, n}^{}\, }^{},\; \exp\; (\, -\, i\; e\int_{\, \lambda}^{}\,\, d^{\,\, 4}\, x
\;\,\, j_{\; \mu}^{}\: (\, x\, )\,\,\, A_{\, -}^{\; \mu\; }(\, x\, )\, )\;\, \Psi_{\, 0\, }^{}\, \rangle\, .
\quad\quad
\end {equation}
\end {corollary}
\begin {proof}
Taking into account (\ref {calcolo ampiezza emissione un fotone}), we find
\begin {eqnarray}\label {correzione n fotoni FGB}
\langle\; {\Psi\, }_{f_{\, 1}^{}\, ...\; f_{\, n}^{}\, }^{},\; \exp\; (\, -\, i\; e\int_{\, \lambda}^{}\,\, d^{\,\, 4}\, x\;\,\, 
j_{\; \mu}^{}\: (\, x\, )\,\,\, A_{\, -}^{\; \mu\; }(\, x\, )\, )\;\, \Psi_{\, 0\, }^{}\, \rangle\, 
\quad\quad\quad\quad\quad\quad\quad\quad\quad\quad\quad\quad\quad\nonumber\\
=\,\, (\, -\, 1\, )^{\; n}\,\, \Pi_{\; j\; =\; 1}^{\; n}\int_{\; \mathbf {k_{\, j}^{}}\, >\, \lambda}\; \frac {d^{\,\, 3}\, k_{\, j}^{}}
{(\, 2\; \pi\, )^{\, 3\, /\, 2}\, \sqrt {\; 2\,\, \mathbf {k}_{\, j}^{}\, }}\,
\sum_{r\;\, =\;\, 1}^{2}\, \frac {\,\, \eta_{\,\, r}^{}\; e\;\, \tilde {\rho\; }(\, \mathbf {k}_{\, j}^{}\, )
\;\, u_{\: r\, ,\,\, \mu_{\, j}^{}}^{}\,\, f_{\: j}^{\,\, \mu_{\, j}^{}}\, (\: \mathbf {k}_{\, j}^{}\, )\, }
{u_{\; r}^{}\cdot\, k_{\: j}^{}\, }\;\, ,
\end {eqnarray}
with $\, u_ {\; 1}^{}\equiv\, u_{\, in}^{}\, ,\: u_ {\; 2}^{}\equiv\, u_{\, out\, }^{},\, \eta_ {\; 2}^{}=\, 1=-\, \eta_ {\; 1}^{}\, .$
\end {proof} 
We wish now to obtain an operator formulation of the factorization property of the contributions associated 
to soft-photon emission with respect to the radiative corrections involving low-energy virtual photons, 
as discussed for example in \cite {Wein}, $\S\, $13.1.  
\begin {theorem}\label {proposizione 7}
In the $\, FGB\, $ gauge, the overall contribution given by the soft-photon radiative corrections to the 
process $\alpha\rightarrow\beta\, $ and by the corrections 
associated to the emission of $\, n$ photons in the state 
$\, \Psi_{\, f_{\, 1}^{}\, ...\, f_{\, n}^{}}$ is reproduced by the 
matrix element
\begin {equation}
\langle\; {\Psi\, }_{f_{\, 1}^{}\, ...\; f_{\, n}^{}\, },\; S_{\;  \lambda\, ,\; u_{\, out}^{}\: u_{\, in}^{}}^{\; (FGB)}
\,\, \Psi_{\, 0\; }^{}\rangle\, .\quad\quad
\end {equation}
\end {theorem}
\begin {proof}
With the aid of formula (\ref {caso particolare formula BH}), we can write
\begin {eqnarray}
S_{\;  \lambda\, ,\; u_{\, out}^{}\: u_{\, in}^{}}^{\; (FGB)}=\;\, \exp\; (\, -\, \frac {\,\, e^{\: 2}} {2\, }\, 
\int_{\, \lambda}^{}\,\, d^{\,\, 4}\, x\;\; d^{\,\, 4}\, y\;\,\, j_{\: \mu}^{}\: (\, x\, )\;\, [\; A_{\, +\; }^{\; \mu}
(\, x\, )\, ,\, A_{\, -}^{\; \nu}\; (\, y\, )\; ]\,\,\,  j_{\; \nu}^{}\: (\, y\, )\, )
\nonumber\\
\times\; \exp\; (\, -\, i\; e\int_{\, \lambda}^{}\,\, d^{\,\, 4}\, x\;\,\, j_{\: \mu}^{}\: (\, x\, )\,\,\, 
A_{\, -}^{\; \mu\; }(\, x\, )\, )\;\, \exp\; (\, -\, i\; e\int_{\, \lambda}^{}\,\, d^{\,\, 4}\, x\;\,\, 
j_{\: \mu}^{}\: (\, x\, )\,\,\, A_{\, +}^{\; \mu\; }(\, x\, )\, )
\end {eqnarray}
on $\mathscr {H}_{\; phys\, }^{},\, $ hence 
\begin {eqnarray}
\langle\; {\Psi\, }_{f_{\, 1}^{}\, ...\; f_{\, n}^{}\, }^{},\; S_{\; \lambda\, ,\; u_{\, out}^{}\: u_{\, in}^{}}^{\; (FGB)}\, 
\Psi_{\, 0}^{}\, \rangle\, =\;\, \exp\; (\,\, \frac {\; i\; e^{\: 2}} {2\,\, }\int_{\, \lambda}^{}\; d^{\; 4}\, x\;\; 
d^{\; 4}\, y\;\,\, j_{\; \mu\; }^{}(\, x\, )\,\,\, \Delta^{\, +\; }(\; x\; -\; y\; )\,\,\, j^{\; \mu}\; (\, y\, )\, )
\;\, \nonumber\\ 
\times\; \langle\; {\Psi\, }_{f_{\, 1}^{}\, ...\; f_{\, n}^{}\, }^{},\,\, \exp\; (\, -\, i\; e\int_{\, \lambda}^{}\; d^{\,\, 4}\, x\,\,\, 
j_{\; \mu\; }^{}(\, x\, )\;\, A_{\, -}^{\: \mu}\; (\, x\, )\, )\,\, \Psi_{\, 0}^{}\: \rangle\, .
\end {eqnarray}
By Lemma \ref {Lemma 1} and Corollary \ref {corollario 1}, 
\begin {eqnarray}\label {ampiezza totale}
\langle\; {\Psi\, }_{f_{\, 1}^{}\, ...\; f_{\, n}^{}\, },\; S_{\; \lambda\, ,\; u_{\, out}^{}\: u_{\, in}^{}}^{\; (FGB)}
\,\, \Psi_{\, 0}^{}\, \rangle\, =\;\, \exp\; (\; e^{\, 2}\, \mathscr {M}_{\; \lambda\, ,\,\, \beta\, \alpha}^{\; (FGB)}\, )
\;\, (\, -\, 1\, )^{\, n}\,\, \Pi_{\; j\,\, =\; 1}^{\; n}\int_{\; \mathbf {k_{\, j}^{}}\, >\, \lambda}\; \frac {d^{\,\, 3\; }k_{\, j}^{}}
{(\, 2\; \pi\, )^{\, 3\, /\, 2}\, \sqrt {\; 2\,\, \mathbf {k}_{\, j}^{}\, }}
\;\;\; \nonumber\\
\times\sum_{r\;\, =\;\, 1\, }^{2}\, \frac {\,\, \eta_{\,\, r}^{}\,\, e\;\, \tilde {\rho\; }(\, \mathbf {k}_{\, j}\, )\;\, 
u_{\: r\, ,\,\, \mu_{\, j}^{}}^{}\; f_{\: j}^{\,\, \mu_{\, j}^{}}\, (\: \mathbf {k}_{\, j}^{}\, )\, }
{u_{\: r}^{}\cdot\, k_{\: j}^{}}\; \cdot\;
\end {eqnarray}
The factorization property discussed above is thus recovered; moreover,
the result (\ref {ampiezza totale}) agrees with the expression obtained in 
perturbation theory (equation $(13.3.1)$ in \cite {Wein}). 
\end {proof}
\ni
The infrared phases occurring in the transition amplitude for a process involving at least two charged 
particles, in either the initial or the final state, can also be recovered by explicit calculations, in terms 
of M\"{o}ller operators corresponding to the sum of one-particle Hamiltonians.  
For the sake of conciseness, we do not report the details.

We conclude the comparison of the $\, BN$ model with the perturbative treatment of infrared $\, QED$ 
in the $\, FGB$ gauge with a number of observations.
First, the proof of the exponentiation of the low-energy corrections arising from on-shell renormalization 
prescriptions is quite streamlined within the Hamiltonian framework, while it is more cumbersome 
within the diagrammatic 
approach, requiring in particular the identification of relevant terms and the application of Ward's identity 
(\cite {JR}, $\S\, $16.1).
Likewise, the proof of the factorization property of the contributions due to the emission of low-energy 
photons with respect to the soft-photon radiative corrections, while straightforward within an operator 
formulation, requires a careful treatment in perturbation theory; as pointed out in \cite {Wein}, $\S\, $ 
13.1, one needs in fact to show that certain on-shell amplitudes are not affected by radiative 
corrections.
\par
\bigskip
The rest of this Section is devoted to establish an operator formulation of infrared $\, QED\, $ 
in the Coulomb gauge.
As in (\ref {fattorizzazione e rin}), let us factor out the contributions of the renormaization counterterms 
to the scattering operator (\ref {matrice S Coulomb BN adiabatica}), 
\begin {equation}\label {fattorizzazione e rin Coulomb}
S_{\; \lambda\, ,\,\, u_{\, out}^{}\: u_{\, in}^{}}^{\; (Coul)\, ,\; (\, \epsilon\, )}\, =\;\, \exp\; (\: i\; e^{\, 2}\; 
\mathbf {u}_{\, out}^{\, 2}\; \frac {\; z_{\; 1\; }^{}(\, u_{\, out}^{}\, )\, } {2\,\, \epsilon\; }\,\, )
\,\,\, S_{\; \lambda\, ,\,\, u_{\, out}^{}\: u_{\, in}^{}\, ,\,\, unr}^{\; (Coul)\, ,\; (\, \epsilon\, )}
\,\, \exp\; (\; i\; e^{\, 2}\; \mathbf {u}_{\, in}^{\, 2}\; \frac {\; z_{\; 1\; }^{}(\, u_{\, in}^{}\, )\, } 
{2\,\, \epsilon\; }\,\, )\; ,
\end {equation}
with $\, S_{\; \lambda\, ,\,\, u_{\, out}^{}\: u_{\, in}^{}\, ,\,\, unr}^{\; (Coul)\, ,\; (\, \epsilon\, )}\, $ 
the corresponding unrenormalized  scattering matrix.
Proceeding as for the derivation of (\ref {matrice S BN spaziale}), the $\, S$-matrix 
(\ref {matrice S Coulomb BN}) can be written as
\begin {equation}\label {matrice S BN Coulspaziale}
S_{\; \lambda\, ,\,\, u_{\, out}^{}\: u_{\, in}^{}}^{\; (Coul)}=\,\,\, \exp\; (\; i\; e\int_{\, \lambda}^{}\,\, d^{\,\, 4}\, x
\;\,\, \mathbf {j}_{\; C}^{}\: (\, x\, )\cdot\mathbf {A}_{\, C}^{}\: (\, x\, )\, )\, \equiv\,\, S_{\; \lambda}^{}\; 
[\;\, \mathbf {j}_{\; C}^{}\; ]\, =\;\, s\, -\lim_{\epsilon\,\, \rightarrow\; 0}\; 
S_{\; \lambda}^{}\; [\:\, \mathbf {j}_{\; C}^{\: (\, \epsilon\, )}\; ]\; .\;
\end {equation}

The Coulomb-gauge Feynman rules which are relevant for the computation of the soft-photon corrections
to the process $\, \alpha\rightarrow\beta\, $ are the following:\footnote {In the case of a process involving a 
single charged particle, to be considered below, the Coulomb interaction merely provides an ultraviolet 
divergent self-energy contribution, which can be removed with the aid of a suitable renormalization 
counterterm and is in any case infrared finite; for this reason we have not introduced the 
associated term in the Coulomb-gauge Hamiltonian and shall not need the corresponding
diagrammatic rule.} A factor $\, i\: e\; \tilde {\rho\: }(\, \mathbf {k}\, )\: u^{\, r}$ has 
to be associated to a vertex with space index $\, r\, ,\, $ occurring on an external charged line carrying 
four-velocity $\, u\, ,$ and a contribution $\, i\; \Delta_{\, F}^{}\, (\, k\, )\:\, P^{\,\, r\, s}\, (\, \mathbf {k}\, )\, $ 
must be supplied for a photon line carrying 
four-momentum $\, k\, $ and joining two vertices with indices $\, r\, $ and $\, s\, .\, $ 
We use the symbol
\begin {equation}\label {propagatore Coulomb}
i\; {}_{\, tr\, }\Delta_{\, F}^{\; r\, s\; }(\, x\, )\, \equiv\,\, 
\frac {i} {(\, 2\; \pi\, )^{\: 4}\, }\int\,\, d^{\,\, 4\: }k\;\,\, e^{\, -\, i\,\, k\, \cdot\; x}
\,\,\, \Delta_{\, F\; }^{}(\, k\, )\,\,\, P^{\,\, r\, s\; }(\, \mathbf {k}\, )\; 
\end {equation}
for the space-time expression of the photon propagator.
The diagrammatic rule pertaining to the electron propagator is the same as in the $FGB\, $ gauge. 
In the sequel, we state and prove the main results.
\begin {lemma}\label {Lemma 4}
The order-$e^{\, 2}$ soft-photon radiative corrections to the process $\, \alpha\rightarrow\beta\, ,$
obtained in the Coulomb gauge with the aid of the infrared Feynman rules, can be written as
\begin {equation}\label {correzioni radiative secondo ordine Coulomb}
e^{\, 2}\; \mathscr {M}_{\; \lambda\, ,\,\, \beta\, \alpha}^{\; (Coul)}\, =\; -\, \frac {\:\, i\; e^{\: 2}} {2\,\, }
\, \int_{\, \lambda}^{}\,\, d^{\; 4}\, x\;\; d^{\; 4}\, y\;\,\, \Delta^{\, +\; }(\; x\, -\, y\; )\;\;\, 
\mathbf {j}_{\; C}^{}\; (\, x\, )\, \cdot\; \mathbf {j}_{\; C}^{}\; (\, y\, )\; .\, 
\end {equation}
\end {lemma}
\begin {proof}
It suffices to prove that  
\begin {equation}\label {correzioni radiative secondo ordine Coulomb Fourier}
e^{\, 2}\, \mathscr {M}_{\; \lambda\, ,\,\, \beta\, \alpha}^{\; (Coul)}\, =\; -\, \frac {\:\, i\; e^{\: 2}} {2\,\, }
\, \int_{\; \mathbf {k}\, >\, \lambda}\,\, \frac {\; d^{\,\, 4}\, k\;\, } {(\, 2\; \pi\, )^{\: 4}\, }\;\,\, 
\Delta^{\, +\; }(\, k\, )\;\;\, \tilde\mathbf {j}_{\; C\; }^{}(\, -\, k\, )\, \cdot\; \tilde
\mathbf {j}_{\; C\; }^{}(\, k\, )\; ,\quad\quad 
\end {equation}
with $\, \tilde\mathbf {j}_{\; C\; }^{}(\, k\, )\, $
the Fourier transform of the current (\ref {corrente trasversa}).
In the evaluation of the Feynman amplitude $\, \mathscr {M}_{\; \lambda\, ,\,\, \beta\, \alpha}^{\; (Coul)}\, ,\, $ the infrared (Coulomb-gauge) diagrammatic rules corresponding to the vertex and to the electron propagator 
and the projection $\, P^{\; r\, s}\, (\, \mathbf {k}\, )\, ,\, $ associated to the tensor structure 
of the photon propagator, contribute a term $\, -\, e^{\, 2}
\;\, \tilde\mathbf {j}_{\; C}^{\; (\, \epsilon\, )}\, (\, -\, \mathbf {k}\, )\, \cdot\; \tilde\mathbf {j}_{\; C}^{\; (\, \epsilon\, )}\, (\, \mathbf {k}\, )\, ,$ where
$\, \tilde\mathbf {j}_{\; C}^{\: (\, \epsilon\, )}\, (\, \mathbf {k}\, )\equiv\;
\tilde\mathbf {j}_{\; C}^{\: (\, \epsilon\, )}\, (\, k_{\: 0}=\, \mathbf {k}\, ,\, 
\mathbf {k}\, )\, ,$
\begin {equation}
\tilde\mathbf {j}_{\; C}^{\, (\, \epsilon\, )\; }(\, k\, )\, =\;\, i\;\, \tilde {\rho\; }(\, \mathbf {k}\, )\;\, P^{\; r\, s}\, (\, \mathbf {k}\, )
\;\, (\; \frac {\mathbf {u}_{\, in}^{\: s}} {\, -\, u_{\, in}^{}\cdot\, k\, +\, i\; \epsilon}+\frac {\mathbf {u}_{\, out}^{\: s}} 
{\, u_{\, out}^{}\cdot\, k\, +\, i\; \epsilon\, }\; )\; 
\end {equation}
being the Fourier transform of the current $\, \mathbf {j}_{\; C}^{\: (\, \epsilon\, )\, }(\, x\, )
\, .\, $
Recalling Remark \ref {remark 2} and (\ref {supporto commutatore Pauli-Jordan}) and taking 
the adiabatic limit, we obtain (\ref {correzioni radiative secondo ordine Coulomb Fourier}).
\end {proof}
The exponentiation of (\ref {correzioni radiative secondo ordine Coulomb}) is recovered within an 
operator formulation through the following 
\begin {theorem}\label {proposizione 8} 
The vacuum expectation value of the unrenormalized Coulomb-gauge $\, S$-matrix for fixed 
$\, \epsilon\, $ and $\, \lambda\, $ is given by 
\begin {eqnarray}\label {elemento di matrice non rin Coul}
(\; \Psi_{\, F}^{}\, ,\: S_{\; \lambda\, ,\,\, u_{\, out}^{}\: u_{\, in\, }^{},\,\, unr}^{\; (Coul)\, ,\; (\, \epsilon\, )}
\,\, \Psi_{\, F}^{}\, )\, =\;\, \exp\; (\, -\, \frac {\; i\; e^{\: 2}} {2\,\, }\int_{\, \lambda}^{}\,\, d^{\,\, 4}\, x
\,\,\, d^{\; 4}\, y
\;\;\, \mathbf {j}_{\; C}^{\; (\, \epsilon\, )\; r}\, (\, x\, )
\;\, {}_{\, tr\, }\Delta_{\, F}^{\; r\, s}\; (\, x\, -\, y\: )\;\, 
\nonumber\\ 
\times\,\, \mathbf {j}_{\; C}^{\; (\, \epsilon\, )\: s}\; (\, y\, )\, )\: .
\end {eqnarray}
For each value of the cutoff $\, \lambda\, ,$ the overall soft-photon radiative corrections to the 
process $\, \alpha\rightarrow\beta\, $ in the same gauge are reproduced by the adiabatic limit of 
the vacuum expectation value of the renormalized $\, S$-matrix (\ref {fattorizzazione e rin Coulomb}):
\begin {equation}\label {correzioni radiative hamiltoniane Coul}
\lim_{\epsilon\; \rightarrow\; 0}\;\, (\; \Psi_{\, F\; }^{},\: 
S_{\; \lambda\, ,\,\, u_{\, out}^{}\: u_{\, in}^{}}^{\; (Coul)\, ,\; (\, \epsilon\, )}\;\, \Psi_{\, F}^{}\, )
\, =\;\, \exp\; (\; e^{\, 2}\, \mathscr {M}_{\, \lambda\, ,\,\, \beta\, \alpha}^{\; (Coul)}\; )\: .
\end {equation}
\end {theorem}
\begin {proof}
The proof of (\ref {elemento di matrice non rin Coul}) follows the same steps as that of 
(\ref {elemento di matrice non rin}). 
In order to establish (\ref {correzioni radiative hamiltoniane Coul}), it suffices  
to recall (\ref {matrice S BN Coulspaziale}) and to proceed as for the proof of 
(\ref {calcolo correzioni radiative hamiltoniane}).
\end {proof}

We turn now to the corrections due to soft-photon emission.
Let $\, {\Phi\; }_{\mathbf {f}}^{}\equiv\, a^{\, *}\, (\, \mathbf {f}\, )\; \Psi_{\, F}^{}$ be the vector belonging to the 
Coulomb-gauge space of free photon states $\mathscr {\, H}_{\, C}^{}$ and indexed by the test function
$\, \mathbf {f}\, (\, \mathbf {k}\, )\in L^{\, 2}\, (\, \mathbb {R}^{\, 3}\, )\, ,$
$\, \mathbf {k}\, \cdot\, \mathbf {f}\, (\, \mathbf {k}\, )=\, 0\, .\, $
As in the analysis carried out in the $\, FGB\, $ gauge, we suppose
that $\, \Phi_{\; \mathbf {f}}^{}\, $ describes a (free) soft-photon state.
The contribution associated to the emission of one photon in such a state is recovered, in the presence 
of a low-energy cutoff $\, \lambda\, $ on the photon momentum, through the following
\begin {lemma}\label {Lemma 5}
The Coulomb-gauge correction to the process $\,\alpha\rightarrow\beta\, $ due to the emission 
of one photon in the state $\, \Phi_{\; \mathbf {f}}^{}$ is reproduced by the matrix element
\begin {equation}\label {ampiezza emissione un fotone Coulomb}
(\; \Phi_{\; \mathbf {f}\: }^{},\,\, \exp\; (\; i\; e\int_{\, \lambda}^{}\,\, d^{\,\, 4}\, x\;\,\, 
\mathbf {j}_{\; C}^{}\: (\, x\, )\, \cdot\, \mathbf {A}_{\, C\, ,\,\, -}^{}\: (\, x\, )\, )\;\, 
\Psi_{\, F}^{}\, )\: .
\end {equation}
\end {lemma}
\begin {proof}
By means of the $\, CCR\, ,$ we obtain
\begin {equation}\label {calcolo ampiezza emissione un fotone Coulomb} 
(\; \Phi_{\; \mathbf {f}\: }^{},\,\, \exp\; (\; i\; e\int_{\, \lambda}^{}\,\, d^{\,\, 4}\, x\;\,\, \mathbf {j}_{\; C}^{}\; (\, x\, )\, \cdot
\, \mathbf {A}_{\, C\, ,\,\, -}^{}\: (\, x\, )\, )\;\, \Psi_{\, F}^{}\, )\, =\; \int_{\; \mathbf {k}\, >\, \lambda}\,\, d^{\,\, 3}\, k\;\;\, 
\mathbf {\Gamma}_{\, C}^{\,\, r}\, (\, \mathbf {k}\, )\,\,\, \frac {\bar {\mathbf {f}}^{\,\, r}\, (\, \mathbf {k}\, )} 
{(\, 2\: \pi\, )^{\; 3\, /\, 2}\; \sqrt {\; 2\,\, \mathbf {k}\, }}\,\, \cdot
\end {equation}
The coefficient function 
$$\, \mathbf {\Gamma}_{\, C}^{\,\, r}\, (\, \mathbf {k}\, )\, \equiv\;\, i\; e\;\, \mathbf {j}_{\; C}^{\,\, r}\, (\, k\, )
\, |_{\; k^{\, 0}\, =\,\, \mathbf {k}\, }$$ 
reproduces the infrared contribution due to the new vertex and to the extra electron propagator of
$\, \mathcal {D}_{\, 1\, }^{},$ according to the Coulomb-gauge diagrammatic rules discussed above.
Taking into account the function $\, \bar {f}^{\: \mu}\, (\, \mathbf {k}\, )\, /\, [\, (\, 2\: \pi\, )^{\; 3\, /\, 2}
\, \sqrt {\; 2\; \mathbf {k}\, }\: ]$ associated to the emitted photon and carrying out the 
momentum-space integration, we recover the perturbative expression.
\end {proof}
Letting $\, \Phi_{\: \mathbf {f}_{\: 1}\, ...\,\, \mathbf {f}_{\: n}}^{}\equiv\, a^{\, *}\, (\, \mathbf {f}_{\; 1}^{}\, )\, ...\: a^{\, *}\, (\, \mathbf {f}_{\; n}^{}\, )\; \Psi_{\, F}^{}\in\mathscr {H}_{\; C}^{}$ be a state describing $\, n\, $ soft photons, the correction due to the emission of photons 
in such a state is reproduced by the following
\begin {corollary}\label {corollario 2}
The Coulomb-gauge correction to the process $\, \alpha\rightarrow\beta$ associated to the emission 
of $\, n$ photons in the state $\, {\Phi\: }_{\mathbf {f}_{\; 1}^{}\, ...\,\, \mathbf {f}_{\, n}^{}}^{}$ is given by the 
matrix element
\begin {equation}
(\: \Phi_{\: \mathbf {f}_{\: 1}^{}\, ...\,\, \mathbf {f}_{\, n}^{}\, }^{},\; \exp\; (\; i\; e\int_{\, \lambda}^{}\,\, d^{\,\, 4}\, x
\;\,\, \mathbf {j}_{\; C}^{}\: (\, x\, )\, \cdot\, \mathbf {A}_{\, C\, ,\,\, -}^{}\: (\, x\, )\, )\,\, \Psi_{\, F}^{}\, )\; .
\quad\quad\quad
\end {equation}
\end {corollary}
\begin {proof}
Taking into account (\ref {calcolo ampiezza emissione un fotone Coulomb}), we find
\begin {eqnarray}\label {correzione n fotoni Coulomb}
(\: \Phi_{\: \mathbf {f}_{\; 1}^{}\, ...\; \mathbf {f}_{\, n}^{}}^{},\; \exp\; (\; i\; e\int_{\, \lambda}^{}\,\, d^{\,\, 4}\, x
\;\,\, \mathbf {j}_{\; C}^{}\: (\, x\, )\, \cdot\, \mathbf {A}_{\, C\, ,\,\, -}^{}\: (\, x\, )\, )\,\, \Psi_{\, F}^{}\, )
\;\;\; \quad\quad\quad\quad\quad\quad\nonumber\\
=\,\, \Pi_{\; j\; =\; 1}^{\; n}\int_{\; \mathbf {k_{\, j}^{}}\, >\, \lambda}\; \frac {d^{\,\, 3}\, k_{\, j}^{}}
{(\, 2\; \pi\, )^{\, 3\, /\, 2}\, \sqrt {\; 2\,\, \mathbf {k}_{\, j}^{}\, }}\,
\sum_{r\;\, =\;\, 1\, }^{2}\, \frac {\,\, \eta_{\,\, r}^{}\,\, e\;\, \tilde {\rho\; }(\, \mathbf {k}_{\, j}\, )
\,\,\, \mathbf {u}_{\; r}^{}\cdot\, \mathbf {f}_{\; j\; }^{}(\: \mathbf {k}_{\, j}^{}\: )\, }
{u_{\; r}^{}\cdot\, k_{\: j}^{}\, }\,\, ,
\end {eqnarray}
with $\, \mathbf {u}_ {\; 1}^{}\equiv\, \mathbf {u}_{\, in}^{}\, ,\: \mathbf {u}_ {\; 2}^{}\equiv
\, \mathbf {u}_{\, out\, }^{},\, \eta_ {\; 2}^{}=\, 1=-\, \eta_ {\; 1\, }^{}.$
\end {proof} 
Finally, we state the Coulomb-gauge counterpart of Proposition \ref {proposizione 7}.
\begin {theorem}\label {proposizione 9}
In the Coulomb gauge, the overall contribution given by the soft-photon radiative corrections to the 
process $\, \alpha\rightarrow\beta\, $ and by the corrections associated to the emission of $\, n$ 
photons in the state $\, \Phi_{\: \mathbf {f}_{\; 1}^{}\, ...\,\, \mathbf {f}_{\: n}^{}\, }$ is reproduced by the 
matrix element
\begin {equation}
(\: \Phi_{\: \mathbf {f}_{\: 1}^{}\, ...\,\, \mathbf {f}_{\, n}^{}\, }^{},\: S_{\; \lambda\, ,\,\, u_{\, out}^{}
\: u_{\, in}^{}}^{\; (Coul)}\,\, \Psi_{\, F}^{}\, )\: .\quad\quad
\end {equation}
\end {theorem}
\begin {proof}
The proof is the same as that of Proposition \ref {proposizione 7}.
\end {proof}
It is straightforward to check that if we the vector $\, \Psi_{f_{\, 1}^{}\, ...\; f_{\, n}^{}}
\in\mathscr {H}_{\; phys}^{}$ in Corollary \ref {corollario 1} is chosen to be indexed by
(square-integrable) smearing functions
$\, f_{\: j}^{\; \mu}\, (\, \mathbf {k}_{\, j}^{}\, )=(\: 0\, ,\: \mathbf {f}_{\: j}^{}\, (\, \mathbf {k}_{\, j}\, )\, )\, ,
\, j\, =\, 1\, ,...\, ,\, n\, ,$
the correction (\ref {correzione n fotoni FGB}) equals the Coulomb-gauge expression (\ref {correzione 
n fotoni Coulomb}).
A more intrinsic formulation of the invariance properties of the soft-photon corrections to the process 
$\, \alpha\rightarrow\beta\, $ with respect to the gauge employed in their calculations, not relying in 
particular upon a specific choice of test functions, is indeed possible and will be given in the 
forthcoming Section.

\section {Infrared Approximations And Invariance Properties}
\label {sect:3}

This Section is devoted to the discussion of the invariance properties of the soft-photon corrections
with respect to the gauge chosen for their evaluation.
For definiteness, we shall compare the expressions obtained in the $FGB$ gauge with those calculated
in the Coulomb gauge; by a similar treatment it is possible to include other local and 
covariant formulations, such as those employing $\, \xi$-gauges 
(discussed for example in \cite {BLOT}, $\S\, $ 10.2). 

Next Proposition contains the main result of this Section; relying on the existence of an (isometric) 
isomorphism $\, T\, $ between $\mathscr {H}_{\; phys}$ and $\mathscr {H}_{\; C}^{}$ (Lemma \ref {Lemma 11}, Appendix \ref {app:3}), we prove that the restriction of the photon 
scattering operator (\ref  {matrice S BN}) to $\mathscr {H}_{\; phys}$ is unitarily equivalent to the 
Coulomb-gauge $\, S$-matrix (\ref {matrice S Coulomb BN}).
\begin {theorem}\label {proposizione 10}
For each value of the infrared cutoff $\, \lambda\, ,$ the equality
\begin {equation}\label {uguaglianza matrici S}
{S_{\: \lambda\, ,\,\, u_{\, out}^{}\; u_{\, in}^{}}^{\; (FGB)}}\, |_{\, \mathscr {H}_{\, phys}^{}}^{}=\;\, 
T^{\; -\, 1}\;\, S_{\: \lambda\, ,\,\, u_{\, out}^{}\; u_{\, in}^{}}^{\; (Coul)}\,\, T
\quad\quad\quad\quad\quad\quad
\end {equation}
holds.
\end {theorem}
\begin {proof}
By virtue of (\ref {matrice S BN Coulspaziale}) and of the transversality of the (free)
Coulomb-gauge vector potential (\ref {potenziale vettore Coulomb}), the r.h.s.
of (\ref {uguaglianza matrici S}) can be expressed as 
\begin {equation}\label {matrice S isomorfa}
T^{\; -\, 1}\;\, S_{\: \lambda\, ,\,\,u_{\, out}^{}\; u_{\, in}^{}}^{\; (Coul)}\,\, T\; =\;\, \exp\; (\; i\; e
\int_{\, \lambda}^{}\,\, d^{\,\, 4}\, x\;\,\, \mathbf {j}\; (\, x\, )\, \cdot\, \mathbf {A}_{\, T}^{}\: 
(\, x\, )\, )\; ,
\end {equation}
with $\, \mathbf {A}_{\, T}^{}\, (\, x\, )\equiv\; T^{\; -\, 1}\; \mathbf {A}_{\, C\: }^{}(\, x\, )\,\,\, T\, .\, $
Lemma \ref {Lemma 12} in Appendix \ref {app:3} (choosing the components of $\, j^{\; \mu},$ defined in
(\ref {quadricorrente classica}), as smearing functions in (\ref {potenziale trasformato})) and 
Stone's theorem\footnote {Stone's theorem can be applied since $\, \mathscr {H}_{\; phys}^{}$ 
is a Hilbert space.} yield 
\begin {eqnarray}\label {relazione matrici S}
T^{\; -\, 1}\;\, S_{\: \lambda\, ,\:\, u_{\, out}^{}\; u_{\, in}^{}}^{\; (Coul)}\,\, T\; =\;\, 
\exp\; (\, -\, i\; e\int_{\, \lambda}^{}\,\, d^{\,\, 4}\, x\,\,\, j_{\; \mu\; }^{}(\, x\, )
\;\, [\; A^{\; \mu\; }(\, x\, )-\; \partial^{\; \mu}\int\, d^{\,\, 3}\, y\;\,\, 
G_{\, \mathbf {y}}^{}\, (\, \mathbf {x}\, )\;\, 
\nonumber\\
\times\: (\: \partial_{\: m}^{}\; A^{\; m}\, )\, (\:  x_{\: 0}^{}\, ,\; \mathbf {y}\: )\: ]\, )
\, |_{\; \mathscr {H}_{\; phys}^{}}^{}\, .
\end {eqnarray}
Making use of the continuity equation for $\,  j^{\: \mu}\, (\, x\, )\, $ the Proposition is proved.
\end {proof}
As a straightforward consequence of Propositions \ref {proposizione 6}, \ref {proposizione 8}, 
\ref {proposizione 10}, we obtain the following
\begin {corollary}\label {corollario 3}
The overall soft-photon radiative corrections to the process $\, \alpha\rightarrow\beta\, $ take the 
same value in the $\, FGB$ gauge and in the Coulomb gauge, for each value of the infrared 
cutoff $\, \lambda\, .$ 
\end {corollary}
In order to give an unambiguous formulation of the invariance properties of low-energy corrections
including the effect of soft-photon emission, it is convenient to introduce a notion of corresponding 
states. 
\begin {definition}
$\, \Phi\in\mathscr {H}_{\; C}^{}$ is the state corresponding to 
$\, \Psi\in\mathscr {H}_{\; phys}^{}$ if $\; \Phi\, =\; T\,\, \Psi\, .$
\end {definition}
Propositions \ref {proposizione 7}, \ref {proposizione 9}, \ref {proposizione 10} 
lead then to the following
\begin {corollary}\label {corollario 4}
The $\, FGB$-gauge contribution associated to the overall soft-photon radiative corrections to the 
process $\, \alpha\rightarrow\beta$ and to the corrections due to the emission of $\, n$ photons in 
the state $\, \Psi_{\, f_{\, 1}^{}\, ...\, f_{\, n}^{}}\in\mathscr {H}_{\; phys\, }^{}$ equals the Coulomb-gauge 
contribution given by the overall soft-photon radiative corrections to the same process and by 
the corrections associated to photon emission in the state corresponding to 
$\, \Psi_{\, f_{\, 1}^{}\, ...\, f_{\, n\, }^{}}.$
\end {corollary}
We are in a position to discuss the lack of invariance of the soft-photon corrections, in the presence of
the approximations of the model studied in Section \ref {sect:1}.
As a consequence of Proposition \ref {proposizione 10} we obtain the following result, which will be 
relevant in the subsequent analysis.
\begin {corollary}\label {corollario 5}
Let $\, f^{\: \mu}=\, (\, f^{\; 0}\, ,\; \mathbf {f}\, )\, $ be chosen in such a way 
that the exponential operators
\begin {equation}
\mathcal {O}_{\, \lambda}^{}\; [\,\, f\,\, ]\, \equiv\;\, \exp\; (\, -\, i\; e\int_{\, \lambda}^{}\; d^{\,\, 4}\, x
\,\,\, f_{\: \mu\; }^{}(\, x\, )\;\, A^{\; \mu\; }(\, x\, )\, )\: ,
\end {equation}
\begin {equation}
\mathcal {O}_{\, \lambda}^{}\; [\:\, \mathbf {f}_{\; C}^{}\; ]\, \equiv\;\, \exp\; (\; i\; e\int_{\, \lambda}^{}\,\, 
d^{\,\, 4}\, x\;\,\, \mathbf {f}_{\; C}^{}\: (\, x\, )\, \cdot\, \mathbf {A}\: (\, x\, )\, )\: ,\;\,\,
\end {equation}
with
\begin {equation}
\mathbf {f}_{\; C}^{\:\, l}\, (\: t\, ,\, \mathbf {x}\, )\, \equiv\, \int\, d^{\,\, 3}\, y\;\,\, \delta_{\; tr}^{\,\, l\, m}\:
(\, \mathbf {x}\, -\, \mathbf {y}\, )\,\,\, \mathbf {f}^{\; m\; }(\: t\, ,\, \mathbf {y}\, )\: ,\quad\,\, 
\end {equation}
are well-defined, respectively on $\, \mathscr {G}$ and on $\, \mathscr {H}_{\,\, C}^{}\, ,$ 
for each value of the infrared cutoff $\, \lambda\, .$
The equality
\begin {eqnarray}
\mathcal {O}_{\, \lambda}^{}\; [\; f\,\, ]\; =\,\,\, T^{\; -\, 1}\;\, \mathcal {O}_{\, \lambda}^{}\; [\:\, \mathbf {f}_{\; C}^{}\; ]
\;\,\, T\,\,\, \exp\; (\, -\, i\; e\int_{\, \lambda}^{}\,\, d^{\,\, 4}\, x\,\,\, f_{\: \mu}^{}\: (\, x\, )\;\,\, \partial^{\; \mu}
\int\, d^{\,\, 3}\, y\;\,\, G_{\, \mathbf {y}}^{}\: (\, \mathbf {x}\, )\, 
\nonumber\\
\times\, (\: \partial_{\: m}^{}\; A^{\; m}\, )\, (\: x_{\: 0}^{}\, ,\; \mathbf {y}\: )\, )
\end {eqnarray}
holds on $\, \mathscr {H}_{\; phys\, }^{}.$
\end {corollary}
\begin {proof}
By the same reasoning leading to equation (\ref {relazione matrici S}) and taking into account the 
commutation relations
\begin {equation}
[\,\, A^{\; \mu}\, (\, x\, )\, ,\; \partial^{\; \nu}\, (\: \partial_{\: m}^{}\; A^{\; m}\, )\, (\: x_{\: 0}^{}\, ,\mathbf {\; y}\: )\; ]
\, =\,\, 0\; ,
\end {equation}
which result from the $\, CCR\, ,$ we obtain
\begin {eqnarray}\label {relazione matrici S fattorizzata}
T^{\; -\, 1}\;\, \mathcal {O}_{\, \lambda}^{}\; [\:\, \mathbf {f}_{\; C}^{}\; ]\,\,\, T\; =\;\, \exp\; (\, -\, i
\; e\int_{\, \lambda}^{}\,\, d^{\,\, 4}\, x\;\, f_{\: \mu}^{}\, (\, x\, )\;\, A^{\; \mu\; }(\, x\, )\, )\;
|_{\, \mathscr {H}_{\, phys}^{}}^{}\,\, \exp\; (\; i\; e\int_{\, \lambda}^{}\,\, d^{\,\, 4}\, x\,\,\, 
f_{\: \mu}^{}\; (\, x\, )\;\;
\nonumber\\
\times\; \partial^{\; \mu}\int\, d^{\,\, 3}\, y\;\,\, G_{\, \mathbf {y}}^{}\, (\, \mathbf {x}\, )\,\, 
(\: \partial_{\: m}^{}\; A^{\; m}\, )\, (\: x_{\: 0}^{}\, ,\; \mathbf {y}\: )\, )\, .
\end {eqnarray}
\end {proof}
In the sequel, besides the standard low-energy assumptions employed in the diagrammatic treatment 
of infrared $\, QED\, ,$ 
we suppose that the electron is non relativistic and introduce a dipole approximation.
The resulting (infrared) Feynman rules are as follows.
In the $FGB$ gauge, a term $-\, i\: e\; \tilde {\rho}\, (\, \mathbf {k}\, )\: \tilde {v}^{\: \mu}$ has to be associated to a vertex 
with index $\, \mu\, $ on a charged-particle line carrying energy-momentum 
$\, p\, ;$  
the Coulomb-gauge expression to be supplied for a vertex with index $\, l\, $ on the same 
line is $\; i\; e\; \tilde {\rho\: }(\, \mathbf {k}\, )\; \mathbf {p}^{\; l}/\, m\, .\, $ 
In both gauges, an electron propagator carrying four-momentum $\, p+k\, $ contributes a term 
$\, i\: /\, (\, \mathbf {k}\, +\, i\; \epsilon\, )\, ,$ with $\, \mathbf {k}\, $ the 
overall photon contribution to the energy carried by the propagator line.\footnote {The peculiar 
functional form of the electron propagator is due to the fact that, in the presence of a dipole 
approximation, the electron momentum is conserved while the total one is not.}

First, we compute the resulting expressions for the radiative corrections at order $e^{\, 2}.$
\begin {lemma}\label {Lemma 6}
In the presence of a dipole approximation, the Coulomb-gauge soft-photon radiative corrections to the 
process $\, \mathbf {p}_{\, in}^{}\rightarrow\mathbf {\, p}_{\, out}^{}$ at second order can be written as
\begin {equation}\label {correzioni radiative secondo ordine dipolare Coulomb}
e^{\: 2}\: \mathscr {M}_{\; \lambda\, ,\:\, \mathbf {p}_{\, out}^{}\,\, \mathbf {p}_{\, in}^{}}^{\; (Coul)\, ,\,\, (dip)}\, =\, -\, 
\frac {\,\, i\; e^{\: 2}} {2\; }\int_{\, \lambda}^{}\,\, d^{\,\, 4}\, x\;\; d^{\,\, 4}\, y\;\,\, \Delta^{\, +\; }(\; x\, -\, y\; )
\;\,\, \mathbf {j}_{\; C\, ,\,\, nr}^{\, (dip)}\: (\, x\, )\, \cdot\; \mathbf {j}_{\; C\, ,\,\, nr}^{\, (dip)}\: (\, y\, )\; ,
\end {equation}
where $\: \mathbf {j}_{\; C\, ,\,\, nr}^{\: (dip)}\, (\, x\, )\, $ is the current (\ref {corrente dipolare Coulomb non rel}).
The $\, FGB$-gauge expression for the same corrections is
\begin {equation}\label {correzioni radiative secondo ordine dipolare FGB}
e^{\: 2}\: \mathscr {M}_{\; \lambda\, ,\:\, \mathbf {p}_{\, out}^{}\; \mathbf {p}_{\, in}^{}}^{\; (FGB)\, ,\,\, (dip)}\, =\, 
-\, \frac {\; i\; e^{\: 2}} {2\; }\int_{\, \lambda}^{}\,\, d^{\,\, 4}\, x\;\; d^{\,\, 4}\, y\;\,\, \Delta^{\, +\; }(\, x\, -\, y\: )
\;\,\, \mathbf {j}_{\; nr}^{\, (dip)}\, (\, x\, )\, \cdot\; \mathbf {j}_{\; nr}^{\, (dip)}\, (\, y\, )\; ,\quad\;\,\,\,
\end {equation}
with $\: \mathbf {j}_{\; nr}^{\: (dip)}\, (\, x\, )\, $ the current (\ref {corrente dipolare non rel}).
\end {lemma}
\begin {proof}
We first prove that
\begin {equation}\label {correzioni radiative secondo ordine dipolare Coulomb Fourier}
e^{\, 2}\, \mathscr {M}_{\; \lambda\, ,\,\, \beta\, \alpha}^{\; (Coul)\, ,\,\, (dip)}\, =
\; -\, \frac {\:\, i\; e^{\: 2}} {2\,\, }\, \int_{\; \mathbf {k}\, >\, \lambda}\; 
\frac {\; d^{\,\, 4}\, k\;\, } {(\, 2\; \pi\, )^{\: 4}\, }\;\,\, \Delta^{\, +\; }(\, k\, )
\;\;\, \tilde\mathbf {j}_{\; C\, ,\,\, nr\; }^{\; (dip)}(\, -\, k\, )\, \cdot
\; \tilde\mathbf {j}_{\; C\, ,\,\, nr\; }^{\; (dip)}(\, k\, )\; ,
\end {equation}
with $\, \tilde\mathbf {j}_{\; C\, ,\,\, nr\; }^{\; (dip)}(\, k\, )\, $ the Fourier transform of 
the current (\ref {corrente dipolare Coulomb non rel}).
In the evaluation of the Feynman amplitude $\, e^{\: 2}\mathscr {M}_{\; \lambda\, ,\:\, \mathbf {p}_{\, out}^{}
\; \mathbf {p}_{\, in}^{}}^{\; (Coul)\, ,\; (dip)}\, ,\, $ the infrared diagrammatic rules corresponding to the 
vertex and to the electron propagator and the projection $\, P^{\; r\, s}\, (\, \mathbf {k}\, )\, $ 
associated to the Coulomb-gauge photon propagator provide a contribution $-\, e^{\: 2}\,\, \tilde\mathbf {j}_{\; C\, ,\,\, nr}^{\: (dip)\, ,\; (\, \epsilon\, )}\, (\, -\, \mathbf {k}\, )\, \cdot\: \tilde\mathbf {j}_{\; C\, ,\,\, nr\; }^{\: (dip)\, ,\; (\, \epsilon\, )}\, (\, \mathbf {k}\, )\, ,$
with $\; \tilde\mathbf {j}_{\; C\, ,\,\, nr\; }^{\: (dip)\, ,\; (\, \epsilon\, )}\, (\, \mathbf {k}\, )
\equiv\; \tilde\mathbf {j}_{\; C\, ,\,\, nr}^{\; (dip)\, ,\; (\, \epsilon\, )}\, (\, k_{\: 0}=
\vert\, \mathbf {k}\, \vert\, ,\, \mathbf {k}\, )\, ,\, $ 
\begin {equation}\label {trasformata di Fourier corrente dipolare Coulomb}
\tilde\mathbf {j}_{\; C\, ,\,\, nr}^{\; (dip)\; l}\, (\, k\, )\, =
\;\, i\;\, \frac {\; \tilde {\rho\; }(\, \mathbf {k}\, )\, } 
{m\; (\: k_{\: 0}+\, i\; \epsilon\, )}\,\,\, P^{\,\, l\: r}\, (\, \mathbf {k}\, )\;\, (\, -\, \mathbf {p}_{\, in}^{\, r}+\, \mathbf {p}_{\, out}^{\, r}\, ) 
\end {equation}
being the Fourier transform of (\ref {corrente dipolare epsilon}).
Proceeding as in the proof of Lemma \ref {Lemma 4} and recalling 
(\ref {supporto commutatore Pauli-Jordan}), we obtain 
(\ref {correzioni radiative secondo ordine dipolare 
Coulomb Fourier}).\\
In order to establish (\ref {correzioni radiative secondo ordine dipolare FGB}), we show that
\begin {equation}\label {correzioni radiative secondo ordine dipolare FGB Fourier}
e^{\, 2}\, \mathscr {M}_{\; \lambda\, ,\,\, \beta\, \alpha}^{\; (FGB)\, ,\,\, (dip)}\, =\; -\, \frac {\:\, i\; e^{\: 2}} 
{2\,\, }\, \int_{\; \mathbf {k}\, >\, \lambda}\,\, \frac {\; d^{\,\, 4}\, k\;\, } {(\, 2\; \pi\, )^{\: 4}\, }\;\,\, 
\Delta^{\, +\; }(\, k\, )\;\;\, \tilde\mathbf {j}_{\,\, nr}^{\, (dip)}\, (\, -\, k\, )\, \cdot\; 
\tilde\mathbf {j}_{\,\, nr}^{\, (dip)}\, (\, k\, )\: ,
\end {equation}
with $\, \tilde\mathbf {j}_{\,\, nr}^{\, (dip)}\, (\, k\, )\, $ the Fourier 
transform of the current (\ref {corrente dipolare non rel}). 
In the calculation of
$\, e^{\, 2}\, \mathscr {M}_{\; \lambda\, ,\:\, \mathbf {p}_{\, out}^{}
\; \mathbf {p}_{\, in}^{}}^{\; (FGB)\, ,\,\, (dip)}\, ,\, $ the (infrared)
diagrammatic rules pertaining to the vertex and to the electron propagator 
and the coefficient $-\, g^{\: \mu\, \nu}$ associated to the photon 
propagator yield a term $\, e^{\, 2\; }\tilde {j\, }_{nr\; \mu}^{\: (dip)\, ,\; (\, \epsilon\, )\; }(\, -\, \mathbf {k}\, )\;\, \tilde {j\, }_{nr}^{\: (dip)\, ,\; (\, \epsilon\, )\,\, \mu}\, (\, \mathbf {k}\, )\, ,$
with $\, \tilde {j\, }_{nr}^{\: (dip)\, ,\; (\, \epsilon\, )\,\, \mu}\, (\, \mathbf {k}\, )\equiv\, 
\tilde {j\, }_{nr}^{\: (dip)\, ,\; (\, \epsilon\, )\,\, \mu}\, (\, k_{\: 0}= 
\vert\, \mathbf {k}\, \vert\, ,\, \mathbf {k}\, )
\, ,\, $
\begin {equation}\label {trasformata di Fourier quadricorrente dipolare}
\tilde {j\; }_{nr}^{\, (dip)\, ,\; (\, \epsilon\, )\,\, \mu\, }(\, k\, )\, =\:\, i\;\, 
\tilde {\rho\; }(\, \mathbf {k}\, )\,\,\, (\; \frac {\tilde {v}_{\, in}^{\; \mu}} 
{\, -\, k_{\: 0}+\, i\; \epsilon\, }+\frac {\tilde {v}_{\, out}^{\; \mu}} 
{\, k_{\: 0}+\, i\; \epsilon\, }\; )\quad\quad\quad\quad
\end {equation}
being the Fourier transform of the four-current (\ref {quadricorrente dipolare epsilon}).
Recalling 
(\ref {supporto commutatore Pauli-Jordan}) and taking the 
adiabatic limit we obtain (\ref {correzioni radiative secondo ordine dipolare FGB Fourier}). 
\end {proof}
\begin {remark}\label {remark 5}
Notice that (\ref {correzioni radiative secondo ordine Coulomb}), (\ref {correzioni radiative secondo ordine 
dipolare Coulomb}) have the same functional dependence upon the respective Coulomb-gauge currents. 
In contrast, while (\ref {correzioni radiative secondo ordine}) depends upon the four-current 
(\ref {quadricorrente classica}), the amplitude (\ref {correzioni radiative secondo ordine 
dipolare FGB}) only involves the space components of (\ref {corrente dipolare 
quadrivettoriale}).
\end {remark}
With the aid of the Feynman rules discussed above, accounting for the non-relativistic electronic 
motion and for the introduction of a dipole approximation, and proceeding as in the 
analysis carried out in \cite {YFS61}, one obtains the overall soft-photon radiative 
corrections to the process $\, \mathbf {p}_{\, in}^{}\rightarrow\mathbf {\, p}_{\, out\, }^{}$ 
in the form of exponentiation of the second-order results (\ref {correzioni 
radiative secondo ordine dipolare Coulomb}), (\ref {correzioni radiative secondo ordine dipolare 
FGB}).
In order to show that such corrections admit an operator formulation, we need the following
preliminar result: 
\begin {lemma}\label {Lemma 7}
The scattering matrices (\ref {matrice S Coul dip}), (\ref {matrice di scattering PFBR}) can be expressed 
respectively as
\begin {equation}\label {relazione matrice S Coulomb}
S_{\: \lambda\, ,\,\, \mathbf {p}_{\, out}^{}\; \mathbf {p}_{\, in}^{}}^{\; (Coul)\, ,\,\, (dip)}\, =\;\, \exp\; (\; i\; e
\int_{\, \lambda}^{}\,\, d^{\,\, 4}\, x\;\;\, \mathbf {j}_{\; C\, ,\,\, nr}^{\: (dip)}\, (\, x\, )\, \cdot\,
\mathbf {A}_{\, C}^{}\: (\, x\, )\, )\; ,\,\, 
\end {equation}
\begin {equation}\label {relazione matrice S Feynman}
S_{\: \lambda\, ,\,\, \tilde {v\, }_{out}^{}\; \tilde {v\, }_{in}^{}}^{\; (FGB)\, ,\; (dip)}\, =\;\, \exp\; (\; i\; e\int_{\, \lambda}^{}
\,\, d^{\,\, 4}\, x\;\,\, \mathbf {j}_{\; nr}^{\, (dip)}\: (\, x\, )\, \cdot\, \mathbf {A}\, (\, x\, )\, )\; .
\quad\quad
\end {equation}
\end {lemma}
\begin {proof}
In order to get (\ref {relazione matrice S Coulomb}) it suffices to follow the same steps 
leading to (\ref {matrice S BN Coulspaziale}).\\ 
Concerning the proof of (\ref {relazione matrice S Feynman}), the explicit 
expressions (\ref {Moller Feynman PFBR}), (\ref {f Moll PFBR}) yield
\begin {equation}
S_{\: \lambda\, ,\,\, \tilde {v\, }_{out}^{}\; \tilde {v\, }_{in}^{}}^{\; (FGB)\, ,\; (dip)}\, =
\;\, \exp\,\, (\; i\; e\,\, [\,\, a_{\, \lambda}^{\; r\,\, \dagger}\: 
(\, f_{\,\, \tilde {v\, }_{out}^{}}^{\,\, r}-f_{\,\, \tilde {v\, }_{in}^{}}^{\,\, r}\, )\, +
\, a_{\, \lambda}^{\; r}\; (\, \overline {f\, }_{\tilde {v\, }_{out}^{}}^{\; r}-
\overline {f\, }_{\tilde {v\, }_{in}^{}}^{\; r}\, )\, ]\, )\: ,
\end {equation}
whence we obtain (\ref {relazione matrice S Feynman}), again by inverse Fourier transform.
\end {proof}
\begin {remark}\label {remark 6}
A proof of (\ref {relazione matrice S Feynman}) in which the role of functional form of the 
four-current (\ref {corrente dipolare quadrivettoriale}) is more clearly displayed is as 
follows.
Proceeding as for the derivation of formula (\ref {matrice S BN spaziale adiabatica}), we get 
\begin {eqnarray}
S_{\: \lambda\, ,\,\, \tilde {v\, }_{out}^{}\; \tilde {v\, }_{in}^{}}^{\; (FGB)\, ,\; (dip)\, ,\; (\, \epsilon\, )}\, =
\;\, \tilde {c}_{\; \tilde {v\, }_{out}^{}\, ,\,\, \tilde {z}}^{\; (\, \epsilon\, )\, -\, 1}\;\,\, 
\tilde {c}_{\; \tilde {v\, }_{in}^{}\, ,\,\, \tilde {z}}^{\; (\, -\, \epsilon\, )}\,\,\, 
\exp\; (\, -\ i\; e^{\, 2}\; \tilde {v\, }_{out}^{\, 2}\,\, 
\frac {\: \tilde {z}\, } {2\,\, \epsilon\; }\,\, )
\;\, \nonumber\\
\times\; \exp\; (\, -\ i\; e^{\, 2}\; \tilde {v\, }_{in}^{\, 2}
\,\, \frac {\: \tilde {z}\, } {2\,\, \epsilon\; }\,\, )\;\, 
S_{\; \lambda}^{}\; [\,\, j_{\; nr}^{\: (dip)\, ,\; (\, \epsilon\, )}\; ]\; .
\end {eqnarray}
Let
\begin {equation}
I_{\; \lambda}^{\; (\, \epsilon\, )}\, \equiv\,\, \int_{\, \lambda}^{}\,\, d^{\,\, 4}\, x
\;\,\, j_{\; nr}^{\: (dip)\, ,\; (\, \epsilon\, )\,\, \mu\, =\, 0}\; (\, x\, )
\,\,\, A^{\,\, 0\,\, }(\, x\, )\, .
\end {equation}
It is straightforward to show that
\begin {equation}\label {relazione limite}
\tau_{\; w}^{}-\lim_{\, \epsilon\,\, \rightarrow\; 0\, }\,\, \int\,\, d\,\, t\;\,\, 
e^{\, -\, \epsilon\; \vert\, t\, \vert}\;\, A^{\,\, 0\,\, }(\, \mathbf {x}\, ,
\; t\, )\, =\; 0\, .\quad
\end {equation}
Since $\, j_{\; nr}^{\: (dip)\,\, \mu\, =\, 0}\, (\, x\, )=
\rho\, (\, \mathbf {x}\, )=j_{\; nr}^{\: (dip)\,\, \mu\, =\, 0}\, (\, \mathbf {x}\, )
\, ,$ (\ref {relazione limite})
implies $\, \tau_{\; w}^{}-\, \lim_{\, \epsilon\, \rightarrow\, 0}\,\, I_{\; \lambda}^{\; (\, \epsilon\, )}=\, 0\, ,$ hence (\ref {relazione matrice S Feynman}) is proven.
\end {remark}
By means of the same calculations leading to the proof of (\ref {correzioni radiative hamiltoniane 
Coul}) and by Lemma \ref {Lemma 7} we immediately obtain the following
\begin {corollary}\label {corollario 6}
In the presence of a dipole approximation, the overall soft-photon radiative corrections to the process 
$\, \mathbf {p}_{\, in}^{}\rightarrow\mathbf {\, p}_{\, out}^{}$ in the Coulomb gauge are reproduced by 
the vacuum expectation
\begin {equation}\label {correzioni radiative hamiltoniane dipolari Coul}
\langle\; \Psi_{\, F\; }^{},\: S_{\: \lambda\, ,\:\, \mathbf {p}_{\, out}^{}\; \mathbf {p}_{\, in}^{}}^{\; (Coul)\, ,\,\, (dip)}\,\, \Psi_{\, F}^{}\: \rangle\, =\lim_{\epsilon\,\, \rightarrow\; 0}\,\, \langle\; \Psi_{\, F\; }^{},\: S_{\: \lambda\, ,\:\, \mathbf {p}_{\, out}^{}\; \mathbf {p}_{\, in}^{}}^{\; (Coul)\, ,\,\, (dip)\, ,\; (\, \epsilon\, )}\,\, \Psi_{\, F}^{}\: \rangle\, =\;\, \exp\; (\; e^{\: 2}\, \mathscr {M}_{\; \lambda\, ,\:\, \mathbf {p}_{\, out}^{}\; \mathbf {p}_{\, in}^{}}^{\; (Coul)\, ,\,\, (dip)}\; )\: ,
\end {equation}
for each value of the low-energy cutoff $\, \lambda\, .$ 
\end {corollary}
Likewise, proceeding as for the proof of (\ref {correzioni radiative hamiltoniane}) and taking again
into account Lemma  \ref {Lemma 7}, we obtain the corresponding result in the $\, FGB\, $ gauge:
\begin {corollary}\label {corollario 7}
In the presence of a dipole approximation, the overall $\, FGB$-gauge soft-photon radiative corrections 
to the process $\, \mathbf {p}_{\, in}^{}\rightarrow\mathbf {\, p}_{\, out}^{}$ are given by the 
vacuum expectation
\begin {equation}\label {correzioni radiative hamiltoniane dipolari}
\langle\; \Psi_{\, 0\; }^{},\: S_{\: \lambda\, ,\:\, \tilde {v\, }_{out}^{}\; \tilde {v\, }_{in}^{}}^{\; (FGB)\, ,\,\, (dip)}\,\, \Psi_{\, 0}^{}\: \rangle\,
=\lim_{\epsilon\,\, \rightarrow\; 0}\,\, \langle\; \Psi_{\, 0\; }^{},\: S_{\: \lambda\, ,\:\, \tilde {v\, }_{out}^{}\; \tilde {v\, }_{in}^{}}^{\; (FGB)\, ,\,\, (dip)\, ,\; (\, \epsilon\, )}\,\, \Psi_{\, 0}^{}\: \rangle\,
 =\;\, \exp\; (\; e^{\: 2}\, \mathscr {M}_{\; \lambda\, ,\:\, \mathbf {p}_{\, out}^{}\; \mathbf {p}_{\, in}^{}}^{\; (FGB)\, ,\,\, (dip)}\; )\; ,
\end {equation}
for each value of the infrared cutoff $\, \lambda\, .$ 
\end {corollary}
Finally, we state a Lemma concerning the discrepancy between the vacuum expectations 
(\ref {correzioni radiative hamiltoniane dipolari Coul}), (\ref {correzioni radiative hamiltoniane dipolari}). 
\begin {lemma}\label {Lemma 8}
The following relation holds:
\begin {eqnarray}\label {discrepanza correzioni}
\langle\; \Psi_{\, 0\; }^{},\: S_{\: \lambda\, ,\:\, \tilde {v\, }_{out}^{}\; \tilde {v\, }_{in}^{}}^{\; (FGB)\, ,\,\, (dip)}\,\, \Psi_{\, 0}^{}\: 
\rangle\, =\,\, \langle\; \Psi_{\, F\; }^{},\: S_{\: \lambda\, ,\:\, \mathbf {p}_{\, out}^{}\; \mathbf {p}_{\, in}^{}}^{\; (Coul)\, ,\,\, (dip)}
\,\, \Psi_{\, F}^{}\: \rangle\,\,\, \exp\; (\; i\; e\int_{\, \lambda}^{}\,\, d^{\,\, 4}\, x\;\,\, \xi\, (\, x\, )\;\,\,
\nonumber\\
\times\int\, d^{\,\, 3}\, y\;\,\, G_{\, \mathbf {y}}^{}\: (\, \mathbf {x}\, )\;\, 
(\: \partial_{\: m}^{}\; A^{\; m}\, )\, (\: x_{\: 0}^{}\, ,\; \mathbf {y}\: )\, )\; ,
\end {eqnarray}
\begin {equation}\label {discrepanza quantitativa}
\xi\: (\, x\, )\, \equiv\;\, \partial_{\; l}^{}\;\, \mathbf {j}_{\; nr}^{\: (dip)\,\, l}\: (\, x\, )\, =
\,\, \partial_{\: \mu}^{}\; j_{\; nr}^{\: (dip)\; \mu}\: (\, x\, )\: .\quad\;\,
\end {equation}
\end {lemma}
\begin {proof}
Taking into account Lemma \ref {Lemma 7} and the transversality of $\, \mathbf {j}_{\; C\, ,\,\, nr\, }^{\: (dip)},$ 
the $S$-matrix (\ref {matrice S Coul dip}) can be written as
\begin {equation}\label {relazione matrice S Coulomb riscritta}
S_{\; \lambda\, ,\,\, \mathbf {p}_{\, out}^{}\; \mathbf {p}_{\, in}^{}}^{\: (Coul)\, ,\,\, (dip)}\, =\;\, \exp\; (\, i\; e
\int_{\, \lambda}^{}\,\, d^{\,\, 4}\, x\;\;\, \mathbf {j}_{\; C\, ,\,\, nr}^{\: (dip)}\, (\, x\, )\, \cdot\,
\mathbf {A}\: (\, x\, )\, )\; .\quad\quad
\end {equation}
Corollary \ref {corollario 5} can thus be applied to the scattering matrices (\ref {matrice S Coul dip}), 
(\ref {matrice di scattering PFBR}); as a result, we obtain
\begin {eqnarray}\label {relazione matrici S dipolari}
S_{\: \lambda\, ,\:\, \tilde {v\, }_{out}^{}\; \tilde {v\, }_{in}^{}}^{\; (FGB)\, ,\,\, (dip)}\, |_{\, \mathscr {H}_{\, phys}^{}}^{}=
\,\,\, T^{\; -\, 1}\,\, S_{\: \lambda\, ,\:\, \mathbf {p}_{\, out}^{}\; \mathbf {p}_{\, in}^{}}^{\; (Coul)\, ,\,\, (dip)}\,\,\, T\,\,\, 
\exp\; (\; i\; e\int_{\, \lambda}^{}\,\, d^{\,\, 4}\, x\;\,\, \xi\: (\, x\, )\;\,\,
\nonumber\\
\times\int\, d^{\,\, 3}\, y\;\,\, G_{\, \mathbf {y}}^{}\, (\, \mathbf {x}\, )\;\, 
(\; \partial_{\: m}^{}\; A^{\; m}\, )\, (\: x_{\: 0}^{}\, ,\; \mathbf {y}\: )\, )\; ,
\end {eqnarray}
hence (\ref {discrepanza correzioni}) follows.
\end {proof}

We wish to give a few comments about the above results, already stated without 
proof and briefly discussed at the end of Section \ref {sect:1}.

The first statement in Remark \ref {remark 5} and the exponentiation of the order-$e^{\, 2}$ corrections 
(\ref {correzioni radiative secondo ordine dipolare Coulomb}) show that the 
dipole approximation is compatible with the low-energy assumptions made in 
the perturbation-theoretic treatment of \emph {Coulomb-gauge} infrared $\, QED\, ;\, $
further, the exponentiation of (\ref {correzioni radiative secondo ordine dipolare Coulomb}) 
can be recovered within a Hamiltonian framework (Corollary \ref {corollario 6}).
The overall $\, FGB$-gauge soft-photon radiative corrections also admit an operator 
formulation (Corollary \ref {corollario 7}); however, a discrepancy arises between 
the explicit expression of such corrections, equation (\ref {correzioni radiative 
hamiltoniane dipolari}), and the 
Coulomb-gauge result (\ref {correzioni radiative hamiltoniane dipolari Coul}). 

The second statement in Remark \ref {remark 5} suggests that such a discrepancy should 
be due to issues arising in the $FGB\, $ gauge formulation. By Remark \ref {remark 6}, 
it is in fact related to the functional form of the four-vector current (\ref {corrente 
dipolare quadrivettoriale}), specifically, by virtue of Lemma \ref {Lemma 8}, to 
its non-conservation, 

The violation of the continuity equation in the presence of a dipole approximation was already  
pointed out in \cite {HiSu09} and shown to cause difficulties with the Gupta-Bleuler condition 
in a specific non-relativistic model; to the best of our knowledge, its implications for the 
lack of invariance of the soft-photon corrections have however not been stressed before.

\section* {Outlook}

\ni
By exploiting the Hamiltonian control of the soft-photon contributions to the Feynman-Dyson expansion 
of $\, QED\, ,$ achieved through the four-vector Bloch-Nordsieck model, the problem of the removal of 
the infrared cutoff in the perturbative expressions can also be addressed.
In particular, it should be possible to outline the extent and limitations of the recipe leading to 
infrared-finite cross-sections, by exploiting the fact that the approach here developed 
allows to reproduce the results of the order-by-order diagrammatic treatment in a compact 
way.
We plan to report on these problems in a future work.

\section* {Acknowledgments}

\ni
The content of this work is a development of part of my Ph.D. thesis at the department of Physics 
of the University of Pisa.
I am grateful to Giovanni Morchio and Franco Strocchi for having suggested to me the problem 
and for having shared with me ideas stemming from their preliminary analysis of Hamiltonian 
models, devoted to a better understanding of the role of local and covariant formulations in 
the treatment of the infrared problem in Quantum Electrodynamics.
I am indebted to G. Morchio for extensive discussions on these topics, for his helpful advice 
and for a number of comments on the manuscript. 
The suggestions of an anonymous referee are also gratefully acknowledged.

\begin{appendices}

\section {Essential Self-Adjointness Of The Pauli-Fierz Hamiltonian}
\label {app:1}

Let $\, b_{\: s}^{}\, (\, f\, )\, ,\, f\in L^{\: 2\, },$ stand for either $\, a_{\, s\: }^{}(\, f\, )$ or $\, a_{\, s\: }^{\: *}(\, f\, )\, ;\, $ 
by virtue of the standard Fock-space estimate $\Vert\, b_{\: s}^{}\, (\, f\, )\, \Psi\, \Vert\leq\Vert\, f\, 
\Vert_{\: 2}^{}\; \Vert\, (\, N+\; \mathbb {I}\, )^{\: 1/\, 2}\,\, \Psi\, \Vert\, ,$ $\forall\; \Psi\in Dom\, 
(\, N^{\: 1 /\, 2}\, )\, ,$ one obtains
\begin {equation}\label {bound potenziale}
\Vert\, \mathbf {A}_{\, C\, ,\; \lambda}^{\, i}\; (\, \rho\, ,\, \mathbf {x}=\, 0\; )\,\, \Psi\, \Vert\, \leq\, c\, (\, \rho\: )\,\, 
\Vert\, (\, N +\, \mathbb {I}\, )^{\; 1 /\, 2}\;\, \Psi\, \Vert\; ,\; \forall\; \Psi\in D_{\, F_{\: 0}^{}}^{},
\end {equation}
with $\, c\, (\, \rho\, )\, $ a (positive) constant, for a given form factor $\, \rho\, .$

The relation $\, 2\; \Vert\, (\, A\, \otimes\, B\, )\; \Phi\, \Vert\leq\Vert\, (\, A^{\: 2}\otimes\, \mathbb {I}\, +\, 
\mathbb {I}\, \otimes\, B^{\: 2}\, )\; \Phi\, \Vert\, ,\, \forall\, \Phi\in D_{\; 0}^{}\, ,$ applied to the operators 
$A=\mathbf {p}\, ,\, B=\, (\, N+\mathbb {\; I}\, )^{\; 1/\, 2}\, ,$ and 
the estimate
$\Vert(\, N+\mathbb {\; I}\, )\,\, \Psi\, \Vert\leq\Vert\, (\, \lambda^{\, -\, 1\; }H_{\: 0\, ,\,\, C\; }^{\; e.\, m.}+
\mathbb {\; I}\, )\; \Psi\, \Vert\, ,$ $\forall\, \Psi\in D_{\, F_{\, 0}^{}},$
yield the bound
\begin {equation}\label {bound autoaggiunzione}
\Vert\, H_{\, \lambda}^{\; (PFB)}\,\, \Phi\, \Vert\, \leq\, d\; (\, e\, ,\, \lambda\, ;\, \rho\, )
\,\, \Vert\, (\, H_{\; 0}^{}+\, \mathbb {I}\, )\,\, \Phi\, \Vert\: ,\; \forall\; \Phi\in D_{\; 0}\, ,
\end {equation}
for a suitable $\, d\: (\, e\, ,\, \lambda\, ;\, \rho\, )\, .\, $
Further, with the aid of (\ref {bound autoaggiunzione}) and of the $CCR$, one finds $\, g\, 
(\, e\, ,\, \lambda\, ;\, \rho\, )\, $ such that $\vert\, (\, \Phi\, ,\, [\, H_{\, \lambda}^{\: (PFB)},
\, H_{\; 0\; }^{}]\: \Phi\, )\, \vert\leq g\, (\, e\, ,\, \lambda\, ;\, \rho\, )\; \Vert(\, H_{\; 0}^{}+
\mathbb {\: I}\, )^{\: 1/\, 2}\,\, \Phi\, \Vert\, ,$ $\forall\; \Phi\in D_{\; 0\; }^{};\, $ 
$H_{\: \lambda}^{\: (PFB)}$ is thus e.s.a. on $\, D_{\; 0}^{}\, ,$ for all 
$\, \vert\: e\: \vert\, ,\, \lambda>0\, ,\, $ by Nelson's commutator 
theorem in the formulation given in \cite {FL74}.

\section {Dynamics Of Solvable Models On Indefinite-Metric Spaces}
\label {app:2}

For models employing four-vector potentials, lack of positivity of the scalar product raises 
substantial questions on selfadjointness and existence and uniqueness of time evolution.
In \cite {HiSu09} such problems have been treated on a slightly different version of the model, and a 
general framework has been provided for existence and uniqueness of the Heisenberg time-evolution
within a Hilbert-space formulation.

Since, however, the \emph {formal} evolution operators defined by the Hamiltonians 
(\ref {hamiltoniano FGB dipolare}), (\ref {hamiltoniano Feynman cl}) are exponentials 
of the canonical variables of the photon field, we adopt a more pragmatic approach; in 
particular, the evolution operators will be defined in terms of Weyl exponentials of 
fields, introduced starting from their algebraic relations, on a suitable invariant 
vector space.

Given a linear \emph {non-positive} and normalized functional $\, \omega\, $ on 
a unital ${}^{*}$-algebra $\, \mathcal {A}\, ,$ fulfilling the hermiticity property $\, \omega\, (A^{\, *})=\, 
\overline {\omega\, (A\, )\, },\, \forall\, A\in\mathcal {A\, },\, $ it is possible to apply a generalized 
$\, GNS$ reconstruction procedure (\cite {MS00}); as a result, one obtains a non-degenerate indefinite 
vector space carrying a ${}^{*}$ representation of $\, \mathcal {A}\, ,$ with the 
involution represented by the indefinite-space adjoint $\, {}^{\dagger\, },$ and expectations 
over a cyclic vector representing $\, \omega\, .$ 

Let now $\, \mathscr {A}_{\,\, ext}^{\:\, e.\, m.}\, $ be the unital ${}^{*}$-algebra generated by the photon 
canonical variables and by variables (Weyl operators in momentum space) 
$\, W\, (\, g\, ,\, h\, )\, ,\, $ indexed by four-vector real-valued functions in 
$\, L^{\, 2\; }(\, \mathbb {R}^{\, 3}\, )\, ,\, $ fulfilling
\begin {equation}\label {proprieta' W}
W\, (\, g\; ,\: h\, )^{\, *}=\; W\, (\, -\, g\: ,-\, h\, )\; ,\; 
W\, (\, g\; ,\: h\, )^{\, *}\;\, W\, (\, g\; ,\: h\, )\, =\, 1\: ,
\quad\quad\quad\quad\quad\quad\quad\quad\;
\end {equation}
\begin {equation}\label {commutatore W}
W\, (\, g\; ,\: h\, )\;\, W\, (\; l\; ,\, m\, )\, =\;\, \exp\; (\; i\,\, [\, \langle\, g\; ,\, m\; \rangle-
\langle\, h\: ,\: l\; \rangle\, ]\, )\;\,\, W\, (\; l\; ,\, m\, )\;\, W\, (\, g\; ,\: h\, )\, ,\;\;
\end {equation}
\begin {equation}\label {commutatore a W}
[\; a\, (\, \overline {f\, }\, )\, ,\, W\, (\, g\; ,\: h\, )\, ]\, =\; \frac {\, i} {\sqrt{\, 2\, }}\;\, 
\langle\, f\, ,\: n\; \rangle\;\,\, W\, (\, g\; ,\: h\, )\, ,\; n\, \equiv\,\, g\, +\, i\; h\; ,
\quad\quad\quad\quad\quad\quad\;
\end {equation}
with the symbol $\, {}^{*}\, $ standing for the algebra involution and with
\begin {equation}\label {prodotto indefinito}
\langle\, f\, ,\, g\; \rangle\, \equiv\,\, (\, f^{\; 0},\, g^{\; 0}\, )\, -\, \sum_{i\, }\,\,\, (\, f^{\; i\, },\, g^{\; i}\, )\: .
\quad\quad\quad\quad
\end {equation}
Let also $\, \mathscr {A}_{\,\, ext\, ,\; \mathscr {S}}^{\:\, e.\, m.}\, $ be the algebra generated by the photon 
variables introduced above, smeared with functions in $\, \mathscr {S}\, (\, \mathbb {R}^{\, 3}\, )\, ,\, $
and denote by $\, \omega_{\; F}^{}\, $ a hermitian linear (non-positive) functional on 
$\, \mathscr {A}_{\,\, ext}^{\,\, e.\, m.},\, $$\mathscr {A}_{\,\, ext\, ,\; \mathscr {S}}^{\,\, e.\, m.}\, ,$
with Fock-type expectations
\begin {equation}\label {propagatore Fock}
\omega_{\; F}^{}\; (\; a\, (\, \overline {f\, }_{1}\, )\;\; a^{\, *}\, (\, f_{\; 2}^{}\, ))\, =
\, -\, \langle\, f_{\; 1}^{}\: ,\, f_{\; 2}^{}\; \rangle\, ,
\quad\quad\quad\quad\quad\quad\quad
\end {equation}
\begin {equation}\label {aspettazione esponenziale}
\omega_{\; F}^{}\; (\, W\, (\, g\: ,\, h\; ))\, \equiv\;\, \exp\; (\,\, \frac {1} {4}
\:\, (\langle\, g\: ,\, g\; \rangle+\langle\; h\, ,\, h\; \rangle)\, )\: .\;\;\;\,
\quad\quad
\end {equation}
The functional $\, \omega_{\; F}^{}$ is identified by (\ref {propagatore Fock}), (\ref {aspettazione esponenziale}),
since expectations of monomials of $\, a\, $ and $\, a^{\, *\, }$ can be expressed in terms of 
(\ref {propagatore Fock}) with the aid of Wick's theorem (\cite {Wick50}), while those 
of monomials of $\, W\, $ are fixed by (\ref {aspettazione esponenziale}) up to a 
phase factor, given by (\ref {commutatore W}), and the other ones follow from 
(\ref {commutatore a W}).

The spaces $\, \mathscr {G}$ and $\, \mathscr {G}_{\; 0}^{}$ are obtained via generalized $\, GNS$ 
constructions, applied respectively to $\, \omega_{\; F}^{}\, (\, \mathscr {A}_{\,\, ext}^{\:\, e.\, m.}\, )\, $ and 
$\, \omega_{\; F}^{}\, (\, \mathscr {A}_{\,\, ext\, ,\; \mathscr {S}}^{\:\, e.\, m.}\, )\, .\, $
The space $\, \mathscr {G}_{\,\, 0}^{}$ 
is weakly dense in $\, \mathscr {G}$ 
by density of the Schwartz space in $\, L^{\, 2 }$ and by the Cauchy-Schwartz 
inequality, applied component-wise to explicit expression of the inner 
product.

Alternatively, starting from representation spaces $\, \mathscr {D}$ obtained by applying 
the generalized $\, GNS\, $ theorem to polynomial ${}^{*}$-algebras, we might have tried to identify 
completions of $\, \mathscr {D}$ containing Weyl exponentials, with the help of a suitable 
topology.
In this respect, we recall that an indefinite-metric space resulting from a (generalized) $\, GNS\, $ construction
does not admit a unique completion, besides not being complete.
In some generality, complete spaces can be constructed as Hilbert-space completions of vector spaces 
obtained via a $\, GNS$ procedure and Hilbert-space structures can be relevant for the existence and 
control of limits (the role of such structures in models of indefinite-metric quantum field theories has 
been discussed in \cite {MS80}).

A different choice, which is the one adopted in this paper, is to employ a vector space.
In the case of models only involving polynomials and exponentials of fields, such a choice looks 
simpler and even more intrinsic; in particular, a strong topology is solely required in order to 
formulate and settle uniqueness of time-evolution operators.

In order to prove uniqueness of the evolution operators, we need a preliminary result.
\begin {lemma}\label {Lemma 9}
Isometries $\, U$ of a non-degenerate indefinite space $\, \mathscr {Z}$ are identified by their 
restriction to a weakly dense subspace $\, \mathscr {Y}.\, $
\end {lemma}
\begin {proof}
One has in fact
$$\langle\: y\, ,\, U\, x\, \rangle\, =\; \langle\; U^{\, -\, 1}\; y\, ,\,  x\, \rangle\, =
\;\, \lim_{n\, }\,\, \langle\; U^{\, -\, 1}\; y\, ,\,  x_{\, n}^{}\, \rangle\, =\;\, 
\lim_{n\, }\,\, \langle\, y\, ,\, U\; x_{\, n}^{}\, \rangle\, ,$$
with $\, y\in\mathscr {Z}$ and $\, x_{\, n\, }^{}$ a sequence of elements of $\, \mathscr {Y}$ (weakly) 
converging to $\, x\, .$ 
\end {proof}
Uniqueness is then a consequence of the following observation, only requiring the 
existence of a positive scalar product $\, (.\, ,.)\, $ majorizing the indefinite product:
\begin {lemma}\label {Lemma 10}
Two one-parameter families of isometries $\, U\, (\, a\, )\, ,\, V\, (\, a\, )\, $ of a vector space $\, V_{\,\, 0}^{}\, ,$ 
endowed with a non-degenerate indefinite inner product $\, \langle\, .\, ,.\, \rangle\, $ admitting a majorizing 
positive scalar product, coincide if they are differentiable on $\, V_{\,\, 0}^{}$ in the corresponding strong 
topology, with the same derivative $\, -\, i\; H\, (\, a\, )\, .$ 
\end {lemma}
\begin {proof} 
One has, $\, \forall\; x\, ,\, y\, \in\, V_{\,\, 0}^{}\, $,  
\begin {eqnarray}
\frac {\; d} {d\; a\, }\;\, \langle\, x\, ,\; V\, (\, -\, a\, )\;\, U\, (\, a\, )\,\, y\, \rangle\, = 
\; \frac {\; d} {d\; a\, }\;\, \langle\; V\, (\, a\, )\,\, x\, , \, U\, (\, a\, )\,\, y\, \rangle
=\, -\; i\;\, \langle\; V\, (\, a\, )\,\, x\, ,\,  H\, (\, a\, )\;\, U\, (\, a\, )\,\, y\, \rangle\,
\nonumber\\
+\; i\;\, \langle\, H\, (\, a\, )\;\, V\, (\, a\, )\,\, x\, ,\; U\, (\, a\, )\,\, y\, \rangle\, =\; 0\; ,
\nonumber
\end {eqnarray}
since strong differentiability implies strong continuity and hermiticity of $\, H\, (\, a\, )\, $ on $\, V_{\,\, 0}^{}\, ,$ 
which by assumption is invariant under $\, U\, (\, a\, )\, ,\, V\, (\, a\, )\, .$
\end {proof}

\section {Gupta-Bleuler Quantization Of The Free Electromagnetic Field}
\label {app:3}

Within the Gupta-Bleuler quantization of Electrodynamics (\cite {Gup50, Ble50}), physical states are 
selected by the linear subsidiary condition\footnote {The decomposition $\, (\, \partial\, \cdot\, A\, )\, 
(\, x\, )=\, (\, \partial\, \cdot\, A\, )_{\, +}^{}\, (\, x\, )+(\, \partial\, \cdot\, A\, )_{\, -}^{}\, (\, x\, )$ is well 
defined, since $\, (\, \partial\, \cdot\, A\, )\, (\, x\, )\, $ is a free field in the $\, FGB\, $ gauge.} 
\begin {equation}\label {condizione Gupta}
(\, \partial\: \cdot\, A\, )_{\, -}^{}\, (\, x\, )\;\, \Psi\, =\,\, 0\; .
\end {equation}
In the free-field case, the solutions of the subsidiary condition (\ref {condizione Gupta}) 
in $\, \mathscr {G}$ define a subspace
\begin {equation}\label {spazio fisico ampliato}
\mathscr {H}\; '\, =\,\, \mathscr {A}_{\,\, ext\, ,\,\, tr}^{\,\, e.\, m.}\,\, \Psi_{\, 0}^{}\, ,\quad\;
\end {equation}
with $\mathscr {A}_{\,\, ext\, ,\,\, tr}^{\,\, e.\, m.}$ the algebra generated by the canonical photon variables 
and their Weyl exponentials, smeared with four-vector test functions 
$\, f^{\, \mu\; }(\, \mathbf {k}\, )\in L^{\, 2\; }(\, \mathbb {R}^{\, 3\, })\, $ 
obeying the transversality condition 
\begin {equation}\label {funzioni test trasverse}
\overline {k}^{\,\, \mu\, }f_{\, \mu}^{}\, (\, \mathbf {k}\, )\, =\; 0\, ,\, \overline {k}^{\,\, \mu}
\equiv\; (\, \vert\, \mathbf {k}\, \vert\, ,\, \mathbf {k}\; )\: .
\end {equation}
$\mathscr {\, H}\; '\, $ contains a null space
\begin {equation}\label {spazio nullo}
\mathscr {H\; }''=\,\, (\, \Psi\in\mathscr {H}\; '\, ,\; \langle\, \Psi\, ,\, \Psi\, \rangle=\; 0\; )\: ,\;\;
\end {equation}
whose vectors result from the application to $\Psi_{\, 0}^{}$ of those elements of 
$\, \mathscr {A}_{\,\, ext\, ,\; tr}^{\; e.\, m.}$ which are indexed by test functions 
$\, f^{\, \mu}\, (\, \mathbf {k}\, )=\; \overline {k}^{\,\, \mu}\, h\, (\, \mathbf {k}\, )\, .\, $ 

Since $\mathscr {\, H}\; '\, $ is endowed with a positive-semidefinite product 
$\, \langle\, .\, ,.\, \rangle_{\; -}^{}\equiv\, -\, \langle\, .\, ,.\, \rangle\, ,\, $
the space $\, \mathscr {H}\, '/\, \mathscr {H}\, '',\, $ obtained by introducing 
equivalence classes in $\mathscr {\, H}\; ',\, $ is a pre-Hilbert space with scalar product 
$\, \langle\, .\, ,.\, \rangle_{\; -}^{}\, .$\\ 
We use the symbol
\begin {equation}\label {spazio fisico}
\mathscr {H}_{\; phys\, }^{}\equiv\;\, \overline {\mathscr {H}\: '\, /\; \mathscr {H}\; ''\; }
\quad\quad\quad\quad
\end {equation}
for the Hilbert space of photon states obtained by completion in the topology of 
$\, \langle\, .\, ,.\, \rangle_{\; -\, }^{}.\, $ 
The states of $\, \mathscr {H}_{\; phys}^{}\, $ have a simple characterization.
\begin {lemma}\label {Lemma 11}
$\, \mathscr {H}_{\; phys}^{}$ is isomorphic as a Hilbert space to $\, \mathscr {H}_{\; C\, }^{},$ the 
Coulomb-gauge space of free photon states. 
\end {lemma}
\begin {proof}
Let $\, g^{\, \mu\, }(\, \mathbf {k}\, )\in L^{\, 2}\, (\, \mathbb {R}^{\, 3}\, )\, ,$  
$\overline {\, k}_{\, \mu}^{}\; g^{\: \mu}\, (\, \mathbf {k}\, )=\, 0\, ,$ 
$g_{\: C}^{\; \mu}\, (\, \mathbf {k}\, )\equiv\, g^{\: \mu}\, (\, \mathbf {k}\, )-(\; \overline {k}^{\; \mu}/\; \overline {k}^{\; 0}\, )
\,\, g^{\; 0}\, (\, \mathbf {k}\, )\, ,$
denote the equivalence class of $\, \Psi_{\, g}=\, a^{\, \dagger}\, (\, g\, )
\; \Psi_ {\, 0}^{}$ by $[\: \Psi_{\, g}\: ]\, $ and let $\, (\, .\, ,.\, )\, $ be the 
scalar product of $\, \mathscr {H}_{\; C}^{}\, .\, $
It is straightforward to check that
\begin {equation}
(\: \Phi_{\; \mathbf {g}_{\, C\, }^{}},\: \Phi_{\: \mathbf {h}_{\, C\, }^{}}^{})\; =\,\, \langle
\; \tilde {\Psi}_{\, g}^{}\, ,\, \tilde {\Psi}_{\, h}^{}\, \rangle_{\; -}^{}\, ,\quad\;\;
\end {equation}
where $\, \tilde {\Psi}_{\, g}\equiv\, a^{\: \dagger}\, (\, g_{\: C}^{}\, )\,\, \Psi_ {\, 0\, }^{}$ is a representative 
of $\, [\, \Psi_{\, g}^{}\: ]\, $ and $\, \Phi_{\: \mathbf {g}_{\, C}^{}}\equiv\; a^{\, *}\, (\; \mathbf {g}_{\: C}^{}\, )
\,\, \Psi_ {\, F}^{}\, $ is the vector of $\, \mathscr {H}_{\; C}^{}\, $ corresponding to the (transverse) 
vector test function $\, \mathbf {g}_{\, C}^{}\, ;\, $ hence
\begin {equation}\label {invarianza Hilbert}
(\: \Phi_{\mathbf {\; g}_{\, C\, }^{}},\: \Phi_{\: \mathbf {h}_{\, C\, }^{}}^{})\;
=\,\, \langle\, [\; \Psi_{\, g}^{}\; ]\: ,\; [\; \Psi_{\, h}^{}\; ]\, \rangle_{\; -}^{}\, .
\end {equation}
With the same notations as for the one-photon states, let
$$
\Psi_{\, g_{\; 1}^{}\, ...\,\, g_{\; n}^{}}^{}\equiv\; a^{\, \dagger}\, (\, g_{\; 1}^{}\, )\, ...\; a^{\, \dagger}\, (\, g_{\; n}^{}\, )\; \Psi_{\, 0}^{}\, , \; \Phi_{\mathbf {\; g}_{\, C\, ,\,\, 1}^{}\, ...\,\, \mathbf {g}_{\, C\, ,\,\, n}^{}}\equiv\; a^{\, *}\, (\: \mathbf {g}_{\: C\, ,\,\, 1}^{}\, )\, ...\; a^{\, *}\, (\: \mathbf {g}_{\: C\, ,\,\, n}^{}\, )\,\, \Psi_ {\, F}^{}
$$
and define the linear map 
\begin {equation}\label {isomorfismo Hilbert n fotoni}
T_{\; 0}^{}\,\, [\,\, \Psi_{\, g_{\; 1}^{}\, ...\,\, g_{\: n}^{}}^{}\; ]\; =\,\, 
\Phi_{\mathbf {\; g}_{\, C\, ,\,\, 1}^{}\, ...\,\, \mathbf {g}_{\, C\, ,\,\, n}^{}}\, ,
\quad\quad\quad\;\;\,
\end {equation}
\begin {equation}
T_{\; 0}^{}\,\, [\; \Psi_{\, 0}^{}\; ]\; =\,\, \Psi_{\, F}^{}\, .\quad\quad\quad\quad\quad\quad
\end {equation}
In order to show that the action of $\, T_{\; 0}^{}$ does not depend upon the choice of a representative 
in each equivalence class, it is enough to notice that $\, \mathbf {g}_{\, C}^{}\, (\, \mathbf {k}\, )=\, 0\, $ 
for square-integrable four-vector functions of the form $\, g^{\, \mu}\, (\, \mathbf {k}\, )=\, \overline {k}^{\,\, \mu}\,
h\, (\, \mathbf {k}\, )\, ,$ 
which index the state vectors belonging to the null space of 
$\, \mathscr {G}\, .$\\ 
The equality (\ref {invarianza Hilbert}) for one-photon states implies 
\begin {equation}\label {invarianza Hilbert n fotoni}
(\; T_{\; 0}^{}\,\, [\; \Psi_{\, g_{\; 1}^{}\, ...\,\, g_{\: n}^{}}^{}\; ]\; ,\; T_{\; 0}^{}\,\, [\; \Psi_{\, h_{\, 1}^{}\, ...\; h_{\, n}^{}}^{}\; ]\, )\, =
\; \langle\, [\; \Psi_{\, g_{\: 1}^{}\, ...\,\, g_{\: n}^{}}^{}\; ]\; ,\; [\; \Psi_{\, h_{\, 1}^{}\, ...\; h_{\, n}^{}}^{}\; ]\, \rangle_{\; -}^{}\, .
\end {equation}
By linearity, $\, T_{\; 0}^{}$ can be extended to $B_{\; 0}^{}\, ,$ the dense set of $\mathscr {H}_{\; phys\, }$ 
spanned by the (equivalence classes of) finite-particle vectors of $\, \mathscr {G}$ which describe
transverse photons (namely, indexed by test functions fulfilling (\ref {funzioni test trasverse})); further, 
by \ref {invarianza Hilbert n fotoni}) such an extension is isometric.

Finally, by virtue of the $\, B.\, L.\, T.\, $ theorem (\cite {RS72a}, Theorem I.7), $T_{\; 0}^{}\, $ 
can be uniquely extended to an unitary operator $\, T\, $ from $\mathscr {H}_{\; phys}^{}$ to 
$\, \mathscr {H}_{\; C\, }^{}.$
\end {proof}
\ni
The smeared field $\, \mathbf {A}_{\, T}^{}\, (\, \mathbf {f}\, )\equiv\; T^{\; -\, 1}\; \mathbf {A}_{\, C\: }^{}(\, \mathbf {f}\, )
\;\, T:\, T^{\; -\, 1}\, Dom\, (\, \mathbf {A}_{\, C}^{}\, (\, \mathbf {f}\, )\, )\rightarrow\mathscr {H}_{\; phys}$ fulfills the 
Coulomb-gauge (transversality) condition
\begin {equation}\label {condizione Coulomb}
(\; \partial_{\,\, l}^{}\; \mathbf {A}_{\, T}^{\: l}\, )\, (\, h\, )\, =\; 0\: .
\end {equation}
Next Lemma shows how to express $\mathbf {A}_{\, T}^{}\, (\, \mathbf {f}\, )$ in terms of the smeared (free)
$FGB$-gauge four-vector potential.
\begin {lemma}\label {Lemma 12}
On $\, B_{\; 0\, }^{},$ $\mathbf {A}_{\, T}^{}\, (\, \mathbf {f}\, )\, $ is essentially self-adjoint and fulfills the 
equality
\begin {equation}\label {potenziale trasformato}
\mathbf {A}_{\, T}^{}\, (\mathbf {\; f}\, )\, =\, -\: A_{\: g}^{}\, (\, f\, )\: ,
\end {equation}
where, following \cite {Sym71},
\begin {equation}\label {trasformazione potenziale}
A_{\: g}^{\; \mu\; }(\, x\, )\, \equiv\; A^{\; \mu\; }(\, x\, )\, -\: \partial^{\; \mu}\int\, d^{\:\, 3}\, y
\;\,\, G_{\, \mathbf {y}}^{}\, (\, \mathbf {x}\, )\;\, (\; \partial_{\; m}^{}\; A^{\; m}\, )\,
(\: x_{\: 0}^{}\, ,\: \mathbf {y}\: )\: .
\end {equation}
\end {lemma}
\begin {proof}
Since $\mathbf {\, A}_{\, C}^{}\, (\, \mathbf {f}\, )\, $ is e.s.a. on $\, F_{\; 0}\, $ and, by Lemma \ref {Lemma 11}, 
$B_{\; 0}=\, T^{\, -\, 1\; }F_{\,\, 0}\, ,$ 
$\mathbf {A}_{\, T}^{}\, (\, \mathbf {f}\, )\, $ is e.s.a. on $B_{\; 0\, }.$ 
A necessary condition for (\ref {potenziale trasformato}) to make sense is 
$\, A_{\: g}\, (\, f\, )\, (\, B_{\; 0}\, )\subset\mathscr {H}_{\; phys\, };\, $ such a requirement is 
indeed fulfilled, as $\, B_{\; 0}\subset Dom\: (\, A_{\: g}^{}\, (\, f\, )\, )\, $ and 
$\, A_{\; g}^{\; \mu}\, $ commutes with $\, \partial\, \cdot A\, $
by virtue of the canonical commutation relations (\cite {Sym71}). 
Maxwell's equations and (\ref {funzione di Green Poisson}) yield
\begin {equation}
A_{\: g}^{\,\, 0\; }(\, f_{\; 0}^{}\; )\, =\; A^{\; 0\; }(\, f_{\; 0}^{}\; )\, -\: \int\, d^{\,\, 4}\, x\;\, d^{\:\, 3}\, y
\;\,\, G_{\, \mathbf {y}}^{}\, (\, \mathbf {x}\, )\;\, (\; \partial_{\; m}^{}\; \partial^{\; m}\, A^{\; 0}\, )\,
(\: x_{\: 0}^{}\, ,\: \mathbf {y}\: )\;\, f_{\; 0}^{}\; (\, x\, )\, =\,\, 0\: ,
\end {equation}
whence $A_{\: g}^{}\, (\, f\, )=-\, \mathbf {A}_{\, g}^{}\, (\, \mathbf {f}\, )\, ;\, $
it is therefore enough to prove that $\mathbf {A}_{\, g}\, (\mathbf {\; f}\, )
=\mathbf {\, A}_{\, T}^{}\, (\, \mathbf {f}\, )\, .\, $
We first show that $\, \mathbf {A\, }_{g}^{}\, (\mathbf {\; f}_{\, tr}^{}\, )=\mathbf {A}_{\, T}^{}\, (\mathbf {\; f}_{\, tr}^{}\, )\, $ on 
$B_{\; 0}^{}\, ,$ for test functions $\, \mathbf {f}_{\; tr}^{}\, (\, x\, )\, $ fulfilling $\, \partial_{\; l}^{}\; \mathbf {f}_{\; tr}^{\; l}\, (\, x\, )=\, 0\, ;$  
on $B_{\; 0}^{}$ one in fact has 
$\mathbf {A}_{\, g}^{}\, (\mathbf {\; f}_{\, tr}^{}\, )
=\mathbf {\, A}\, (\mathbf {\; f}_{\, tr}^{}\, )\, ,$
since $\mathbf {\, f}_{\, tr}^{}$ is transverse,
and $\, \mathbf {A}\, (\, \mathbf {f}_{\; tr}^{}\, )=\mathbf {\, A}_{\, T}^{}
\, (\, \mathbf {f}_{\; tr}^{}\, )\, ,$ 
by Lemma \ref {Lemma 11}.
Finally, it follows from (\ref {funzione di Green Poisson}) that
\begin {equation}\label {verifica Coulomb}
(\: \partial_{\,\, l}^{}\; A_{\, g}^{\;\, l\; })\, (\, h\, )\, =\; (\; \partial_{\,\, l}^{}\; A^{\,\, l}\, )\, (\, h\, )\, +
\int\, d^{\,\, 4}\, x\;\, d^{\:\, 3}\, y\;\, \Laplace\; G_{\, \mathbf {y}}^{}\, (\, \mathbf {x}\, )\;\, 
(\; \partial_{\: m}^{}\; A^{\; m}\, )\, (\: x_{\: 0}^{}\, ,\: \mathbf {y}\: )\;\, h\, (\, x\, )\, =\,\, 0\: .
\end {equation}
The Lemma is thus proved.
\end {proof}

\end{appendices}

\end {document}